\documentclass[11pt]{article}
\usepackage{epsfig} 
\newcommand{\sect}[1]{ \section{#1} \setcounter{equation}{0} }

\newcommand{\MSbar}{\overline{\mbox{MS}}} 
 
\newcommand{\Nf}{N_{\!f}}

\begin{document}
\title{Three loop $\MSbar$ operator correlation functions for deep inelastic
scattering in the chiral limit}
\author{J.A. Gracey, \\ Theoretical Physics Division, \\ 
Department of Mathematical Sciences, \\ University of Liverpool, \\ P.O. Box 
147, \\ Liverpool, \\ L69 3BX, \\ United Kingdom.} 
\date{} 
\maketitle 

\vspace{5cm} 
\noindent 
{\bf Abstract.} We compute a variety of operator-operator correlation functions
to third order in the $\MSbar$ scheme in the chiral limit. These include
combinations of quark bilinear currents with gauge invariant operators such as
moments $n$~$=$~$2$ and $3$ of the flavour non-singlet Wilson operators of deep
inelastic scattering and moment $n$~$=$~$2$ of the transversity operator, as 
well as the correlation functions of the latter operators with themselves. The 
explicit values of these gauge independent correlation functions are required 
to assist with non-perturbative matching to lattice regularized calculations of
the same quantities. As part of the computation we determine the mixing matrix 
of renormalization constants of these non-singlet currents with their 
associated total derivative operators at the same twist to three loops in 
$\MSbar$. Such operators are crucial in extracting renormalization constants 
for the operator-operator correlation functions which are consistent with the 
corresponding renormalization group equation. As a by-product we deduce the 
$R$-ratio for the tensor current to third order in the $\MSbar$ scheme. 

\vspace{-20cm}
\hspace{13.5cm}
{\bf LTH 827}

\newpage

\sect{Introduction.}

Lattice regularization of non-abelian gauge theories has provided many 
insights into the non-perturbative properties of the quantum field theory
underlying the strong force, known as Quantum Chromodynamics (QCD). For
instance, now that dynamical quarks can be treated via the use of powerful
supercomputers, the meson and hadron spectra are on the whole in remarkable
agreement with experimental data. One current problem of interest for lattice
regularization is the measurement of Green's functions relevant for, say, deep
inelastic scatttering. For instance, whilst the perturbative renormalization of
the underlying twist-$2$ flavour non-singlet and singlet operators are known to
three loop accuracy in the $\MSbar$ scheme, \cite{1,2,3,4,5,6,7}, the
associated matrix elements are also required but they can only be fully
computed using non-perturbative methods. These are crucial to fully 
understanding the structure functions of the original hadrons and mesons which 
are broken up in deep inelastic experiments. Such matrix elements are, however,
accessible via lattice regularization with notable progress through the years 
via collaborations such as QCDSF, \cite{8,9,10,11,12,13}, and others, 
\cite{14,15,16,17}. However, such measurements of Green's functions on the 
lattice need a variety of techniques to allow comparison with continuum 
results. Therefore, we will first briefly discuss some of the relevant issues 
in a general context before formulating the aims for the current article. 

The first issue is that the lattice computations necessarily renormalize their 
operators and Green's functions using a renormalization scheme which is not the
standard $\MSbar$ one. The general name for the scheme we refer to in this 
context is regularization invariant (RI), \cite{18}. Though in practice the 
main scheme used is referred to as RI${}^\prime$. (In some articles this is 
synonymous with RI-MOM.) Therefore, one needs a means to convert lattice 
results from RI type renormalization schemes to the reference scheme of 
$\MSbar$, \cite{18,19}. For certain classes of Green's functions the continuum 
definition and use of RI${}^\prime$ (and RI) have been given in three and four 
loop renormalization of (massless) QCD in arbitrary covariant gauges, 
\cite{20,21,22,23}. The second main issue to deal with is that of matching 
lattice regularized results for the Green's functions with the corresponding 
continuum results. There are two main approaches for this. The first is the 
Schr\"{o}dinger functional method, \cite{24,25}, upon which we make no further 
comment. The second is the matching to explicit perturbative QCD results for 
the {\em same} Green's function which is what we concentrate on here. In this 
approach, \cite{18,19,20,21,22,23}, the main ethos is to compute the Green's 
function to as high a loop order as possible in perturbation theory as a 
function of the gauge coupling constant, $g$. Then the lattice computation 
should match on to the continuum behaviour in the {\em same} renormalization 
scheme in the high energy limit. Having a perturbative result to as high a loop
order as is calculationally feasible will in principle lead to a more accurate 
extraction of numerical values for the Green's function in the non-perturbative
range of interest for, say, nucleon structure functions.

As already indicated, previous work in the continuum concentrated on a variety
of gauge invariant twist-$2$ flavour non-singlet operators, ${\cal O}$, and the
perturbative evaluation of the flavour non-singlet Green's function $\langle 
\psi(p) {\cal O}(0) \bar{\psi}(-p) \rangle$ in the chiral limit,
\cite{20,21,22,23}, where $p$ is the momentum flowing through the Green's 
function and $\psi$ is the quark field. The operators considered for this 
Green's function were the Wilson and the transversity operators to and 
including moment $n$~$=$~$3$ and $4$ respectively and various quark bilinear 
operators such as the tensor current, \cite{21,22,23}. However, whilst such 
results were useful in many ways, they suffer from one major drawback. This is 
simply stated by noting that although each of the operators considered was 
gauge invariant, the Green's function itself was gauge dependent. Although 
ultimately one was only interested in the Landau gauge, the results of 
\cite{21,22,23} were provided in an arbitrary linear covariant gauge. Whilst 
this was not too problematic for continuum calculations, from the point of view
of lattice regularization one has also to fix the Landau gauge. However, this 
is a tough exercise in itself and in principle could open up reliability issues
to do with say ensuring the Gribov problem was avoided. To circumvent these 
lattice regularization gauge fixing issues another approach has been 
devised\footnote{The author is grateful to Dr P.E.L. Rakow and Dr R. Horsley 
for their patient enlightment on this point.}. Briefly to extract the 
appropriate renormalization constants for the operators and hence determine the
finite parts of the Green's functions, the approach is to consider gauge 
independent correlation functions of gauge invariant operators. In this way the
potential gauge fixing ambiguity never becomes an issue in the first place 
since the gauge does not then need to be fixed on the lattice. More 
specifically the appropriate correlation function to consider is $\langle 
{\cal O}(p) {\cal O}(-p) \rangle$. However, for, say, high moment Wilson 
operators the increase in the number of covariant derivatives may lead to too 
noisy a numerical signal for extraction of meaningful values of the Green's 
function. Hence, rather than consider this diagonal correlation function, a 
simple proposal would be to analyse an off-diagonal correlator such as $\langle
{\cal O}^1(p) {\cal O}^2(-p) \rangle$ where ${\cal O}^1(p)$ is a Wilson 
operator, say, and ${\cal O}^2(p)$ is a simple quark bilinear current operator.
This has to be chosen in such a way that the Green's function is not simply 
trivial in the chiral limit. However, information on the Wilson operator
renormalization constant can still be derived. 

Having reviewed the background and key issues we now indicate the main aim of 
this article. It is to simply to provide the explicit values of the appropriate
operator correlation functions, $\langle {\cal O}^1(p) {\cal O}^2(-p) \rangle$,
relevant to the lattice problem to as many loop orders in perturbation theory 
that is calculationally feasible. Specifically we will focus on three sets of 
flavour non-singlet operators used in deep inelastic scattering. These are the 
Wilson operators with moments $2$ and $3$ and moment $2$ of the set of 
transversity operators. Several correlations of the operator with itself will 
be provided as well as the appropriate non-zero off-diagonal one. We will 
present results to third order or three loops in the $\MSbar$ scheme as a 
function of the momentum flowing through the $2$-point function. We note that 
whilst we compute three loop Feynman diagrams, since it is clear that the 
leading diagram of the correlator is independent of the strong coupling 
constant, $a$, then the results will be to $O(a^2)$ inclusive where 
$a$~$=$~$g^2/(16\pi^2)$. However, it will also become evident that it is not 
possible to consider correlators of higher moment operators and expect to 
evaluate the correlation functions to the same three loop order. Whilst our 
main motivation is to provide the finite parts of these correlation functions 
several technical issues need to be addressed to obtain the correct answers. 
For instance, there is an assumption that the flavour non-singlet Wilson and
transversity operators do not mix under renormalization. It will turn out that 
this observation needs to be clarified within the present context. Their three 
loop $\MSbar$ anomalous dimensions are available, \cite{1,2,3,4,5,6,7}, and the 
lower loop results were originally obtained by considering the renormalization 
of $\langle \psi(p) {\cal O}(0) \bar{\psi}(-p) \rangle$. In this momentum 
configuration the mixing is not relevant. However, in the momentum 
configuration for the correlators of the present article $\langle {\cal O}^1(p)
{\cal O}^2(-p) \rangle$, since a momentum flows through the operator the mixing
is relevant and {\em cannot} be neglected. Suffice to say at this point that 
the additional operators are gauge invariant but total derivatives. Therefore, 
as part of our correlator renormalization programme we have had to compute the 
relevant anomalous dimension mixing matrices to allow us to extract 
consistently renormalized correlation functions. Such results will no doubt be 
important for other areas of deep inelastic scattering such as generalized 
parton distribution function analyses. 

The article is organised as follows. Section $2$ introduces the notation,
operators and general features of the correlation functions we consider
throughout. The general renormalization properties of the underlying 
correlation functions are discussed in section $3$ including the operator 
mixing issue. Section $4$ is devoted to the very mundane but important exercise
of recording all the results for the Green's functions we have considered. As a
spin-off we record the $R$-ratio for the tensor current to third order in 
section $5$. Finally, after concluding remarks in section $6$, several 
appendices are provided. The first records the Lorentz tensor decomposition of 
several operator correlators. This is necessary since the lattice requires the 
use of operators with uncontracted indices. In the renormalization of the 
matrix elements $\langle \psi(p) {\cal O}(0) \bar{\psi}(-p) \rangle$ of 
\cite{1,2,3}, the Lorentz indices were contracted with a null vector 
$\Delta_\mu$ with $\Delta^2$~$=$~$0$. This was because, for example, the Wilson
operators were traceless and symmetric and this contraction excluded the part 
with metric tensors in order to ease the extraction of renormalization 
constants directly. Here, since we will use a very specific computer algebra 
package and algorithm which only operates on scalar Feynman integrals, we need
to project out the relevant scalar amplitudes with respect to some tensor 
basis, which is, of course, not unique. This is discussed in Appendix A. The 
remaining appendix records the explicit {\em numerical} values of the various 
finite parts of the correlation functions for the colour group $SU(3)$ which 
were originally presented in exact form to $O(a^2)$ in section $4$.

\sect{Preliminaries.}

In this section we define our notation and operators and discuss the operator
correlation functions from a general perspective. Throughout we use the
standard QCD Lagrangian with massless quarks to immediately put us in the 
chiral case and an arbitrary linear covariant gauge fixing which is 
parametrized by the (renormalized) parameter $\alpha$. However, since our
correlation functions involve gauge invariant operators and therefore are gauge
independent, $\alpha$ will never actually appear in any of our final correlator
expressions. Though we stress that at no stage have we set $\alpha$~$=$~$0$ 
internally in our computations. Its {\em natural} cancellation is a strong 
internal consistency check on the construction of, say, our Feynman rules and 
the operator renormalization. Given this we define the general correlation 
function as, \cite{26,27,28},
\begin{equation}
\Pi^{ij}_{\mu_1 \ldots \mu_{n_i} \nu_1 \ldots \nu_{n_j}}(q^2) ~=~ (4\pi)^2 i 
\int \, d^d x \, e^{iqx} \langle 0 | {\cal O}^i_{\mu_1 \ldots \mu_{n_i}}(x) 
{\cal O}^j_{\nu_1 \ldots \nu_{n_j}} (0) | 0 \rangle
\label{opcordef}
\end{equation}
where $q$ is the momentum (with $q^2$~$=$~$-$~$Q^2$) and we have labelled the 
Lorentz indices of the respective constituent operators by a different parent 
Greek letter for clarity. At this point it is worth noting that we closely 
follow the procedures of \cite{26,27,28} which are excellent reviews of 
calculating (\ref{opcordef}) for quark current operators. The Green's function 
itself is illustrated schematically in Figure $1$ with the momentum flow made 
explicit. Our notation needs explanation. We use superscripts $i$ and $j$ to 
denote the left and right operators ${\cal O}$ of the correlation function as 
indicated in Figure $1$ where the momentum flows into the left operator and out
through the right operator. In principle, these operators are different whence 
the two distinct sets of Lorentz indices $\{\mu_i\}$ and $\{\nu_j\}$. Though 
for some cases we will include the quark mass operator which has no Lorentz 
tensor structure and so no such indices will be formally required. The flavour 
indices have been omitted to avoid cluttering the notation further. It is 
understood that there is a flavour generator included within each operator and 
later we will note the internal accounting method used to ensure that we derive
results only for the correlation of flavour non-singlet currents as opposed to 
flavour singlet currents. For the latter there would be an additional but 
different operator mixing problem from that into total derivative operators.
This singlet mixing is already well documented, \cite{1,2,3}. In other words 
flavour singlet quark blinear current operators can mix into purely gluonic 
operators with the same twist and quantum numbers, as well as gauge variant 
operators with the same properties but constructed from, say, Faddeev-Popov 
ghost fields. Moreover, one would also have to handle equation of motion 
operators too. We mention this aspect for completeness and note that as far as 
we are aware there is currently no lattice proposal to examine the flavour 
singlet case and we therefore will not devote any time to it here in the 
analogous continuum problem. 

\begin{figure}[ht]
\hspace{5cm}
\epsfig{file=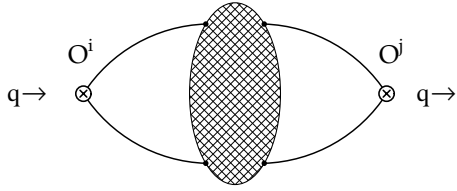,height=2.5cm}
\vspace{0.5cm}
\caption{Operator correlation function $\langle {\cal O}^i(q) {\cal O}^j(-q)
\rangle$.}
\end{figure}

Since we will consider a variety of operators we introduce a shorthand 
notation, akin to \cite{26,27,28}, for the superscripts $i$ and $j$ to indicate
which operator appears in (\ref{opcordef}). These are listed below as 
\begin{eqnarray} 
S & \equiv & \bar{\psi} \psi \nonumber \\ 
V & \equiv & \bar{\psi} \gamma^\mu \psi \nonumber \\ 
T & \equiv & \bar{\psi} \sigma^{\mu\nu} \psi \nonumber \\ 
W_2 & \equiv & {\cal S} \bar{\psi} \gamma^\mu D^\nu \psi \nonumber \\ 
\partial W_2 & \equiv & {\cal S} \partial^\mu \left( \bar{\psi} \gamma^{\nu} 
\psi \right) \nonumber \\ 
W_3 & \equiv & {\cal S} \bar{\psi} \gamma^\mu D^\nu D^\sigma \psi \nonumber \\ 
\partial W_3 & \equiv & {\cal S} \partial^\mu \left( \bar{\psi} \gamma^\nu 
D^\sigma \psi \right) \nonumber \\ 
\partial \partial W_3 & \equiv & {\cal S} \partial^\mu \partial^\nu \left( 
\bar{\psi} \gamma^\sigma \psi \right) \nonumber \\ 
T_2 & \equiv & {\cal S} \bar{\psi} \sigma^{\mu\nu} D^\sigma \psi \nonumber \\ 
\partial T_2 & \equiv & {\cal S} \partial^\mu \left( \bar{\psi} 
\sigma^{\nu\sigma} \psi \right) 
\label{oplist}
\end{eqnarray} 
where ${\cal S}$ denotes the appropriate symmetrization in the Lorentz indices
as well as the operator's tracelessness which we will discuss shortly. This
list already reveals our hand in terms of the operator mixing issue. Moreover,
in choosing this notation we have disguised to a degree what some of the total
derivative operators are. However, we choose to work within certain sectors.
By this we mean, using the Wilson operator of moment $3$ for illustration, that
$W_3$ is the main label for that sector as well as being the parent operator. 
In its renormalization it spawns offspring total derivative operators denoted 
by one or more $\partial$ symbols. These are labelled $\partial W_3$ and 
$\partial \partial W_3$. However, clearly from (\ref{oplist}) these derive from
the parent $W_2$ operator of the preceding or lower sector and the vector 
current of the first sector. It therefore might have been apt to choose the 
respective notation for these to be $\partial W_2$ and $\partial \partial V$ 
for labels. We have chosen not to do so because for the $W_3$ sector the number
of Lorentz indices are the same on each combination of these operators.
Moreover, one would have to count the derivative labels to deduce which sector
the label was associated with. Hence the Lorentz projection tensors used to 
project out the scalar amplitudes we will compute are the same for these 
combinations. So for the $W_3$ sector there is only one basic projector. Whilst
this means we will use $W_3$ as both an operator label and sector indicator, 
this we believe will minimize the confusion in choice of notation when trying 
to ascertain which sector, say, $\partial V$ belongs to especially when 
transversity is dealt with in parallel calculations.

Next in (\ref{oplist}) we choose the symmetrization with respect to Lorentz
indices and the tracelessness conditions in the standard way. For transversity,
which involves the $\gamma$-matrix commutator 
$\sigma^{\mu\nu}$~$=$~$[\gamma^\mu,\gamma^\nu]$, the definition is not the same
as for the Wilson operators, \cite{29,30,31}. For the three moments we 
consider, the explicit definitions of the symmetric traceless operators are, in
$d$-dimensions,  
\begin{eqnarray}
{\cal S} {\cal O}^{W_2}_{\mu\nu} &=& {\cal O}^{W_2}_{\mu\nu} ~+~ 
{\cal O}^{W_2}_{\nu\mu} ~-~ \frac{2}{d} \eta_{\mu\nu} 
{\cal O}^{W_2\,\sigma}_{\sigma} \nonumber \\ 
{\cal S} {\cal O}^{W_3}_{\mu\nu\sigma} &=&
{\cal O}^{W_3}_{S~\mu\nu\sigma} ~-~ \frac{1}{(d+2)} \left[ 
\eta_{\mu\nu} {\cal O}^{W_3~~\rho}_{S~\sigma\rho} ~+~
\eta_{\nu\sigma} {\cal O}^{W_3~~\rho}_{S~\mu\rho} ~+~
\eta_{\sigma\mu} {\cal O}^{W_3~~\rho}_{S~\nu\rho} \right] \nonumber \\
{\cal O}^{W_3}_{S~\mu\nu\sigma} &=&
\frac{1}{6} \left[ {\cal O}^{W_3}_{\mu\nu\sigma} ~+~ 
{\cal O}^{W_3}_{\nu\sigma\mu} ~+~ 
{\cal O}^{W_3}_{\sigma\mu\nu} ~+~ 
{\cal O}^{W_3}_{\mu\sigma\nu} ~+~ 
{\cal O}^{W_3}_{\sigma\nu\mu} ~+~ 
{\cal O}^{W_3}_{\nu\mu\sigma} \right] \nonumber \\
{\cal S} {\cal O}^{T_2}_{\mu\nu\sigma} &=& {\cal O}^{T_2}_{\mu\nu\sigma} ~+~ 
{\cal O}^{T_2}_{\mu\sigma\nu} ~-~ 
\frac{2}{(d-1)} \eta_{\nu\sigma} {\cal O}^{T_2\,\rho}_{\mu\rho} ~+~ 
\frac{1}{(d-1)} \left[ \eta_{\mu\nu} {\cal O}^{T_2\,\lambda}_{\rho\lambda} ~+~ 
\eta_{\mu\rho} {\cal O}^{T_2\,\lambda}_{\nu\lambda} \right]
\end{eqnarray} 
where
\begin{eqnarray}
{\cal O}^{W_2}_{\mu\nu} &=& \bar{\psi} \gamma_\mu D_\nu \psi \nonumber \\ 
{\cal O}^{W_3}_{\mu\nu\sigma} &=& \bar{\psi} \gamma_\mu D_\nu D_\sigma \psi 
\nonumber \\ 
{\cal O}^{T_2}_{\mu\nu\sigma} &=& \bar{\psi} \sigma_{\mu\nu} D_\sigma \psi 
\end{eqnarray} 
which are clearly consistent with the definitions in \cite{21,22,23}. We note 
that ${\cal O}^{W_3}_{S~\mu\nu\sigma}$ is the intermediate definition of the
symmetrized operator. As in \cite{21,22,23} we have derived the $d$-dimensional 
versions. Although the lattice computations are in strictly four dimensions, we
will dimensionally regularize in $d$~$=$~$4$~$-$~$2\epsilon$ dimensions where 
$\epsilon$ plays the role of the regularizing parameter. The renormalization 
constants will have a Laurent series in $\epsilon$ when subtracted in the 
$\MSbar$ scheme. The main calculational tool is the use of the {\sc Mincer} 
algorithm derived in \cite{32}. It is ideal for the present work as it 
evaluates massless $2$-point Feynman diagrams to the finite part at three loops
in dimensional regularization. The correlation function (\ref{opcordef}) 
clearly falls into this category and does not require infrared rearrangement or
external momentum nullification. More practically the {\sc Mincer} algorithm 
has been encoded in the powerful symbolic manipulation language {\sc Form}, 
\cite{33}, in \cite{34}, which therefore allows for a fully automated 
computation. For instance, once the Feynman rules for the operators are derived
consistently then the algorithm is applied to produce the finite parts. Crucial
to this is the electronic generation of the Feynman diagrams via the 
{\sc Qgraf} package, \cite{35}. These are then converted into {\sc Form} input 
notation for application of the {\sc Mincer} algorithm by systematically 
including the Lorentz and colour indices for the gluon, quark and Faddeev-Popov
ghost fields. For the present calculation there are $1$ one loop, $8$ two loop 
and $109$ three loop Feynman graphs to be evaluated in principle for every 
combination of operators in (\ref{oplist}) we consider here. These totals 
include graphs where there are two covariant derivatives in each operator as 
occurs for the $W_3$ sector. In the {\sc Qgraf} generation of graphs we 
restrict the diagrams to the one particle irreducible, no tadpole and no snail
set-up since we are dealing with massless fields. So, for example, there are no
closed gluon loops at the location of an operator. Such graphs are trivially
zero in dimensional regularization but may arise in, say, a lattice
regularization. That aside, when, for example, $\partial \partial W_3$ is part 
of the correlator the majority of the $109$ three loop graphs will in fact be 
trivially absent. For practical purposes it is best to have the most general 
set of diagrams for the full calculation rather than design {\sc Qgraf} 
routines for specific cases and potentially omit graphs which contribute. By 
the same token the presence of the covariant derivatives in the $W_3$ sector 
means that the explicit evaluation is slowed significantly on the available 
computers. Therefore, when this occurred we chose to run each Lorentz 
projection individually in series, which improved run times substantially. 
Given this we note the final aspect of our notation and that is the 
decomposition of the correlation function into the explicit scalar amplitudes, 
$\Pi^{ij}_{(k)}(q)$. These are defined by 
\begin{equation}
\Pi^{ij}_{\mu_1 \ldots \mu_{n_i} \nu_1 \ldots \nu_{n_j}}(q^2) ~=~ 
\sum_{k=1}^{n_{ij}} {\cal P}^{ij}_{(k) \{ \mu_1 \ldots \mu_{n_i} | 
\nu_1 \ldots \nu_{n_j} \} }(q) \, \Pi^{ij}_{(k)}(q) 
\label{opcordec}
\end{equation} 
where ${\cal P}^{ij}_{(k) \{ \mu_1 \ldots \mu_{n_i} | \nu_1 \ldots \nu_{n_j} \}
}(q)$ are the Lorentz projectors. The subscript $(k)$ (and also $(l)$ later) 
label the sector. How these projectors are derived and their explicit forms are
relegated to Appendix A. However, we note the number of projectors for each of 
the eight correlation function sectors we focus on here, $n_{ij}$, is given in 
Table $1$. Clearly the number of projectors increases with the number of free 
Lorentz indices. Given this decomposition then to find each individual scalar 
amplitude $\Pi^{ij}_{(k)}(q)$ we multiply (\ref{opcordec}) by the appropriate 
element of the inverse projection tensor which is defined for each sector in 
Appendix A. It is then this Lorentz scalar object which is put through the 
{\sc Mincer} algorithm. 

{\begin{table}[ht]
\begin{center}
\begin{tabular}{|c||c|c|c|c|c|c|c|c|}
\hline
$ij$ & $S,S$ & $V,V$ & $T,T$ & $V,W_2$ & $V,W_3$ & $W_2,W_2$ & $W_3,W_3$ & 
$T,T_2$ \\ 
\hline
$n_{ij}$ & $1$ & $2$ & $2$ & $2$ & $2$ & $3$ & $4$ & $4$ \\
\hline
\end{tabular}
\end{center}
%\vspace{0.3cm}
\begin{center}
{Table 1. Number of projectors for each correlation function sector.}
\end{center}
\end{table}}

As is usual with a renormalizable quantum field theory each amplitude is 
divergent. To extract the explicit divergence we follow the algorithm of
\cite{36} derived from automatic multiloop computations. In general terms, one 
computes the Green's function as a function of bare parameters. Then the
renormalized variables are introduced by the simple rescaling definition. For
example, for the bare and renormalized coupling constants 
$g_{\mbox{\footnotesize{o}}}$ and $g$ respectively, we use
$g_{\mbox{\footnotesize{o}}}$~$=$~$g Z_g$ where $Z_g$ is the coupling constant
renormalization constant. Though in dimensional regularization we will use
$g_{\mbox{\footnotesize{o}}}$~$=$~$\mu^\epsilon g Z_g$ where $\mu$ is the
arbitrary renormalization scale present due to the regularization. Here the
explicit renormalization constants for the operator correlation functions are
derived and discussed at length in the next section where the associated
renormalization constants are also constructed. It suffices to say at this 
point that there is an extension to the algorithm of \cite{36} in that the bare
operators in the Green's function have to be rescaled to their renormalized 
operator without neglecting the mixing into other operators. Moreover, the 
explicit forms of the gauge independent operator correlation function 
renormalization constants are equally as important as the finite parts of the 
correlators from the point of view of assisting with the matching of lattice 
results to the corresponding continuum values in the high energy limit.

There is one concern with relation to the topology of the graphs which needs to
be addressed. That is that we need to ensure that within our computer
programmes we are in fact calculating the correlation functions of flavour 
non-singlet operators as opposed to flavour singlet ones. For instance, without
the presence of a flavour matrix of some sort graphs which have a closed quark
loop and only include one of ${\cal O}^1$ or ${\cal O}^2$ but not both together
must be set to zero. If not it would be a contribution to a flavour singlet 
operator correlator. Therefore, whilst we have formally omitted flavour indices
in the definition (\ref{opcordef}), our Feynman rules for the operators 
actually include a flavour matrix for each operator. Denoting this by 
$\lambda^i$ where $i$ labels the left operator $1$ or right operator $2$, then 
at an appropriate point of the computation terms with $\mbox{tr} \left( 
\lambda^1 \right) \mbox{tr} \left( \lambda^2 \right)$ are set to zero to only 
leave terms proportional to $\mbox{tr} \left( \lambda^1 \lambda^2 \right)$. 
This is then formally set to unity since it flags the flavour non-singlet 
contribution uniquely.  

Finally, we comment on how we have chosen the correlation functions presented
here. The set we consider is listed in Table $1$. First, our choice is 
motivated by ensuring that the corresponding lattice calculation has a minimal 
set of covariant derivatives to handle. Second, we are constrained by the 
masslessness of the problem. For instance, as is clear from Figure $1$ and 
(\ref{opcordef}) we are dealing with closed quark loops. Therefore, one must 
have an even number of $\gamma$-matrices. As the quarks are massless and each 
quark propagator has exactly one $\gamma$-matrix the sum of $\gamma$-matrices 
present in both operators of (\ref{opcordef}) has to be even. Therefore, whilst
from a lattice point of view it would be simple to have the quark mass operator 
as the off-diagonal element for the $W_2$ and $W_3$ sectors this correlator is
trivially zero. In other words in the presence of quarks with generic mass
$m_q$ then the correlator will vanish as $O(m_q^\xi)$ where $\xi$~$>$~$0$. 
Hence, for $W_2$ and $W_3$ they have to be paired with $V$. Similarly as $T_2$
involves $\sigma^{\mu\nu}$ one requires an even number of $\gamma$-matrices for
the other operator of the correlator. Naively one would assume that this 
natural pairing would be with $S$. However, that leaves free Lorentz indices 
only on one operator and for $T_2$, given the symmetry properties of the 
operator itself via $\sigma^{\mu\nu}$, it is not in fact possible to decompose 
the correlator into Lorentz tensors built from the metric tensor, 
$\eta^{\mu\nu}$, and the momentum, $q^\mu$. Therefore, we have had to pair 
$T_2$ with $T$. Whilst these off-diagonal operators will probably be the ones 
of most interest to lattice computations we have chosen to consider the 
diagonal sectors $\{W_2,W_2\}$ and $\{W_3,W_3\}$ as well. There are several 
reasons for this. With a second avenue available to extract information on all
the $W_2$ and $W_3$ sector renormalization constants used for the lattice, 
these will actually give useful consistency checks provided sufficient 
computation power is available for the lattice calculations. Next, the operator
mixing issue is a novel feature of these correlators and we choose to consider 
them to ensure that we have obtained the correct overall picture of view from a
calculational and renormalization point of view. A final, less firm, motivation
is that the diagonal correlators of quark bilinear currents are useful to 
derive decay rates via the $R$-ratio formalism. (See, for example, 
\cite{26,27,28}.) Whilst those for $W_2$ and $W_3$ are tenuous in this respect, 
and we take them no further than finding the amplitudes, we do evaluate 
$\{T,T\}$ for this reason and construct the corresponding $R$-ratio for the 
tensor current as a by-product of our full computation.  

\sect{Renormalization group.}

In this section we concentrate on general aspects of the renormalization of the
operators we are interested in, (\ref{oplist}), and the construction of the
renormalization group equations satisfied by the renormalized operator
correlation functions. There are essentially two parts to this. The first 
relates to the operator mixing which is a separate exercise unrelated to the
correlation functions, whilst the second is an application of the mixing
property.

For the operator mixing issue we note first that the quark current operators
$S$, $V$ and $T$ are clearly mutiplicatively renormalizable in the chiral limit
and mass independent renormalization schemes. Therefore, we concentrate on the
three sectors $W_2$, $W_3$ and $T_2$ and justify our choice of operator basis. 
The first and last of these are similar, so we will specifically consider $W_2$
and $W_3$. First, for $W_2$ we note that the usual Lorentz symmetric and
traceless twist-$2$ flavour non-singlet operator used in deep inelastic 
scattering is 
\begin{equation} 
{\cal O}^{W_2}_{\mu\nu} ~=~ {\cal S} \bar{\psi} \gamma_\mu D_\nu \psi ~. 
\end{equation} 
This is not independent since one can of course add the independent operator
${\cal S} \left( D_\nu \bar{\psi} \right) \gamma_\mu \psi$ to the set of 
operators with the same symmetry properties. However, we want to make use of
known renormalization results for ${\cal O}^{W_2}_{\mu\nu}$, 
\cite{1,2,3,4,5,6,7,37,38}, and using the latter noncanonical operator, whilst 
not difficult from a technical point of view, is not the only operator 
independent of it. Instead the operator 
\begin{equation} 
{\cal O}^{\partial W_2}_{\mu\nu} ~=~ {\cal S} \partial_\mu \left( \bar{\psi} 
\gamma_\nu \psi \right)
\end{equation} 
is independent of $W_2$ and with $\partial W_2$ and $W_2$ we can obtain
${\cal S} \left( D_\nu \bar{\psi} \right) \gamma_\mu \psi$ as a linear 
combination. Likewise for the $T_2$ sector the analogous basis is
\begin{eqnarray}
{\cal O}^{T_2}_{\mu\nu\sigma} &=& {\cal S} \bar{\psi} \sigma_{\mu\nu} D_\sigma 
\psi \nonumber \\ 
{\cal O}^{\partial T_2}_{\mu\nu\sigma} &=& {\cal S} \partial_\mu \left( 
\bar{\psi} \sigma_{\nu\sigma} \psi \right) ~. 
\end{eqnarray}
At the next sector higher one now has an object with three Lorentz indices and
so one would expect three independent operators. Again we wish to retain the
renormalization properties of the parent operator which means choosing the
basis as $\{W_3, \partial W_3, \partial \partial W_3\}$ where 
\begin{eqnarray}
{\cal O}^{W_3}_{\mu\nu\sigma} &=& {\cal S} \bar{\psi} \gamma_\mu D_\nu D_\sigma
\psi \nonumber \\ 
{\cal O}^{\partial W_3}_{\mu\nu\sigma} &=& {\cal S} \partial_\mu \left( 
\bar{\psi} \gamma_\nu D_\sigma \psi \right) \nonumber \\ 
{\cal O}^{\partial \partial W_3}_{\mu\nu\sigma} &=& {\cal S} \partial_\mu 
\partial_\nu \left( \bar{\psi} \gamma_\sigma \psi \right) ~. 
\end{eqnarray}
Clearly the latter two are total derivatives of the set $\{W_2,\partial W_2\}$
and if one were to extend to the next sector level that set would involve the
parent $W_4$ and the total derivatives of the $W_3$ sector. From the explicit
calculation of the operator anomalous dimensions there is an important
computational advantage from choosing the basis in this way. Though it should 
be stressed that at higher levels the choice of basis is arbitrary and one 
could in principle choose, say, ${\cal S} \partial_\mu \left( \left( D_\nu 
\bar{\psi} \right) \gamma_\sigma \psi \right)$ as an independent member of the 
set. 

With the choices we have detailed for each sector, there is mixing under
renormalization but clearly each mixing matrix of renormalization constants,
$Z^{\cal O}_{ij}$, is upper triangular where
\begin{equation}
{\cal O}_{{\mbox{\footnotesize{o}}}\,i} ~=~ Z^{\cal O}_{ij} {\cal O}_j
\end{equation}
relates bare operators, denoted by the subscript 
${}_{\mbox{\footnotesize{o}}}$, to their renormalized version. To be explicit 
the matrix for the $W_2$ and $T_2$ sectors is
\begin{equation} 
Z^{\cal O}_{ij} ~=~ \left(
\begin{array}{cc}
Z^{\cal O}_{11} & Z^{\cal O}_{12} \\
0 & Z^{\cal O}_{22} \\
\end{array}
\right)
\label{mixmat2}
\end{equation}
and that for $W_3$ is 
\begin{equation} 
Z^{\cal O}_{ij} ~=~ \left(
\begin{array}{ccc}
Z^{\cal O}_{11} & Z^{\cal O}_{12} & Z^{\cal O}_{13} \\
0 & Z^{\cal O}_{22} & Z^{\cal O}_{23} \\
0 & 0 & Z^{\cal O}_{33} \\
\end{array}
\right) ~.
\label{mixmat3}
\end{equation}
Here we have chosen to simplify our notation by using numbers to denote the
mixing matrix elements rather than the more clumsy $\{W_2, \partial W_2 \}$,
$\{W_3, \partial W_3, \partial \partial W_3 \}$ or $\{T_2, \partial T_2 \}$ as
subscripts. Given these matrices we then define our anomalous dimension mixing 
matrix elements, $\gamma^{\cal O}_{ij}(a)$, formally as  
\begin{equation}
\gamma^{\cal O}_{ij} ~=~ \mu \frac{d ~}{d \mu} 
\ln Z^{\cal O}_{ij}
\label{mixmatdef}
\end{equation}
where 
\begin{equation}
\mu \frac{d~}{d\mu} ~=~ \beta(a) \frac{\partial ~}{\partial a} ~+~
\alpha \gamma_\alpha(a,\alpha) \frac{\partial ~}{\partial \alpha} ~. 
\label{muderiv}
\end{equation} 
Here $\beta(a)$ is the $\beta$-function and $\gamma_\alpha(a,\alpha)$ is the
anomalous dimension of the gauge parameter where we follow the conventions 
used in \cite{21} to define its renormalization. Although all our 
renormalization constants will in fact be independent of $\alpha$ we have 
included it in (\ref{muderiv}) as it technically appears as a formal parameter
in the QCD Lagrangian. For a renormalization where there is operator mixing,
(\ref{mixmatdef}) is invariably given as the formal definition of the anomalous
dimensions. However, for practical purposes in the derivation of the operator
correlation function anomalous dimensions it is more appropriate to give the
explicit consequences of (\ref{mixmatdef}). For sectors $W_2$ and $T_2$ we have 
\begin{eqnarray}
0 &=& \gamma^{\cal O}_{11}(a) Z^{\cal O}_{11} ~+~ 
\mu \frac{d ~}{d \mu} Z^{\cal O}_{11} \nonumber \\ 
0 &=& \gamma^{\cal O}_{11}(a) Z^{\cal O}_{12} ~+~ 
\gamma^{\cal O}_{12}(a) Z^{\cal O}_{22} ~+~ 
\mu \frac{d ~}{d \mu} Z^{\cal O}_{12} \nonumber \\ 
0 &=& \gamma^{\cal O}_{22}(a) Z^{\cal O}_{22} ~+~ 
\mu \frac{d ~}{d \mu} Z^{\cal O}_{22} 
\label{mix2}
\end{eqnarray}
and for $W_3$ we have similar relations, 
\begin{eqnarray}
0 &=& \gamma^{\cal O}_{11}(a) Z^{\cal O}_{11} ~+~ 
\mu \frac{d ~}{d \mu} Z^{\cal O}_{11} \nonumber \\ 
0 &=& \gamma^{\cal O}_{11}(a) Z^{\cal O}_{12} ~+~ 
\gamma^{\cal O}_{12}(a) Z^{\cal O}_{22} ~+~ 
\mu \frac{d ~}{d \mu} Z^{\cal O}_{12} \nonumber \\ 
0 &=& \gamma^{\cal O}_{11}(a) Z^{\cal O}_{13} ~+~ 
\gamma^{\cal O}_{12}(a) Z^{\cal O}_{23} ~+~ 
\gamma^{\cal O}_{13}(a) Z^{\cal O}_{33} ~+~ 
\mu \frac{d ~}{d \mu} Z^{\cal O}_{13} \nonumber \\ 
0 &=& \gamma^{\cal O}_{22}(a) Z^{\cal O}_{22} ~+~ 
\mu \frac{d ~}{d \mu} Z^{\cal O}_{22} \nonumber \\ 
0 &=& \gamma^{\cal O}_{22}(a) Z^{\cal O}_{23} ~+~ 
\gamma^{\cal O}_{23}(a) Z^{\cal O}_{33} ~+~ 
\mu \frac{d ~}{d \mu} Z^{\cal O}_{23} \nonumber \\ 
0 &=& \gamma^{\cal O}_{33}(a) Z^{\cal O}_{33} ~+~ 
\mu \frac{d ~}{d \mu} Z^{\cal O}_{33} ~. 
\label{mix3}
\end{eqnarray}

\begin{figure}[ht]
\hspace{6cm}
\epsfig{file=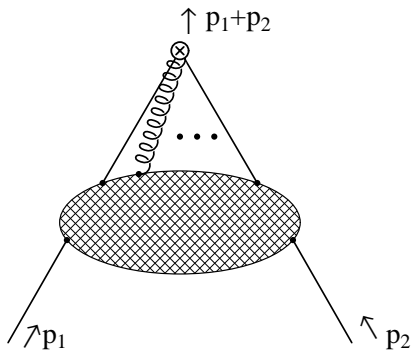,height=5cm}
\vspace{0.5cm}
\caption{Green's function, $\langle \psi(p_1) {\cal O}^i(-p_1-p_2)
\bar{\psi}(p_2) \rangle$, used to renormalize the operators ${\cal O}^i$.}
\end{figure}

We now turn to the practical problem of evaluating the anomalous dimensions
explicitly to the loop order necessary for the operator correlation function
renormalization at three loops. Whilst this is actually $O(a^2)$ we have 
determined the mixing matrices to $O(a^3)$ partly for completeness and checking
reasons but also because of their potential use in phenomenological problems 
where a momentum flows out through the operator itself. For the mixing matrix 
anomalous dimensions the main issue is the determination of the off-diagonal 
elements. In the original approach of \cite{1,2,3} the operators were inserted 
in the Green's function $\langle \psi(p) {\cal O}(0) \bar{\psi}(-p) \rangle$ 
whose more general version is illustrated graphically in Figure $2$. As the 
operator is a zero momentum insertion the contribution to the renormalization 
of the off-diagonal part of $Z^{\cal O}_{ij}$ cannot be deduced as the total 
derivative operator insertions vanish trivially for this momentum 
configuration. So only all the diagonal elements of our mixing matrix can be 
deduced via this momentum routing. There are two remaining choices for routing 
momenta with one nullification, which is necessary if we wish to apply 
{\sc Mincer} as our tool of computation. With the choice $\langle \psi(p) 
{\cal O}(-p) \bar{\psi}(0) \rangle$ it is clear that one can access the 
elements $Z^{W_2}_{12}$, $Z^{T_2}_{12}$ and $Z^{W_3}_{23}$ once the respective 
values for $Z^{W_2}_{11}$, $Z^{T_2}_{11}$ and $Z^{W_3}_{22}$ have been included
from the original results of \cite{1,2,3,4,5,6,7,39,40,41}. The determination 
of $Z^{W_3}_{12}$ and $Z^{W_3}_{13}$ is more difficult. This is because there 
is only {\em one} momentum choice which makes contact with both these 
renormalization constants at the same time. We circumvented this difficulty to 
two loops by not omitting the terms involving $\ln(p^2/\mu^2)$ which derive 
from the dimensionality of the loop integrals at each loop order. One 
ordinarily ignores such terms in a renormalization since they cancel trivially 
in a renormalizable theory. Retaining them means that divergences such as 
$\frac{1}{\epsilon} \ln(p^2/\mu^2)$, $\frac{1}{\epsilon^2} \ln(p^2/\mu^2)$ and 
$\frac{1}{\epsilon} \left( \ln(p^2/\mu^2) \right)^2$ are present at various
loop orders but some of their coefficients involve counterterm parts of 
$Z^{W_3}_{12}$ and $Z^{W_3}_{13}$ in their coupling constant series and Laurent
expansion in $\epsilon$. They give extra constraints on $Z^{W_3}_{12}$ and 
$Z^{W_3}_{13}$ which allowed us to decipher the $W_3$ sector mixing matrix to 
{\em two} loops. At three loops we can only derive a simple relation between 
the simple poles of each renormalization constant. However, this is not in fact 
necessary for our operator correlator results. Again the main tool is the 
{\sc Mincer} package written in {\sc Form}, \cite{35,36,37}. The graphs are 
again generated by {\sc Qgraf}, \cite{38}. For the maximum number of possible 
covariant derivative terms that can arise, there are $3$ one loop, $37$ two 
loop and $684$ three loop Feynman graphs to evaluate. As before these are one
particle irreducible graphs without snails or tadpoles.  

Given these considerations it is evident that sector $W_4$ would be more
difficult to deduce completely to three loops due to a similar momentum routing
issue. However, it could be determined to two loops in principle by performing
a four loop calculation with $\ln(p^2/\mu^2)$ terms included. If there were a
four loop {\sc Mincer} routine available then this would be a viable 
proposition. Therefore, in principle, the procedure to determine the mixing
matrix for the non-singlet sectors $W_n$ is available but in practice requires 
the computational machinery to evaluate the off-diagonal mixing matrix 
elements. All that remains is to record the explicit values which are all given
in the $\MSbar$ scheme. First, for completeness and for comparing with the 
conventions of previous calculations the vector and tensor current anomalous 
dimensions, \cite{39,40,41}, are
\begin{eqnarray} 
\gamma^V(a) &=& O(a^4) \nonumber \\
\gamma^T(a) &=& C_F a ~+~ [ 257 C_A ~-~ 171 C_F ~-~ 52 T_F \Nf ] 
\frac{C_F a^2}{18} \nonumber \\ 
&& +~ \left[ 13639 C_A^2 ~-~ 4320 \zeta(3) C_A^2 ~+~ 
12096 \zeta(3)C_A C_F \right. \nonumber \\ 
&& \left. ~~~~-~ 20469 C_A C_F ~-~ 1728 \zeta(3) C_A T_F \Nf ~-~ 
4016 C_A T_F \Nf \right. \nonumber \\ 
&& \left. ~~~~-~ 6912 \zeta(3) C_F^2 ~+~ 6570 C_F^2 ~+~ 
1728 \zeta(3) C_F T_F \Nf \right. \nonumber \\ 
&& \left. ~~~~+~ 1176 C_F T_F \Nf ~-~ 144 T_F^2 C_F^2) \right] 
\frac{C_F a^3}{108} ~+~ O(a^4)  
\end{eqnarray}  
where $\zeta(z)$ is the Riemann zeta function, $C_A$ and $C_F$ are the usual 
colour group Casimirs and $T_F$ is defined by $\mbox{Tr}\left( T^a T^b 
\right)$~$=$~$T_F \delta^{ab}$ where $T^a$ are the generators of the colour 
group. For the two sectors with two operators we have
\begin{eqnarray}
\gamma^{W_2}_{11}(a) &=& \frac{8}{3} C_F a ~+~ \frac{1}{27} \left[ 376 C_A C_F
- 112 C_F^2 - 128 C_F T_F \Nf \right] a^2 \nonumber \\ 
&& +~ \frac{1}{243} \left[ \left( 5184 \zeta(3) + 20920 \right) C_A^2 C_F
- \left( 15552 \zeta(3) + 8528 \right) C_A C_F^2 \right. \nonumber \\
&& \left. ~~~~~~~~~~-~ \left( 10368 \zeta(3) + 6256 \right) C_A C_F T_F \Nf 
+ \left( 10368 \zeta(3) - 560 \right) C_F^3 \right. \nonumber \\
&& \left. ~~~~~~~~~~+~ \left( 10368 \zeta(3) - 6824 \right) C_F^2 T_F \Nf
- 896 C_F T_F^2 \Nf^2 \right] a^3 ~+~ O(a^4) \nonumber \\
\gamma^{W_2}_{12}(a) &=& -~ \frac{4}{3} C_F a ~+~ \frac{1}{27} 
\left[ 56 C_F^2 - 188 C_A C_F + 64 C_F T_F \Nf \right] a^2 \nonumber \\ 
&& +~ \frac{1}{243} \left[ \left( 7776 \zeta(3) + 4264 \right) C_A C_F^2
- \left( 2592 \zeta(3) + 10460 \right) C_A^2 C_F \right. \nonumber \\
&& \left. ~~~~~~~~~~+~ \left( 5184 \zeta(3) + 3128 \right) C_A C_F T_F \Nf 
- \left( 5184 \zeta(3) - 280 \right) C_F^3 \right. \nonumber \\
&& \left. ~~~~~~~~~~-~ \left( 5184 \zeta(3) - 3412 \right) C_F^2 T_F \Nf
+ 448 C_F T_F^2 \Nf^2 \right] a^3 ~+~ O(a^4) \nonumber \\
\gamma^{W_2}_{22}(a) &=& O(a^4) 
\end{eqnarray} 
and 
\begin{eqnarray}
\gamma^{T_2}_{11}(a) &=& 3 C_F a ~+~ \frac{1}{2} \left[ 35 C_A C_F
- 9 C_F^2 - 12 C_F T_F \Nf \right] a^2 \nonumber \\ 
&& +~ \frac{1}{108} \left[ 12553 C_A^2 C_F - 7479 C_A C_F^2 + 1782 C_F^3
- \left( 5184 \zeta(3) + 4168 \right) C_A C_F T_F \Nf \right. \nonumber \\
&& \left. ~~~~~~~~~~+~ \left( 5184 \zeta(3) - 3240 \right) C_F^2 T_F \Nf
- 368 C_F T_F^2 \Nf^2 \right] a^3 ~+~ O(a^4) \nonumber \\ 
\gamma^{T_2}_{12}(a) &=& -~ C_F a ~+~ \frac{1}{18} 
\left[ 28 C_F T_F \Nf - 45 C_F^2 - 29 C_A C_F \right] a^2 \nonumber \\ 
&& +~ \frac{1}{108} \left[ \left( 6048 \zeta(3) - 6495 \right) C_A C_F^2
- \left( 2160 \zeta(3) - 543 \right) C_A^2 C_F \right. \nonumber \\ 
&& \left. ~~~~~~~~~~+~ \left( 1728 \zeta(3) + 76 \right) C_A C_F T_F \Nf 
- \left( 3456 \zeta(3) - 2394 \right) C_F^3 \right. \nonumber \\
&& \left. ~~~~~~~~~~-~ \left( 1728 \zeta(3) - 2208 \right) C_F^2 T_F \Nf
+ 112 C_F T_F^2 \Nf^2 \right] a^3 ~+~ O(a^4) \nonumber \\ 
\gamma^{T_2}_{22}(a) &=& C_F a ~+~ \frac{1}{18} 
\left[ 257 C_A C_F - 171 C_F^2 - 52 C_F T_F \Nf \right] a^2 \nonumber \\
&& +~ \frac{1}{108} \left[ \left( 13639 - 4320 \zeta(3) \right) C_A^2 C_F
+ \left( 12096 \zeta(3) - 20469 \right) C_A C_F^2 \right. \nonumber \\
&& \left. ~~~~~~~~~~-~ \left( 1728 \zeta(3) + 4016 \right) C_A C_F T_F \Nf 
- \left( 6912 \zeta(3) - 6570 \right) C_F^3 \right. \nonumber \\
&& \left. ~~~~~~~~~~+~ \left( 1728 \zeta(3) + 1176 \right) C_F^2 T_F \Nf
- 144 C_F T_F^2 \Nf^2 \right] a^3 ~+~ O(a^4) ~.  
\end{eqnarray} 
Given the issues with computing the full set of anomalous dimensions to three
loops for the $W_3$ sector we record that the results are
\begin{eqnarray}
\gamma^{W_3}_{11}(a) &=& \frac{25}{6} C_F a ~+~ \frac{1}{432} \left[ 
8560 C_A C_F - 2035 C_F^2 - 3320 C_F T_F \Nf \right] a^2 \nonumber \\ 
&& +~ \frac{1}{15552} \left[ \left( 285120 \zeta(3) + 1778866 \right) C_A^2 C_F
- \left( 855360 \zeta(3) + 311213 \right) C_A C_F^2 \right. \nonumber \\
&& \left. ~~~~~~~~~~~~~-~ \left( 1036800 \zeta(3) + 497992 \right) 
C_A C_F T_F \Nf + \left( 570240 \zeta(3) - 244505 \right) C_F^3 
\right. \nonumber \\
&& \left. ~~~~~~~~~~~~~+~ \left( 1036800 \zeta(3) - 814508 \right) 
C_F^2 T_F \Nf - 82208 C_F T_F^2 \Nf^2 \right] a^3 ~+~ O(a^4) \nonumber \\
\gamma^{W_3}_{12}(a) &=& -~ \frac{3}{2} C_F a ~+~ \frac{1}{144} \left[ 
81 C_F^2 - 848 C_A C_F + 424 C_F T_F \Nf \right] a^2 ~+~ O(a^3) \nonumber \\ 
\gamma^{W_3}_{13}(a) &=& -~ \frac{1}{2} C_F a ~+~ \frac{1}{144} \left[ 
103 C_F^2 - 388 C_A C_F + 104 C_F T_F \Nf \right] a^2 ~+~ O(a^3) \nonumber \\ 
\gamma^{W_3}_{22}(a) &=& \frac{8}{3} C_F a ~+~ \frac{1}{27} \left[ 376 C_A C_F
- 112 C_F^2 - 128 C_F T_F \Nf \right] a^2 \nonumber \\ 
&& +~ \frac{1}{243} \left[ \left( 5184 \zeta(3) + 20920 \right) C_A^2 C_F
- \left( 15552 \zeta(3) + 8528 \right) C_A C_F^2 \right. \nonumber \\
&& \left. ~~~~~~~~~~-~ \left( 10368 \zeta(3) + 6256 \right) C_A C_F T_F \Nf 
+ \left( 10368 \zeta(3) - 560 \right) C_F^3 \right. \nonumber \\
&& \left. ~~~~~~~~~~+~ \left( 10368 \zeta(3) - 6824 \right) C_F^2 T_F \Nf
- 896 C_F T_F^2 \Nf^2 \right] a^3 ~+~ O(a^4) \nonumber \\
\gamma^{W_3}_{23}(a) &=& -~ \frac{4}{3} C_F a ~+~ \frac{1}{27} 
\left[ 56 C_F^2 - 188 C_A C_F + 64 C_F T_F \Nf \right] a^2 \nonumber \\ 
&& +~ \frac{1}{243} \left[ \left( 7776 \zeta(3) + 4264 \right) C_A C_F^2
- \left( 2592 \zeta(3) + 10460 \right) C_A^2 C_F \right. \nonumber \\
&& \left. ~~~~~~~~~~+~ \left( 5184 \zeta(3) + 3128 \right) C_A C_F T_F \Nf 
- \left( 5184 \zeta(3) - 280 \right) C_F^3 \right. \nonumber \\
&& \left. ~~~~~~~~~~-~ \left( 5184 \zeta(3) - 3412 \right) C_F^2 T_F \Nf
+ 448 C_F T_F^2 \Nf^2 \right] a^3 ~+~ O(a^4) \nonumber \\
\gamma^{W_3}_{33}(a) &=& O(a^4) 
\end{eqnarray} 
where those for $\gamma^{W_3}_{12}(a)$ and $\gamma^{W_3}_{13}(a)$ are only
given to two loops. Clearly the diagonal anomalous dimensions of each sector,
including those with total derivatives, are the same as the corresponding
non-total derivative operator. Equally the mixing of the second row of $W_3$
is trivially related to that of the first row of $W_2$. This is a reassuring 
observation. As further checks on the results, since we have used the 
algorithm of \cite{36} to determine the renormalization constants, therefore, 
the double and triple poles in $\epsilon$ are predetermined by the 
renormalization group equations of (\ref{mix2}) and (\ref{mix3}). We note that 
our results are consistent with those constraints.

We now focus on the renormalization structure of the operator correlation
functions. These are Green's functions of operators rather than of fields but
like Green's functions of fields they have an associated renormalization
constant after the constituent bare operators have been replaced by their
renormalized versions, taking into account any mixing. We denote these
additional renormalization constants by $Z^{ij}_{(k)}$ and follow the quark
current correlator renormalization formalism of \cite{26,27,28}. This
renormalization constant appears as a contact term rather than as a canonical
multiplicative renormalization constant that one normally expects in the
renormalization of a Green's function involving only fields. In the case of
operator correlators where there is no mixing of the constituent operators,
the relation between bare and renormalized correlators has been given in, for
example, \cite{26,27,28}. For completeness, we give the form for the tensor
current correlation function. It is  
\begin{equation}
\Pi^{T,T}_{(i)}(q) ~=~ Z^{T,T}_{(i)} q^2 ~+~ \mu^{2\epsilon} \left( Z^T 
\right)^2 \Pi^{T,T}_{\mbox{\footnotesize{o}}\,(i)}(q) 
\label{pitt}
\end{equation}
where we have included the subscript label deriving from the Lorentz tensor
decomposition since there will in principle be a divergence for each
projection. This relation is the basic form used in our computer algebra setup
and the automatic Feynman diagram renormalization procedure of \cite{36} is 
easy to extend and encode in {\sc Form} for this case. Given (\ref{pitt}) it is
straightforward to derive the renormalization group equation satisfied by the 
renormalized correlation function. Applying (\ref{muderiv}) to (\ref{pitt}) we 
have  
\begin{equation}
0 ~=~ \mu \frac{d~}{d\mu} \Pi^{T,T}_{(i)}(q) ~+~ 
2 \gamma^T(a) \Pi^{T,T}_{(i)}(q) ~-~ q^2 \gamma^{T,T}_{(i)}(a) 
\end{equation}
where the correlation function anomalous dimension is formally given by
\begin{equation}
\gamma^{T,T}_{(i)}(a) ~=~ \left[ -~ \epsilon ~+~ \beta(a) 
\frac{\partial~}{\partial a} ~+~ 2 \gamma^T(a) \right] Z_{(i)}^{T,T} ~. 
\label{gamtt}
\end{equation} 
Although part of our ultimate aim is to provide the finite parts of the 
amplitudes we will also determine these anomalous dimensions to $O(a^2)$
inclusive. Indeed we use the results, such as (\ref{gamtt}) as consistency
checks on the explicit renormalization constants. Aside from being gauge
independent expressions, the double and triple poles in $\epsilon$ are again
determined by the lower order simple poles. Moreover, if one had not taken
the issue of mixing into account this internal consistency check would in fact
fail.

Given this form for the quark current correlators, $S$ and $V$, 
\cite{26,27,28}, and now also $T$, their extension to operator correlators 
where there is mixing is subtle. Therefore, we highlight the structure for two 
cases which are $\{V,W_3\}$ and $\{W_3,W_3\}$. The resulting renormalization 
group functions for the remaining cases, $\{V,W_2\}$, $\{W_2,W_2\}$ and 
$\{T,T_2\}$ can be readily deduced from the final relations by appropriate 
relabelling and ignoring irrelevant terms which, say, do not occur for the 
$W_2$ sector. For each case we consider, the key is to simply write down the 
forms analogous to (\ref{pitt}) but including all possible consistent mixings 
and all possible operator combinations in the correlation functions for that 
sector and ensure the equation is dimensionally consistent. The reason for this
is that the relation between bare and renormalized correlation functions are 
entwined in an intricate way. As the $\{V,W_3\}$ case is simple since only one 
of the constituent operator undergoes mixing, we illustrate this by giving the 
three renormalization definitions explicitly as,   
\begin{eqnarray}
\Pi^{V,W_3}_{(i)}(q) &=& Z^{V,W_3}_{(i)} (q^2)^2 ~+~ \mu^{2\epsilon} Z^V 
\left[  Z^{W_3}_{11} \Pi^{V,W_3}_{\mbox{\footnotesize{o}}\,(i)}(q) \,+\, 
Z^{W_3}_{12} \Pi^{V,\partial W_3}_{\mbox{\footnotesize{o}}\,(i)}(q) \,+\, 
Z^{W_3}_{13} 
\Pi^{V,\partial \partial W_3}_{\mbox{\footnotesize{o}}\,(i)}(q) \right]
\nonumber \\ 
\Pi^{V,\partial W_3}_{(i)}(q) &=& Z^{V,\partial W_3}_{(i)} (q^2)^2 ~+~ 
\mu^{2\epsilon} Z^V \left[ 
Z^{W_3}_{22} \Pi^{V,\partial W_3}_{\mbox{\footnotesize{o}}\,(i)}(q) ~+~ 
Z^{W_3}_{23} 
\Pi^{V,\partial \partial W_3}_{\mbox{\footnotesize{o}}\,(i)}(q) \right]
\nonumber \\ 
\Pi^{V,\partial \partial W_3}_{(i)}(q) &=& Z^{V,\partial \partial W_3}_{(i)} 
(q^2)^2 ~+~ \mu^{2\epsilon} Z^V Z^{W_3}_{33} 
\Pi^{V,\partial \partial W_3}_{\mbox{\footnotesize{o}}\,(i)}(q)
\label{pivw2}
\end{eqnarray} 
where the mixing matrix elements of (\ref{mixmat3}) appear and the factor of
$q^2$ multiplying $Z^{i,j}_{(k)}$ derives from the dimensionality of the actual
correlator in question to ensure a dimensionless renormalization constant. The 
next stage is to apply (\ref{muderiv}) to each of equation and then rewrite the
full set without any bare correlators. This is algebraically tedious but it is 
best to start with the final equation since it is similar to (\ref{pitt}) 
whence 
\begin{equation} 
0 ~=~ \mu \frac{d~}{d\mu} \Pi^{V,\partial \partial W_3}_{(i)}(q) ~+~ 
\left( \gamma^V(a) + \gamma^{W_3}_{33}(a) \right) 
\Pi^{V,\partial \partial W_3}_{(i)}(q) ~-~ 
(q^2)^2 \gamma^{V,\partial \partial W_3}_{(i)}(a) 
\end{equation} 
with 
\begin{equation}
\gamma^{V,\partial \partial W_3}_{(i)}(a) ~=~ \left[ -~ \epsilon ~+~ \beta(a) 
\frac{\partial~}{\partial a} ~+~ \gamma^{V}(a) ~+~ \gamma^{W_3}_{33}(a) 
\right] Z_{(i)}^{V,\partial \partial W_3} ~. 
\end{equation} 
Considering the second equation of (\ref{pivw2}) next, after the application
of (\ref{muderiv}) it is necessary to rewrite both bare correlators which now
occur back in terms of their renormalized versions. Throughout this and all our
other similar manipulations, we always associate terms with powers of the 
momentum $q^2$ as contributing to the correlator anomalous dimension. Hence, we 
have 
\begin{eqnarray} 
0 &=& \mu \frac{d~}{d\mu} \Pi^{V,\partial W_3}_{(i)}(q) ~+~ 
\left( \gamma^V(a) + \gamma^{W_3}_{22}(a) \right) 
\Pi^{V,\partial W_3}_{(i)}(q) \nonumber \\
&& +~ \gamma^{W_3}_{23}(a) \Pi^{V,\partial \partial W_3}_{(i)}(q) ~-~ (q^2)^2 
\gamma^{V,\partial W_3}_{(i)}(a) 
\end{eqnarray} 
and 
\begin{equation}
\gamma^{V,\partial W_3}_{(i)}(a) ~=~ \left[ -~ \epsilon ~+~ \beta(a) 
\frac{\partial~}{\partial a} ~+~ \gamma^{V}(a) ~+~ \gamma^{W_3}_{22}(a) 
\right] Z_{(i)}^{V,\partial W_3} ~+~ \gamma^{W_3}_{23}(a) 
Z_{(i)}^{V,\partial \partial W_3} ~. 
\end{equation} 
Finally, both of the last equations of (\ref{pivw2}) are required to complete
the set of three renormalization group functions for the $\{V,W_3\}$ case. We
have 
\begin{eqnarray}
0 &=& \mu \frac{d~}{d\mu} \Pi^{V,W_3}_{(i)}(q) ~+~ 
\left( \gamma^V(a) + \gamma^{W_3}_{11}(a) \right) \Pi^{V,W_3}_{(i)}(q) ~+~
\gamma^{W_3}_{12}(a) \Pi^{V,\partial W_3}_{(i)}(q) \nonumber \\
&& +~ \gamma^{W_3}_{13}(a) \Pi^{V,\partial \partial W_3}_{(i)}(q) ~-~ (q^2)^2 
\gamma^{V,W_3}_{(i)}(a) 
\end{eqnarray} 
where 
\begin{eqnarray}
\gamma^{V,W_3}_{(i)}(a) &=& \left[ -~ \epsilon ~+~ \beta(a) 
\frac{\partial~}{\partial a} ~+~ \gamma^{V}(a) ~+~ \gamma^{W_3}_{11}(a) 
\right] Z_{(i)}^{V,W_3} \nonumber \\ 
&& +~ \gamma^{W_3}_{12}(a) Z_{(i)}^{V,\partial W_3} ~+~ \gamma^{W_3}_{13}(a) 
Z_{(i)}^{V,\partial \partial W_3} ~.
\end{eqnarray} 

For the $\{W_3,W_3\}$ case the derivation follows parallel lines though the
starting point is a more complicated set of six relations between bare and
renormalized correlators as there is mixing in both inserted operators. These 
are 
\begin{eqnarray}
\Pi^{W_3,W_3}_{(i)}(q) &=& Z^{W_3,W_3}_{(i)} (q^2)^3 \nonumber \\
&& +~ \mu^{2\epsilon} \left[ 
\left( Z^{W_3}_{11} \right)^2 
\Pi^{W_3,W_3}_{\mbox{\footnotesize{o}}\,(i)}(q) ~+~ 
2 Z^{W_3}_{11} Z^{W_3}_{12} 
\Pi^{W_3,\partial W_3}_{\mbox{\footnotesize{o}}\,(i)}(q) \right. \nonumber \\
&& \left. ~~~~~~~~+~ 
2 Z^{W_3}_{11} Z^{W_3}_{13} 
\Pi^{W_3,\partial \partial W_3}_{\mbox{\footnotesize{o}}\,(i)}(q) ~+~ 
\left( Z^{W_3}_{12} \right)^2 
\Pi^{\partial W_3,\partial W_3}_{\mbox{\footnotesize{o}}\,(i)}(q) \right.
\nonumber \\
&& \left. ~~~~~~~~+~ 
2 Z^{W_3}_{12} Z^{W_3}_{13} 
\Pi^{\partial W_3,\partial \partial W_3}_{\mbox{\footnotesize{o}}\,(i)}(q) ~+~ 
\left( Z^{W_3}_{13} \right)^2 
\Pi^{\partial \partial W_3,\partial \partial 
W_3}_{\mbox{\footnotesize{o}}\,(i)}(q) \right]
\nonumber \\ 
\Pi^{W_3,\partial W_3}_{(i)}(q) &=& 
Z^{W_3,\partial W_3}_{(i)} (q^2)^3 \nonumber \\
&& +~ \mu^{2\epsilon} \left[ 
Z^{W_3}_{11} Z^{W_3}_{22} 
\Pi^{W_3,\partial W_3}_{\mbox{\footnotesize{o}}\,(i)}(q) ~+~ 
Z^{W_3}_{12} Z^{W_3}_{22} 
\Pi^{\partial W_3,\partial W_3}_{\mbox{\footnotesize{o}}\,(i)}(q) \right.
\nonumber \\
&& \left. ~~~~~~~~+~ 
Z^{W_3}_{13} Z^{W_3}_{22} 
\Pi^{\partial W_3,\partial \partial W_3}_{\mbox{\footnotesize{o}}\,(i)}(q) ~+~ 
Z^{W_3}_{11} Z^{W_3}_{23} 
\Pi^{W_3,\partial \partial W_3}_{\mbox{\footnotesize{o}}\,(i)}(q) \right.
\nonumber \\
&& \left. ~~~~~~~~+~ 
Z^{W_3}_{12} Z^{W_3}_{23} 
\Pi^{\partial W_3,\partial \partial W_3}_{\mbox{\footnotesize{o}}\,(i)}(q) ~+~ 
Z^{W_3}_{13} Z^{W_3}_{23} 
\Pi^{\partial \partial W_3,\partial \partial 
W_3}_{\mbox{\footnotesize{o}}\,(i)}(q) \right]
\nonumber \\ 
\Pi^{W_3,\partial \partial W_3}_{(i)}(q) &=& 
Z^{W_3,\partial \partial W_3}_{(i)} (q^2)^3 \nonumber \\
&& +~ \mu^{2\epsilon} \left[ 
Z^{W_3}_{11} Z^{W_3}_{33} 
\Pi^{W_3,\partial \partial W_3}_{\mbox{\footnotesize{o}}\,(i)}(q) ~+~ 
Z^{W_3}_{12} Z^{W_3}_{33} 
\Pi^{\partial W_3,\partial
\partial W_3}_{\mbox{\footnotesize{o}}\,(i)}(q) \right. \nonumber \\
&& \left. ~~~~~~~~+~ 
Z^{W_3}_{13} Z^{W_3}_{33} 
\Pi^{\partial \partial W_3,\partial \partial 
W_3}_{\mbox{\footnotesize{o}}\,(i)}(q) \right]
\nonumber \\ 
\Pi^{\partial W_3,\partial W_3}_{(i)}(q) &=& 
Z^{\partial W_3,\partial W_3}_{(i)} (q^2)^3 \nonumber \\
&& +~ \mu^{2\epsilon} \left[ 
\left( Z^{W_3}_{22} \right)^2 
\Pi^{\partial W_3,\partial W_3}_{\mbox{\footnotesize{o}}\,(i)}(q) ~+~ 
2 Z^{W_3}_{22} Z^{W_3}_{23} 
\Pi^{\partial W_3,\partial \partial W_3}_{\mbox{\footnotesize{o}}\,(i)}(q)
\right. \nonumber \\
&& \left. ~~~~~~~~+~ 
\left( Z^{W_3}_{23} \right)^2 
\Pi^{\partial \partial W_3,\partial \partial 
W_3}_{\mbox{\footnotesize{o}}\,(i)}(q) 
\right] \nonumber \\
\Pi^{\partial W_3,\partial \partial W_3}_{(i)}(q) &=& 
Z^{\partial W_3,\partial \partial W_3}_{(i)} (q^2)^3 \nonumber \\
&& +~ \mu^{2\epsilon} \left[ 
Z^{W_3}_{22} Z^{W_3}_{33}
\Pi^{\partial W_3,\partial \partial W_3}_{\mbox{\footnotesize{o}}\,(i)}(q) ~+~ 
Z^{W_3}_{23} Z^{W_3}_{23} 
\Pi^{\partial \partial W_3,\partial \partial 
W_3}_{\mbox{\footnotesize{o}}\,(i)}(q) \right] \nonumber \\
\Pi^{\partial \partial W_3,\partial \partial W_3}_{(i)}(q) &=& 
Z^{\partial \partial W_3,\partial \partial W_3}_{(i)} (q^2)^3 ~+~ 
\mu^{2\epsilon} \left( Z^{W_3}_{33} \right)^2 
\Pi^{\partial \partial W_3,\partial 
\partial W_3}_{\mbox{\footnotesize{o}}\,(i)}(q) ~. 
\label{piw3w3}
\end{eqnarray} 
It is worth noting that our choice of the upper triangular form for the mixing
matrix in fact leads to a simpler renormalization group equation derivation
from the point of view of disentangling the relations to produce equations
without bare correlators. Again for this sector it is best to begin deriving
the full renormalization group equations from the final equation of 
(\ref{piw3w3}) and then systematically move to the row immediately above in
the matrix. As this exercise is equally as straightforward though more tedious
than the $\{V,W_3\}$ case, we merely record that the final renormalization
group equations are  
\begin{eqnarray}
0 &=& \mu \frac{d~}{d\mu} \Pi^{W_3,W_3}_{(i)}(q) ~+~ 
2 \gamma^{W_3}_{11}(a) \Pi^{W_3,W_3}_{(i)}(q) ~+~
2 \gamma^{W_3}_{12}(a) \Pi^{W_3,\partial W_3}_{(i)}(q) \nonumber \\
&& +~ 2 \gamma^{W_3}_{13}(a) \Pi^{W_3,\partial \partial W_3}_{(i)}(q) ~-~ 
(q^2)^3 \gamma^{W_3,W_3}_{(i)}(a) \nonumber \\
0 &=& \mu \frac{d~}{d\mu} \Pi^{W_3,\partial W_3}_{(i)}(q) ~+~ 
\left( \gamma^{W_3}_{11}(a) + \gamma^{W_3}_{22}(a) \right)
\Pi^{W_3,\partial W_3}_{(i)}(q) ~+~
\gamma^{W_3}_{12}(a) \Pi^{\partial W_3,\partial W_3}_{(i)}(q) \nonumber \\
&& +~ \gamma^{W_3}_{23}(a) 
\Pi^{\partial W_3,\partial \partial W_3}_{(i)}(q) ~+~ 
\gamma^{W_3}_{13}(a) 
\Pi^{\partial \partial W_3,\partial \partial W_3}_{(i)}(q) ~-~ 
(q^2)^3 \gamma^{W_3,\partial W_3}_{(i)}(a) \nonumber \\
0 &=& \mu \frac{d~}{d\mu} \Pi^{W_3,\partial \partial W_3}_{(i)}(q) ~+~ 
\left( \gamma^{W_3}_{11}(a) + \gamma^{W_3}_{33}(a) \right)
\Pi^{W_3,\partial \partial W_3}_{(i)}(q) ~+~
\gamma^{W_3}_{12}(a) \Pi^{\partial W_3,\partial \partial W_3}_{(i)}(q) 
\nonumber \\
&& +~ \gamma^{W_3}_{13}(a) 
\Pi^{\partial \partial W_3,\partial \partial W_3}_{(i)}(q) ~-~ 
(q^2)^3 \gamma^{W_3,\partial \partial W_3}_{(i)}(a) \nonumber \\
0 &=& \mu \frac{d~}{d\mu} \Pi^{\partial W_3,\partial W_3}_{(i)}(q) ~+~ 
2 \gamma^{W_3}_{22}(a) \Pi^{\partial W_3,\partial W_3}_{(i)}(q) ~+~
2 \gamma^{W_3}_{23}(a) \Pi^{\partial W_3,\partial \partial W_3}_{(i)}(q) 
\nonumber \\
&& -~ (q^2)^3 \gamma^{\partial W_3,\partial W_3}_{(i)}(a) \nonumber \\
0 &=& \mu \frac{d~}{d\mu} \Pi^{\partial W_3,\partial \partial W_3}_{(i)}(q) ~+~ 
\left( \gamma^{W_3}_{22}(a) + \gamma^{W_3}_{33}(a) \right)
\Pi^{\partial W_3,\partial \partial W_3}_{(i)}(q) ~+~
\gamma^{W_3}_{23}(a) \Pi^{\partial \partial W_3,\partial \partial W_3}_{(i)}(q) 
\nonumber \\
&& -~ (q^2)^3 \gamma^{\partial W_3,\partial \partial W_3}_{(i)}(a) \nonumber \\
0 &=& \mu \frac{d~}{d\mu} 
\Pi^{\partial \partial W_3,\partial \partial W_3}_{(i)}(q) ~+~ 
2 \gamma^{W_3}_{33}(a) 
\Pi^{\partial \partial W_3,\partial \partial W_3}_{(i)}(q) ~-~
(q^2)^3 \gamma^{\partial \partial W_3,\partial \partial W_3}_{(i)}(a) 
\label{w3w3rge}
\end{eqnarray}
where the correlator anomalous dimensions are 
\begin{eqnarray}
\gamma^{W_3,W_3}_{(i)}(a) &=& \left[ -~ \epsilon ~+~ \beta(a) 
\frac{\partial~}{\partial a} ~+~ 2 \gamma^{W_3}_{11}(a) 
\right] Z_{(i)}^{W_3,W_3} \nonumber \\ 
&& +~ 2 \gamma^{W_3}_{12}(a) Z_{(i)}^{W_3,\partial W_3} ~+~ 
2 \gamma^{W_3}_{13}(a) Z_{(i)}^{W_3,\partial \partial W_3} \nonumber \\
\gamma^{W_3,\partial W_3}_{(i)}(a) &=& \left[ -~ \epsilon ~+~ \beta(a) 
\frac{\partial~}{\partial a} ~+~ \gamma^{W_3}_{11}(a) ~+~ \gamma^{W_3}_{22}(a) 
\right] Z_{(i)}^{W_3,\partial W_3} \nonumber \\ 
&& +~ \gamma^{W_3}_{12}(a) Z_{(i)}^{\partial W_3,\partial W_3} ~+~ 
\gamma^{W_3}_{23}(a) Z_{(i)}^{W_3,\partial \partial W_3} ~+~ 
\gamma^{W_3}_{13}(a) Z_{(i)}^{\partial W_3,\partial \partial W_3} \nonumber \\
\gamma^{W_3,\partial \partial W_3}_{(i)}(a) &=& \left[ -~ \epsilon ~+~ \beta(a) 
\frac{\partial~}{\partial a} ~+~ \gamma^{W_3}_{11}(a) ~+~ \gamma^{W_3}_{33}(a) 
\right] Z_{(i)}^{W_3,\partial \partial W_3} \nonumber \\ 
&& +~ \gamma^{W_3}_{12}(a) Z_{(i)}^{\partial W_3,\partial \partial W_3} ~+~ 
\gamma^{W_3}_{13}(a) Z_{(i)}^{\partial \partial W_3,\partial \partial W_3} 
\nonumber \\
\gamma^{\partial W_3,\partial W_3}_{(i)}(a) &=& \left[ -~ \epsilon ~+~ \beta(a) 
\frac{\partial~}{\partial a} ~+~ 2 \gamma^{W_3}_{22}(a) 
\right] Z_{(i)}^{\partial W_3,\partial W_3} ~+~ 2 \gamma^{W_3}_{23}(a) 
Z_{(i)}^{\partial W_3,\partial \partial W_3}  \nonumber \\
\gamma^{\partial W_3,\partial \partial W_3}_{(i)}(a) &=& 
\left[ -~ \epsilon ~+~ \beta(a) \frac{\partial~}{\partial a} ~+~ 
\gamma^{W_3}_{22}(a) ~+~ \gamma^{W_3}_{33}(a) 
\right] Z_{(i)}^{\partial W_3,\partial \partial W_3} ~+~ \gamma^{W_3}_{23}(a) 
Z_{(i)}^{\partial \partial W_3,\partial \partial W_3}  \nonumber \\
\gamma^{\partial \partial W_3,\partial \partial W_3}_{(i)}(a) &=& 
\left[ -~ \epsilon ~+~ \beta(a) \frac{\partial~}{\partial a} ~+~ 
2 \gamma^{W_3}_{33}(a) \right] 
Z_{(i)}^{\partial \partial W_3,\partial \partial W_3} ~. 
\end{eqnarray} 
Comparing these final forms with the original relationships (\ref{piw3w3}) an
evident pattern emerges in the final renormalization group equations. Not all
the original bare operators of (\ref{piw3w3}) appear in the corresponding
equation. However, this is partly because the transformation to (\ref{w3w3rge})
involves the non-trivial entanglement alluded to earlier but in such a way that
no information is lost. In practice there are cancellations in the derivation 
in such a way that the coefficient of certain off-diagonal elements is zero.
Indeed given the upper triangular form of the mixing matrix and the final forms 
(\ref{w3w3rge}), one could have been tempted merely to write these down without
derivation. 

For completeness, we close this section by recording the formal definitions of
the remaining operator correlation function anomalous dimensions. As is
apparent from comparing with their $\{V,W_3\}$ and $\{W_3,W_3\}$ counterparts
there is a consistent correspondence between the terms of the anomalous
dimensions and the form of the renormalization group function itself that means
we only record the anomalous dimensions themselves for brevity and as an aid to
checking the renormalization group equations. We have for those cases involving
$W_2$  
\begin{eqnarray}
\gamma^{V,W_2}(a) &=& \left[ -~ \epsilon ~+~ \beta(a) 
\frac{\partial~}{\partial a} ~+~ \gamma^{V}(a) ~+~ \gamma^{W_2}_{11}(a) 
\right] Z^{V,W_2} ~+~ \gamma^{W_2}_{12}(a) Z^{V,\partial W_2} \nonumber \\ 
\gamma^{V,\partial W_2}(a) &=& \left[ -~ \epsilon ~+~ \beta(a) 
\frac{\partial~}{\partial a} ~+~ \gamma^{V}(a) ~+~ \gamma^{W_2}_{22}(a) 
\right] Z^{V,\partial W_2} 
\end{eqnarray} 
and
\begin{eqnarray}
\gamma_{(i)}^{W_2,W_2}(a) &=& \left[ -~ \epsilon ~+~ \beta(a) 
\frac{\partial~}{\partial a} ~+~ 2 \gamma^{W_2}_{11}(a) \right] 
Z_{(i)}^{W_2,W_2} ~+~ 2 \gamma^{W_2}_{12}(a) Z_{(i)}^{W_2,\partial W_2} 
\nonumber \\
\gamma_{(i)}^{W_2,\partial W_2}(a) &=& \left[ -~ \epsilon ~+~ \beta(a) 
\frac{\partial~}{\partial a} ~+~ \gamma^{W_2}_{11}(a) ~+~ \gamma^{W_2}_{22}(a)
\right] Z_{(i)}^{W_2,\partial W_2} ~+~ \gamma^{W_2}_{12}(a) 
Z_{(i)}^{\partial W_2,\partial W_2} \nonumber \\
\gamma_{(i)}^{\partial W_2,\partial W_2}(a) &=& \left[ -~ \epsilon ~+~ \beta(a) 
\frac{\partial~}{\partial a} ~+~ 2 \gamma^{W_2}_{22}(a) 
\right] Z_{(i)}^{\partial W_2,\partial W_2} ~. 
\end{eqnarray} 
Finally, for $T_2$ we have 
\begin{eqnarray}
\gamma^{T,T_2}_{(i)}(a) &=& \left[ -~ \epsilon ~+~ 
\beta(a) \frac{\partial~}{\partial a} ~+~ \gamma^T(a) ~+~ \gamma^{T_2}_{11}(a) 
\right] Z_{(i)}^{T,T_2} ~+~ \gamma^{T_2}_{12}(a) Z_{(i)}^{T,\partial T_2} 
\nonumber \\
\gamma^{T,\partial T_2}_{(i)}(a) &=& \left[ -~ \epsilon ~+~ 
\beta(a) \frac{\partial~}{\partial a} ~+~ \gamma^T(a) ~+~ \gamma^{T_2}_{22}(a) 
\right] Z_{(i)}^{T,\partial T_2} ~. 
\end{eqnarray} 

\sect{Results.}

We now turn to the mundane task of recording all our results for the operator
correlation functions. These are broken into subsections where the first named
operator of the title corresponds to the operator ${\cal O}^1$ of Figure $1$.
In each section, we provide the finite renormalized amplitudes with respect to 
the various projections and then the associated correlator anomalous dimension.
In recording the finite parts of all our correlators we show the overall 
dimension of the amplitude by explicitly factorizing off the overall $q^2$ 
dependence which is not the same for each sector. We note that we included the
powers of $q^2$ in the contact term of the relation between bare and 
renormalized amplitudes in the basic relation in order to identify the operator 
correlator anomalous dimensions in analysing the renormalization group 
structure. For certain sectors due to the nature of the total derivative 
operators the explicit form of some amplitudes appear in an earlier subsection 
since there is a clear relation to $O(a^2)$ with the explicit value of the 
amplitudes. In certain cases, such as the vector correlator, the vanishing of 
the correlator projection to three loops is actually an all orders feature due 
to symmetry. In other cases where relations hold to $O(a^2)$ this may be valid 
to all orders but we make no assertion beyond the order we have calculated to. 
Finally, for completeness as well as for comparing conventions, we also display
results for $\{S,S\}$ and $\{V,V\}$ which are in agreement with, 
\cite{26,27,28}. With $d(R)$ the dimension of the quark representation and
\begin{equation}
\ell ~=~ \ln \left( \frac{\mu^2}{q^2} \right) 
\end{equation}
we have: 

\subsection{Scalar-Scalar.}

\begin{equation}
\Pi^{S,S}(q) ~=~ q^2 \tilde{\Pi}^{S,S}(a)
\end{equation}

\begin{eqnarray}
\tilde{\Pi}^{S,S}(a) &=& d(R) \left[ 4 + 2\ell ~+~ C_F \left[ \frac{131}{2} 
- 24 \zeta(3) + 34 \ell + 6 \ell^2 \right] a \right. \nonumber \\
&& \left. ~~~~~~~+~ C_F \left[ \left( 64 \zeta(3) - \frac{2044}{9}
- 130 \ell + 32 \zeta(3) \ell - \frac{88}{3} \ell^2 - \frac{8}{3} \ell^3 
\right) T_F \Nf \right. \right. \nonumber \\
&& \left. \left. ~~~~~~~~~~~~~~~~~+~ \left( \frac{14419}{18} - 300 \zeta(3) 
- 18 \zeta(4) - 40 \zeta(5) + \frac{893}{2} \ell \right. \right. \right.
\nonumber \\
&& \left. \left. \left. ~~~~~~~~~~~~~~~~~~~~~~~~-~ 124 \zeta(3) \ell 
+ \frac{284}{3} \ell^2 + \frac{22}{3} \ell^3 \right) C_A \right. \right. 
\nonumber \\
&& \left. \left. ~~~~~~~~~~~~~~~~~+~ \left( \frac{1613}{4} - 384 \zeta(3)
+ 36 \zeta(4) + 240 \zeta(5) + \frac{691}{2} \ell \right. \right. \right.
\nonumber \\
&& \left. \left. \left. ~~~~~~~~~~~~~~~~~~~~~~~~-~ 72 \zeta(3) \ell
+ 105 \ell^2 + 12 \ell^3 \right) C_F \right] a^2 \right] ~+~ O(a^3)
\end{eqnarray}

\begin{eqnarray}
\gamma^{S,S}(a) &=& d(R) \left[ 2 ~+~ 10 C_F a ~+~ \frac{C_F}{2} \left[ 
( 154 - 72 \zeta(3) ) C_A \right. \right. \nonumber \\
&& \left. \left. ~~~~~~~+~ ( 144 \zeta(3) - 119 ) C_F - 32 T_F \Nf \right] a^2 
\right] ~+~ O(a^3) ~.
\end{eqnarray}

\subsection{Vector-Vector.}

\begin{equation}
\Pi^{V,V}_{(i)}(q) ~=~ q^2 \tilde{\Pi}_{(i)}^{V,V}(a)
\end{equation}

\begin{eqnarray}
\tilde{\Pi}^{V,V}_{(1)}(a) &=& -~ 2 \tilde{\Pi}^{V,W_2}_{(1)}(a) ~=~ -~ 
\tilde{\Pi}^{V,\partial W_2}_{(1)}(a) \nonumber \\
&=& d(R) \left[ -~ \frac{20}{9} - \frac{4}{3} \ell ~+~ C_F \left[ 16 \zeta(3) 
- \frac{55}{3} - 4 \ell \right] a \right. \nonumber \\
&& \left. ~~~~~~~+~ C_F \left[ \left( \frac{7402}{81} - \frac{608}{9}
\zeta(3) + \frac{88}{3} \ell - \frac{64}{3} \zeta(3) \ell + \frac{8}{3} \ell^2
\right) T_F \Nf \right. \right. \nonumber \\
&& \left. \left. ~~~~~~~~~~~~~~~~~+~ \left( \frac{1816}{9} \zeta(3) 
+ \frac{80}{3} \zeta(5) - \frac{44215}{162} - 82 \ell + \frac{176}{3} \zeta(3)
\ell - \frac{22}{3} \ell^2 \right) C_A 
\right. \right. \nonumber \\
&& \left. \left. ~~~~~~~~~~~~~~~~~+~ \left( \frac{286}{9} + \frac{296}{3} 
\zeta(3) - 160 \zeta(5) + 2 \ell \right) C_F \right] a^2 \right] ~+~ O(a^3)
\nonumber \\ 
\tilde{\Pi}^{V,V}_{(2)}(a) &=& -~ 2 \tilde{\Pi}^{V,W_2}_{(2)}(a) ~=~ -~ 
\tilde{\Pi}^{V,\partial W_2}_{(2)}(a) ~=~ O(a^3)
\end{eqnarray}

\begin{equation}
\gamma^{V,V}(a) ~=~ d(R) \left[ -~ \frac{4}{3} ~-~ 4 C_F a ~+~ 
\frac{C_F}{9} \left[ 18 C_F - 133 C_A + 44 T_F \Nf \right] a^2 \right] ~+~ 
O(a^3) ~. 
\end{equation}

\subsection{Tensor-Tensor.}

\begin{equation}
\Pi^{T,T}_{(i)}(q) ~=~ q^2 \tilde{\Pi}_{(i)}^{T,T}(a)
\end{equation}

\begin{eqnarray}
\tilde{\Pi}^{T,T}_{(1)}(a) &=& d(R) \left[ -~ \frac{4}{9} - \frac{2}{3} \ell ~+~ 
C_F \left[ 8 \zeta(3) - \frac{491}{54} - \frac{14}{9} \ell + \frac{2}{3} \ell^2
\right] a \right. \nonumber \\
&& \left. ~~~~~~~+~ C_F \left[ \left( \frac{10672}{243} - \frac{1024}{27}
\zeta(3) + \frac{766}{81} \ell - \frac{32}{3} \zeta(3) \ell 
- \frac{8}{9} \ell^2 - \frac{8}{27} \ell^3 \right) T_F \Nf \right. \right. 
\nonumber \\
&& \left. \left. ~~~~~~~~~~~~~~~~~+~ \left( \frac{2732}{27} \zeta(3) 
- \frac{14}{3} \zeta(4) + \frac{40}{3} \zeta(5) - \frac{19427}{162}
\right. \right. \right. \nonumber \\
&& \left. \left. \left. ~~~~~~~~~~~~~~~~~~~~~~~~-~ \frac{1771}{162} \ell 
+ 20 \zeta(3) \ell + \frac{20}{3} \ell^2 + \frac{22}{27} \ell^3 \right) C_A 
\right. \right. \nonumber \\
&& \left. \left. ~~~~~~~~~~~~~~~~~+~ \left( \frac{608}{9} \zeta(3)
+ \frac{28}{3} \zeta(4) - 80 \zeta(5) - \frac{15973}{972} 
- \frac{1075}{54} \ell \right. \right. \right. \nonumber \\
&& \left. \left. \left. ~~~~~~~~~~~~~~~~~~~~~~~~+~ \frac{8}{3} \zeta(3) \ell 
- \frac{43}{9} \ell^2 - \frac{4}{9} \ell^3 \right) C_F \right] a^2 \right] ~+~ 
O(a^3)
\label{pitt1}
\end{eqnarray}

\begin{eqnarray}
\tilde{\Pi}^{T,T}_{(2)}(a) &=& d(R) \left[ \frac{20}{9} + \frac{4}{3} \ell ~+~ 
C_F \left[ \frac{593}{27} - 16 \zeta(3) + \frac{28}{9} \ell 
- \frac{4}{3} \ell^2 \right] a \right. \nonumber \\
&& \left. ~~~~~~~+~ C_F \left[ \left( \frac{2048}{27} \zeta(3)
- \frac{21328}{243} - \frac{1532}{81} \ell + \frac{64}{3} \zeta(3) \ell
+ \frac{16}{9} \ell^2 + \frac{16}{27} \ell^3 \right) T_F \Nf \right. \right. 
\nonumber \\
&& \left. \left. ~~~~~~~~~~~~~~~~~+~ \left( \frac{58075}{243} 
- \frac{5296}{27} \zeta(3) + \frac{28}{3} \zeta(4) - \frac{80}{3} \zeta(5) 
\right. \right. \right. \nonumber \\
&& \left. \left. \left. ~~~~~~~~~~~~~~~~~~~~~~~+~ \frac{1771}{81} \ell 
- 40 \zeta(3) \ell - \frac{40}{3} \ell^2 - \frac{44}{27} \ell^3 \right) C_A 
\right. \right. \nonumber \\
&& \left. \left. ~~~~~~~~~~~~~~~~~+~ \left( \frac{22051}{486} 
- \frac{1328}{9} \zeta(3) - \frac{56}{3} \zeta(4) + 160 \zeta(5)
+ \frac{1075}{27} \ell \right. \right. \right. \nonumber \\
&& \left. \left. \left. ~~~~~~~~~~~~~~~~~~~~~~~-~ \frac{16}{3} \zeta(3) \ell 
+ \frac{86}{9} \ell^2 + \frac{8}{9} \ell^3 \right) C_F \right] a^2 \right] ~+~ 
O(a^3)
\label{pitt2}
\end{eqnarray}

\begin{eqnarray}
\gamma^{T,T}_{(1)}(a) &=& d(R) \left[ -~ \frac{2}{3} ~-~ \frac{22}{9} C_F a ~+~ 
\frac{C_F}{162} \left[ ( 3024 \zeta(3) - 4803 ) C_F \right. \right. 
\nonumber \\
&& \left. \left. ~~~~~~~+~ ( 1574 - 1512 \zeta(3) ) C_A - 16 T_F \Nf \right] 
a^2 \right] ~+~ O(a^3) \nonumber \\
\gamma^{T,T}_{(2)}(a) &=& d(R) \left[ \frac{4}{3} ~+~ \frac{68}{9} C_F a ~+~ 
\frac{C_F}{81} \left[ ( 1512 \zeta(3) + 388 ) C_A \right. \right. \nonumber \\
&& \left. \left. ~~~~~~~+~ ( 3363 - 3024 \zeta(3) ) C_F - 200 T_F \Nf \right] 
a^2 \right] ~+~ O(a^3) ~.
\end{eqnarray}

\subsection{Vector-Wilson $2$.}

\begin{equation}
\Pi^{V,W_2}_{(i)}(q) ~=~ q^2 \tilde{\Pi}_{(i)}^{V,W_2}(a)
\end{equation}

\begin{eqnarray}
\gamma^{V,W_2}(a) &=& d(R) \left[ \frac{2}{3} ~+~ 2 C_F a ~+~ 
\frac{C_F}{18} \left[ 133 C_A - 18 C_F - 44 T_F \Nf \right] a^2 \right] ~+~ 
O(a^3) \nonumber \\
\gamma^{V,\partial W_2}(a) &=& d(R) \left[ \frac{4}{3} ~+~ 4 C_F a ~+~ 
\frac{C_F}{9} \left[ 133 C_A - 18 C_F - 44 T_F \Nf \right] a^2 \right] ~+~ 
O(a^3) ~. 
\end{eqnarray} 

\subsection{Wilson $2$-Wilson $2$.}

\begin{equation}
\Pi^{W_2,W_2}_{(i)}(q) ~=~ (q^2)^2 \tilde{\Pi}_{(i)}^{W_2,W_2}(a)
\end{equation}

\begin{eqnarray}
\tilde{\Pi}^{W_2,W_2}_{(1)}(a) &=& d(R) \left[ \frac{12}{25} + \frac{1}{5} \ell ~-~ 
C_F \left[ \frac{12}{5} \zeta(3) + \frac{17533}{13500} + \frac{473}{225} \ell
+ \frac{8}{15} \ell^2 \right] a \right. \nonumber \\
&& \left. ~~~~~~~+~ C_F \left[ \left( \frac{419327}{303750} 
+ \frac{1816}{135} \zeta(3) + \frac{69266}{10125} \ell \right. \right. \right.
\nonumber \\
&& \left. \left. \left. ~~~~~~~~~~~~~~~~~~~+~ \frac{16}{5} \zeta(3) \ell 
+ \frac{1586}{675} \ell^2 + \frac{32}{135} \ell^3 \right) T_F \Nf 
\right. \right. \nonumber \\
&& \left. \left. ~~~~~~~~~~~~~~~~~+~ \left( \frac{16}{5} \zeta(4) 
- \frac{15838}{675} \zeta(3) - 4 \zeta(5) - \frac{3541817}{1215000}
- \frac{399953}{20250} \ell \right. \right. \right. \nonumber \\
&& \left. \left. \left. ~~~~~~~~~~~~~~~~~~~~~~~-~ \frac{12}{5} \zeta(3) \ell
- \frac{8963}{1350} \ell^2 - \frac{88}{135} \ell^3 \right) C_A 
\right. \right. \nonumber \\
&& \left. \left. ~~~~~~~~~~~~~~~~~+~ \left( \frac{12235087}{455625}
- \frac{2606}{135} \zeta(3) - \frac{32}{5} \zeta(4) + 24 \zeta(5)
+ \frac{383653}{20250} \ell \right. \right. \right. \nonumber \\
&& \left. \left. \left. ~~~~~~~~~~~~~~~~~~~~~~~+~ \frac{1448}{225} \ell^2
+ \frac{128}{135} \ell^3 \right) C_F \right] a^2 \right] ~+~  O(a^3) 
\end{eqnarray}

\begin{eqnarray}
\tilde{\Pi}^{W_2,W_2}_{(2)}(a) &=& d(R) \left[ \frac{92}{225} + \frac{2}{15} \ell ~-~ 
C_F \left[ \frac{8}{5} \zeta(3) + \frac{15683}{20250} + \frac{986}{675} \ell
+ \frac{16}{45} \ell^2 \right] a \right. \nonumber \\
&& \left. ~~~~~~~+~ C_F \left[ \left( \frac{1685066}{1366875}
+ \frac{3632}{405} \zeta(3) + \frac{48404}{10125} \ell \right. \right. \right.
\nonumber \\
&& \left. \left. \left. ~~~~~~~~~~~~~~~~~~~+~ \frac{32}{15} \zeta(3) \ell
+ \frac{1084}{675} \ell^2 + \frac{64}{405} \ell^3 \right) T_F \Nf 
\right. \right. \nonumber \\
&& \left. \left. ~~~~~~~~~~~~~~~~~+~ \left( \frac{32}{15} \zeta(4) 
- \frac{29276}{2025} \zeta(3) - \frac{8}{5} \zeta(5) - \frac{7819793}{2733750} 
- \frac{46547}{3375} \ell \right. \right. \right. \nonumber \\
&& \left. \left. \left. ~~~~~~~~~~~~~~~~~~~~~~~-~ \frac{8}{5} \zeta(3) \ell
- \frac{3061}{675} \ell^2 - \frac{176}{405} \ell^3 \right) C_A 
\right. \right. \nonumber \\
&& \left. \left. ~~~~~~~~~~~~~~~~~+~ \left( \frac{8792588}{455625}
- \frac{5788}{405} \zeta(3) - \frac{64}{15} \zeta(4) + 16 \zeta(5)
+ \frac{44597}{3375} \ell \right. \right. \right. \nonumber \\
&& \left. \left. \left. ~~~~~~~~~~~~~~~~~~~~~~~+~ \frac{9008}{2025} \ell^2
+ \frac{256}{405} \ell^3 \right) C_F \right] a^2 \right] ~+~ O(a^3)
\end{eqnarray}

\begin{eqnarray}
\tilde{\Pi}^{W_2,W_2}_{(3)}(a) &=& d(R) \left[ \frac{17}{225} + \frac{2}{15} \ell ~-~ 
C_F \left[ \frac{8}{5} \zeta(3) - \frac{34439}{6750} - \frac{638}{225} \ell
- \frac{8}{15} \ell^2 \right] a \right. \nonumber \\
&& \left. ~~~~~~~+~ C_F \left[ \left( \frac{464}{135} \zeta(3)
- \frac{6263527}{303750} - \frac{128096}{10125} \ell \right. \right. \right.
\nonumber \\
&& \left. \left. \left. ~~~~~~~~~~~~~~~~~~~+~ \frac{32}{15} \zeta(3) \ell
- \frac{1916}{675} \ell^2 - \frac{32}{135} \ell^3 \right) T_F \Nf 
\right. \right. \nonumber \\
&& \left. \left. ~~~~~~~~~~~~~~~~~+~ \left( \frac{918157}{15000} 
- \frac{17492}{675} \zeta(3) - \frac{16}{5} \zeta(4) - \frac{8}{3} \zeta(5) 
+ \frac{364234}{10125} \ell \right. \right. \right. \nonumber \\
&& \left. \left. \left. ~~~~~~~~~~~~~~~~~~~~~~~-~ \frac{184}{15} \zeta(3) \ell
+ \frac{5389}{675} \ell^2 + \frac{88}{135} \ell^3 \right) C_A \right. \right. 
\nonumber \\
&& \left. \left. ~~~~~~~~~~~~~~~~~+~ \left( 16 \zeta(5) + \frac{32}{5}
\zeta(4) - \frac{436}{135} \zeta(3) - \frac{28297949}{911250}
- \frac{160709}{10125} \ell \right. \right. \right. \nonumber \\
&& \left. \left. \left. ~~~~~~~~~~~~~~~~~~~~~~~-~ \frac{1288}{225} \ell^2
- \frac{128}{135} \ell^3 \right) C_F \right] a^2 \right] ~+~ O(a^3) 
\end{eqnarray}

\begin{eqnarray}
\tilde{\Pi}^{W_2,\partial W_2}_{(1)}(a) &=& \tilde{\Pi}^{W_2,\partial W_2}_{(2)}(a) ~=~ O(a^3) 
\nonumber \\
\tilde{\Pi}^{W_2,\partial W_2}_{(3)}(a) &=& d(R) \left[ \frac{10}{9} 
+ \frac{2}{3} \ell ~+~ C_F \left[ \frac{55}{6} - 8 \zeta(3) + 2 \ell \right] a 
\right. \nonumber \\
&& \left. ~~~~~~~+~ C_F \left[ \left( \frac{304}{9} \zeta(3)
- \frac{3701}{81} - \frac{44}{3} \ell + \frac{32}{3} \zeta(3) \ell
- \frac{4}{3} \ell^2 \right) T_F \Nf \right. \right. \nonumber \\
&& \left. \left. ~~~~~~~~~~~~~~~~~+~ \left( \frac{44215}{324} 
- \frac{908}{9} \zeta(3) - \frac{40}{3} \zeta(5) + 41 \ell 
- \frac{88}{3} \zeta(3) \ell + \frac{11}{3} \ell^2 \right) C_A \right. \right. 
\nonumber \\
&& \left. \left. ~~~~~~~~~~~~~~~~~+~ \left( 80 \zeta(5) 
- \frac{148}{3} \zeta(3) - \frac{143}{9} - \ell \right) C_F \right] a^2 
\right] ~+~ O(a^3) 
\end{eqnarray}

\begin{eqnarray}
\tilde{\Pi}^{\partial W_2,\partial W_2}_{(1)}(a) &=& 
\tilde{\Pi}^{\partial W_2,\partial W_2}_{(2)}(a) ~=~ O(a^3) \nonumber \\
\tilde{\Pi}^{\partial W_2,\partial W_2}_{(3)}(a) &=& 
2 \tilde{\Pi}^{W_2,\partial W_2}_{(3)}(a) ~+~ O(a^3)
\end{eqnarray} 

\begin{eqnarray}
\gamma^{W_2,W_2}_{(1)}(a) &=& d(R) \left[ \frac{1}{5} ~+~ \frac{103}{225} 
C_F a ~+~ \frac{C_F}{40500} \left[ ( 259200 \zeta(3) - 65603 ) C_A 
\right. \right. \nonumber \\
&& \left. \left. ~~~~~~~+~ ( 325498 - 518400 \zeta(3) ) C_F 
+ 22612 T_F \Nf \right] a^2 \right] ~+~ O(a^3) \nonumber \\ 
\gamma^{W_2,W_2}_{(2)}(a) &=& d(R) \left[ \frac{2}{15} ~+~ \frac{18}{25} 
C_F a ~+~ \frac{C_F}{60750} \left[ ( 26507 + 259200 \zeta(3) ) C_A 
\right. \right. \nonumber \\
&& \left. \left. ~~~~~~~+~ ( 345738 - 518400 \zeta(3) ) C_F 
- 7828 T_F \Nf \right] a^2 \right] ~+~ O(a^3) \nonumber \\ 
\gamma^{W_2,W_2}_{(3)}(a) &=& d(R) \left[ \frac{2}{15} ~+~ \frac{62}{225} 
C_F a ~+~ \frac{C_F}{20250} \left[ ( 78919 - 129600 \zeta(3) ) C_A 
\right. \right. \nonumber \\
&& \left. \left. ~~~~~~~+~ ( 259200 \zeta(3) - 184754 ) C_F 
- 26276 T_F \Nf \right] a^2 \right] ~+~ O(a^3) \nonumber \\ 
\gamma^{W_2,\partial W_2}_{(1)}(a) &=& \gamma^{W_2,\partial W_2}_{(2)}(a) ~=~ 0 \nonumber \\
\gamma^{W_2,\partial W_2}_{(3)}(a) &=& d(R) \left[ \frac{2}{3} ~+~ 2 C_F a ~+~ 
\frac{C_F}{18} \left[ 133 C_A - 18 C_F - 44 T_F \Nf \right] a^2 \right] \,+\, 
O(a^3) \nonumber \\ 
\gamma^{\partial W_2,\partial W_2}_{(1)}(a) &=& 
\gamma^{\partial W_2,\partial W_2}_{(2)}(a) ~=~ 0 \nonumber \\
\gamma^{\partial W_2,\partial W_2}_{(3)}(a) &=& d(R) \left[ \frac{4}{3} ~+~ 
4 C_F a ~+~ \frac{C_F}{9} \left[ 133 C_A - 18 C_F - 44 T_F \Nf \right] a^2 
\right] \nonumber \\
&& +~ O(a^3) ~. 
\end{eqnarray} 

\subsection{Vector-Wilson $3$.}

\begin{equation}
\Pi^{V,W_3}_{(i)}(q) ~=~ (q^2)^2 \tilde{\Pi}_{(i)}^{V,W_3}(a)
\end{equation}

\begin{eqnarray}
\tilde{\Pi}^{V,W_3}_{(1)}(a) &=& -~ \tilde{\Pi}^{V,W_3}_{(2)}(a) \nonumber \\ 
&=& d(R) \left[ \frac{31}{675} + \frac{1}{45} \ell ~+~ 
C_F \left[ \frac{2177}{6480} - \frac{4}{15} \zeta(3) + \frac{2}{27} \ell
\right] a \right. \nonumber \\
&& \left. ~~~~~~~+~ C_F \left[ \left( \frac{152}{135} \zeta(3)
- \frac{4070273}{2624400} - \frac{7267}{14580} \ell 
+ \frac{16}{45} \zeta(3) \ell - \frac{11}{243} \ell^2 \right) T_F \Nf 
\right. \right. \nonumber \\
&& \left. \left. ~~~~~~~~~~~~~~~~~+~ \left( \frac{48524449}{10497600} 
- \frac{1364}{405} \zeta(3) - \frac{4}{9} \zeta(5) \right. \right. \right. 
\nonumber \\
&& \left. \left. \left. ~~~~~~~~~~~~~~~~~~~~~~~+~ \frac{4051}{2916} \ell 
- \frac{44}{45} \zeta(3) \ell + \frac{121}{972} \ell^2 \right) C_A 
\right. \right. \nonumber \\
&& \left. \left. ~~~~~~~~~~~~~~~~~+~ \left( \frac{8}{3} \zeta(5)
- \frac{1118783}{2624400} - \frac{140}{81} \zeta(3) 
- \frac{187}{19440} \ell \right) C_F \right] a^2 \right] \nonumber \\
&& +~ O(a^3) \nonumber \\
\tilde{\Pi}^{V,\partial W_3}_{(1)}(a) &=& -~ \tilde{\Pi}^{V,\partial W_3}_{(2)}(a) ~=~  
\frac{1}{2} \tilde{\Pi}^{V,\partial \partial W_3}_{(1)}(a) ~=~ -~ \frac{1}{2} 
\tilde{\Pi}^{V,\partial \partial W_3}_{(2)}(a) \nonumber \\ 
&=& d(R) \left[ \frac{2}{27} + \frac{1}{27} \ell ~+~ 
C_F \left[ \frac{19}{36} - \frac{4}{9} \zeta(3) + \frac{1}{9} \ell
\right] a \right. \nonumber \\
&& \left. ~~~~~~~+~ C_F \left[ \left( \frac{152}{81} \zeta(3)
- \frac{3719}{1458} - \frac{22}{27} \ell + \frac{16}{27} \zeta(3) \ell 
- \frac{2}{27} \ell^2 \right) T_F \Nf \right. \right. \nonumber \\
&& \left. \left. ~~~~~~~~~~~~~~~~~+~ \left( \frac{44437}{5832} 
- \frac{454}{81} \zeta(3) - \frac{20}{27} \zeta(5) \right. \right. \right. 
\nonumber \\
&& \left. \left. \left. ~~~~~~~~~~~~~~~~~~~~~~~+~ \frac{41}{18} \ell 
- \frac{44}{27} \zeta(3) \ell + \frac{11}{54} \ell^2 \right) C_A 
\right. \right. \nonumber \\
&& \left. \left. ~~~~~~~~~~~~~~~~~+~ \left( \frac{40}{9} \zeta(5)
- \frac{8}{9} - \frac{74}{27} \zeta(3) 
- \frac{1}{18} \ell \right) C_F \right] a^2 \right] \nonumber \\
&& +~ O(a^3)
\end{eqnarray}

\begin{eqnarray}
\gamma^{V,W_3}_{(1)}(a) &=& -~ \gamma^{V,W_3}_{(2)}(a) ~=~ d(R) \left[ 
\frac{1}{45} ~+~ \frac{13}{162} C_F a \right. \nonumber \\
&& \left. ~~~~~~~~~~~~~~~~~~~~~~~~~~~~+~ \frac{C_F}{116640} \left[ 
27062 C_A + 239 C_F - 9136 T_F \Nf \right] a^2 \right] \,+\, O(a^3) 
\nonumber \\
\gamma^{V,\partial W_3}_{(1)}(a) &=& -~ \gamma^{V,\partial W_3}_{(2)}(a) ~=~ 
\frac{1}{2} \gamma^{V,\partial \partial W_3}_{(1)}(a) ~=~ -~ \frac{1}{2} 
\gamma^{V,\partial \partial W_3}_{(2)}(a) \nonumber \\
&=& d(R) \left[ \frac{1}{27} ~+~ \frac{1}{9} C_F a + \frac{C_F}{108} \left[ 
37 C_A - 6 C_F - 12 T_F \Nf \right] a^2 \right] ~+~ O(a^3) ~.
\end{eqnarray}

\subsection{Wilson $3$-Wilson $3$.}

\begin{equation}
\Pi^{W_3,W_3}_{(i)}(q) ~=~ (q^2)^3 \tilde{\Pi}_{(i)}^{W_3,W_3}(a)
\end{equation}

\begin{eqnarray}
\tilde{\Pi}^{W_3,W_3}_{(1)}(a) &=& d(R) \left[ \frac{457}{396900} 
+ \frac{1}{3780} \ell ~-~ C_F \left[ \frac{4831049}{200037600} 
+ \frac{1}{315} \zeta(3) + \frac{3}{196} \ell + \frac{1}{378} \ell^2 \right] a
\right. \nonumber \\
&& \left. ~~~~~~~+~ C_F \left[ \left( \frac{254}{8505} \zeta(3)
+ \frac{61767749}{675126900} + \frac{45362}{694575} \ell 
\right. \right. \right. \nonumber \\ 
&& \left. \left. \left. ~~~~~~~~~~~~~~~~~~+~ \frac{4}{945} \zeta(3) \ell 
+ \frac{5339}{357210} \ell^2 + \frac{2}{1701} \ell^3 \right) T_F \Nf 
\right. \right. \nonumber \\
&& \left. \left. ~~~~~~~~~~~~~~~~~+~ \left( \frac{277}{23814} \zeta(3)
+ \frac{1}{63} \zeta(4) - \frac{1}{189} \zeta(5) 
- \frac{27125381251}{108020304000} \right. \right. \right. \nonumber \\
&& \left. \left. \left. ~~~~~~~~~~~~~~~~~~~~~~~-~ \frac{17694461}{100018800} 
\ell + \frac{19}{945} \zeta(3) \ell - \frac{14383}{357210} \ell^2
- \frac{11}{3402} \ell^3 \right) C_A \right. \right. \nonumber \\
&& \left. \left. ~~~~~~~~~~~~~~~~~+~ \left(
\frac{84198061049}{216040608000} + \frac{2}{63} \zeta(5)
- \frac{2}{63} \zeta(4) - \frac{59}{945} \zeta(3) 
\right. \right. \right. \nonumber \\
&& \left. \left. \left. ~~~~~~~~~~~~~~~~~~~~~~~+~
\frac{54952451}{240045120} \ell + \frac{3959}{63504} \ell^2
+ \frac{25}{3402} \ell^3 \right) C_F \right] a^2 \right] ~ +~ O(a^3) 
\nonumber \\
\tilde{\Pi}^{W_3,W_3}_{(2)}(a) &=& d(R) \left[ -~ \frac{599}{132300} 
- \frac{1}{630} \ell ~+~ C_F \left[ \frac{7012477}{133358400} 
+ \frac{2}{105} \zeta(3) + \frac{5851}{158760} \ell + \frac{5}{756} \ell^2 
\right] a \right. \nonumber \\
&& \left. ~~~~~~~+~ C_F \left[ -~ \left( \frac{1034}{8505} \zeta(3)
+ \frac{18993505}{100018800} + \frac{5117891}{33339600} \ell 
\right. \right. \right. \nonumber \\ 
&& \left. \left. \left. ~~~~~~~~~~~~~~~~~~~~~~~+~ \frac{8}{315} \zeta(3) \ell 
+ \frac{2189}{59535} \ell^2 + \frac{5}{1701} \ell^3 \right) T_F \Nf 
\right. \right. \nonumber \\
&& \left. \left. ~~~~~~~~~~~~~~~~~+~ \left( \frac{6793}{59535} \zeta(3)
- \frac{5}{126} \zeta(4) + \frac{2}{63} \zeta(5) 
+ \frac{269779943}{533433600} \right. \right. \right. \nonumber \\
&& \left. \left. \left. ~~~~~~~~~~~~~~~~~~~~~~~+~ \frac{3463261}{8334900} \ell 
- \frac{1}{105} \zeta(3) \ell + \frac{94321}{952560} \ell^2
+ \frac{55}{6804} \ell^3 \right) C_A \right. \right. \nonumber \\
&& \left. \left. ~~~~~~~~~~~~~~~~~+~ \left( \frac{6619}{34020} \zeta(3)
+ \frac{5}{63} \zeta(4) - \frac{4}{21} \zeta(5) - \frac{5595369371}{5761082880} 
\right. \right. \right. \nonumber \\
&& \left. \left. \left. ~~~~~~~~~~~~~~~~~~~~~~~-~
\frac{95885941}{160030080} \ell - \frac{20453}{127008} \ell^2
- \frac{125}{6804} \ell^3 \right) C_F \right] a^2 \right] ~+~ O(a^3) 
\nonumber \\
\tilde{\Pi}^{W_3,W_3}_{(3)}(a) &=& d(R) \left[ \frac{4051}{396900} 
+ \frac{13}{3780} \ell ~-~ C_F \left[ \frac{5103787}{200037600} 
+ \frac{13}{315} \zeta(3) + \frac{63503}{2381400} \ell 
+ \frac{41}{5670} \ell^2 \right] a \right. \nonumber \\
&& \left. ~~~~~~~+~ C_F \left[ \left( \frac{5594}{25515} \zeta(3)
- \frac{9014474417}{270050760000} + \frac{3187379}{41674500} \ell 
\right. \right. \right. \nonumber \\ 
&& \left. \left. \left. ~~~~~~~~~~~~~~~~~~+~ \frac{52}{945} \zeta(3) \ell 
+ \frac{109283}{3572100} \ell^2 + \frac{82}{25515} \ell^3 \right) T_F \Nf 
\right. \right. \nonumber \\
&& \left. \left. ~~~~~~~~~~~~~~~~~+~ \left( \frac{16759291981}{154314720000}
- \frac{364031}{893025} \zeta(3) + \frac{41}{945} \zeta(4) 
- \frac{13}{189} \zeta(5) \right. \right. \right. \nonumber \\
&& \left. \left. \left. ~~~~~~~~~~~~~~~~~~~~~~~-~ \frac{230532091}{1000188000} 
\ell - \frac{61}{945} \zeta(3) \ell - \frac{1261123}{14288400} \ell^2
- \frac{451}{51030} \ell^3 \right) C_A \right. \right. \nonumber \\
&& \left. \left. ~~~~~~~~~~~~~~~~~+~ \left(
\frac{221977783933}{1080203040000} + \frac{26}{63} \zeta(5)
- \frac{82}{945} \zeta(4) - \frac{8251}{25515} \zeta(3) 
\right. \right. \right. \nonumber \\
&& \left. \left. \left. ~~~~~~~~~~~~~~~~~~~~~~~+~
\frac{19415531}{111132000} \ell + \frac{159653}{2381400} \ell^2
+ \frac{521}{51030} \ell^3 \right) C_F \right] a^2 \right] ~+~ O(a^3)
\nonumber \\
\tilde{\Pi}^{W_3,W_3}_{(4)}(a) &=& d(R) \left[ -~ \frac{233}{26460} 
- \frac{1}{252} \ell ~+~ C_F \left[ \frac{1}{21} \zeta(3) 
- \frac{6186559}{666792000} + \frac{6283}{264600} \ell + \frac{31}{3780} \ell^2 
\right] a \right. \nonumber \\
&& \left. ~~~~~~~+~ C_F \left[ \left( \frac{577820077}{5000940000}
- \frac{2144}{8505} \zeta(3) - \frac{2806243}{55566000} \ell 
\right. \right. \right. \nonumber \\ 
&& \left. \left. \left. ~~~~~~~~~~~~~~~~~~-~ \frac{4}{63} \zeta(3) \ell 
- \frac{8921}{297675} \ell^2 - \frac{31}{8505} \ell^3 \right) T_F \Nf 
\right. \right. \nonumber \\
&& \left. \left. ~~~~~~~~~~~~~~~~~+~ \left( \frac{314851}{595350} \zeta(3)
- \frac{31}{630} \zeta(4) + \frac{5}{63} \zeta(5) 
- \frac{5651256311}{17146080000} \right. \right. \right. \nonumber \\
&& \left. \left. \left. ~~~~~~~~~~~~~~~~~~~~~~~+~ 
\frac{28483291}{166698000} \ell + \frac{8}{105} \zeta(3) \ell 
+ \frac{426019}{4762800} \ell^2 + \frac{341}{34020} \ell^3 \right) C_A
\right. \right. \nonumber \\
&& \left. \left. ~~~~~~~~~~~~~~~~~+~ \left( \frac{11023}{34020} \zeta(3)
+ \frac{31}{315} \zeta(4) - \frac{10}{21} \zeta(5) 
+ \frac{95645127727}{720135360000} \right. \right. \right. \nonumber \\
&& \left. \left. \left. ~~~~~~~~~~~~~~~~~~~~~~~+~
\frac{64654361}{4000752000} \ell - \frac{256373}{9525600} \ell^2
- \frac{271}{34020} \ell^3 \right) C_F \right] a^2
\right] \nonumber \\
&& +~ O(a^3) 
\end{eqnarray}

\begin{eqnarray}
\tilde{\Pi}^{W_3,\partial W_3}_{(1)}(a) &=& d(R) \left[ -~ \frac{1}{675} 
- \frac{1}{1620} \ell ~+~ C_F \left[ \frac{1}{135} \zeta(3) 
- \frac{24571}{3499200} - \frac{1}{810} \ell \right] a \right. \nonumber \\
&& \left. ~~~~~~~+~ C_F \left[ \left( \frac{3011809}{94478400} 
- \frac{38}{1215} \zeta(3) + \frac{23999}{2624400} \ell 
\right. \right. \right. \nonumber \\ 
&& \left. \left. \left. ~~~~~~~~~~~~~~~~~~-~ \frac{4}{405} \zeta(3) \ell 
+ \frac{31}{43740} \ell^2 \right) T_F \Nf \right. \right. \nonumber \\
&& \left. \left. ~~~~~~~~~~~~~~~~~+~ \left( \frac{329}{3645} \zeta(3)
+ \frac{1}{81} \zeta(5) - \frac{36851477}{377913600}
\right. \right. \right. \nonumber \\
&& \left. \left. \left. ~~~~~~~~~~~~~~~~~~~~~~~-~ \frac{66631}{2624400} \ell 
+ \frac{11}{405} \zeta(3) \ell - \frac{341}{174960} \ell^2 \right) C_A 
\right. \right. \nonumber \\
&& \left. \left. ~~~~~~~~~~~~~~~~~+~ \left( \frac{151}{3645} \zeta(3)
- \frac{2}{27} \zeta(5) + \frac{175559}{188956800} 
\right. \right. \right. \nonumber \\
&& \left. \left. \left. ~~~~~~~~~~~~~~~~~~~~~~~-~ \frac{37657}{3499200} \ell 
- \frac{8}{3645} \ell^2 \right) C_F \right] a^2 \right] ~+~ O(a^3) \nonumber \\
\tilde{\Pi}^{W_3,\partial W_3}_{(2)}(a) &=& O(a^3) \nonumber \\
\tilde{\Pi}^{W_3,\partial W_3}_{(3)}(a) &=& d(R) \left[ \frac{26}{2025} 
+ \frac{7}{1620} \ell ~-~ C_F \left[ \frac{255937}{17496000} 
+ \frac{7}{135} \zeta(3) + \frac{239}{6075} \ell + \frac{4}{405} \ell^2 
\right] a \right. \nonumber \\
&& \left. ~~~~~~~+~ C_F \left[ \left( \frac{5571143}{2361960000}
+ \frac{1022}{3645} \zeta(3) + \frac{1622549}{13122000} \ell 
\right. \right. \right. \nonumber \\ 
&& \left. \left. \left. ~~~~~~~~~~~~~~~~~~+~ \frac{28}{405} \zeta(3) \ell 
+ \frac{9601}{218700} \ell^2 + \frac{16}{3645} \ell^3 \right) T_F \Nf 
\right. \right. \nonumber \\
&& \left. \left. ~~~~~~~~~~~~~~~~~+~ \left( \frac{170483597}{9447840000}
- \frac{332}{675} \zeta(3) + \frac{8}{135} \zeta(4) 
- \frac{7}{81} \zeta(5) \right. \right. \right. \nonumber \\
&& \left. \left. \left. ~~~~~~~~~~~~~~~~~~~~~~~-~ \frac{4693921}{13122000} \ell
- \frac{29}{405} \zeta(3) \ell - \frac{108491}{874800} \ell^2
- \frac{44}{3645} \ell^3 \right) C_A \right. \right. \nonumber \\
&& \left. \left. ~~~~~~~~~~~~~~~~~+~ \left(
\frac{2525581969}{4723920000} + \frac{14}{27} \zeta(5)
- \frac{16}{135} \zeta(4) - \frac{1598}{3645} \zeta(3) 
\right. \right. \right. \nonumber \\
&& \left. \left. \left. ~~~~~~~~~~~~~~~~~~~~~~~+~
\frac{6610253}{17496000} \ell + \frac{764}{6075} \ell^2
+ \frac{64}{3645} \ell^3 \right) C_F \right] a^2 \right] ~+~ O(a^3) 
\nonumber \\ 
\tilde{\Pi}^{W_3,\partial W_3}_{(4)}(a) &=& 
\tilde{\Pi}^{\partial W_3,\partial W_3}_{(4)}(a) \nonumber \\ 
&=& d(R) \left[ -~ \frac{1}{75} 
- \frac{1}{180} \ell ~+~ C_F \left[ \frac{17533}{486000} 
+ \frac{1}{15} \zeta(3) + \frac{473}{8100} \ell + \frac{2}{135} \ell^2 
\right] a \right. \nonumber \\
&& \left. ~~~~~~~+~ C_F \left[ -~ \left( \frac{419327}{10935000}
+ \frac{454}{1215} \zeta(3) + \frac{34633}{182250} \ell 
\right. \right. \right. \nonumber \\ 
&& \left. \left. \left. ~~~~~~~~~~~~~~~~~~~~~~~+~ \frac{4}{45} \zeta(3) \ell 
+ \frac{793}{12150} \ell^2 + \frac{8}{1215} \ell^3 \right) T_F \Nf 
\right. \right. \nonumber \\
&& \left. \left. ~~~~~~~~~~~~~~~~~+~ \left( \frac{3541817}{43740000}
+ \frac{7919}{12150} \zeta(3) - \frac{4}{45} \zeta(4) 
+ \frac{1}{9} \zeta(5) \right. \right. \right. \nonumber \\
&& \left. \left. \left. ~~~~~~~~~~~~~~~~~~~~~~~+~ \frac{399953}{729000} \ell
+ \frac{1}{15} \zeta(3) \ell + \frac{8963}{48600} \ell^2
+ \frac{22}{1215} \ell^3 \right) C_A \right. \right. \nonumber \\
&& \left. \left. ~~~~~~~~~~~~~~~~~+~ \left( \frac{1303}{2430} \zeta(3) 
- \frac{2}{3} \zeta(5) + \frac{8}{45} \zeta(4) - \frac{12235087}{16402500} 
- \frac{383653}{729000} \ell 
\right. \right. \right. \nonumber \\
&& \left. \left. \left. ~~~~~~~~~~~~~~~~~~~~~~~-~ \frac{362}{2025} \ell^2
- \frac{32}{1215} \ell^3 \right) C_F \right] a^2 \right] ~+~ O(a^3) 
\end{eqnarray}

\begin{eqnarray}
\tilde{\Pi}^{W_3,\partial \partial W_3}_{(1)}(a) &=& -~ 
\tilde{\Pi}^{W_3,\partial \partial W_3}_{(3)}(a) \nonumber \\
&=& d(R) \left[ -~ \frac{2}{675} - \frac{1}{810} \ell ~+~ 
C_F \left[ \frac{2}{135} \zeta(3) 
- \frac{6971}{349920} - \frac{1}{243} \ell \right] a \right. \nonumber \\
&& \left. ~~~~~~~+~ C_F \left[ \left( \frac{4088713}{47239200} 
- \frac{76}{1215} \zeta(3) + \frac{7267}{2624400} \ell 
\right. \right. \right. \nonumber \\ 
&& \left. \left. \left. ~~~~~~~~~~~~~~~~~~-~ \frac{8}{405} \zeta(3) \ell 
+ \frac{11}{4374} \ell^2 \right) T_F \Nf \right. \right. \nonumber \\
&& \left. \left. ~~~~~~~~~~~~~~~~~+~ \left( \frac{682}{3645} \zeta(3)
+ \frac{2}{81} \zeta(5) - \frac{48746669}{188956800}
\right. \right. \right. \nonumber \\
&& \left. \left. \left. ~~~~~~~~~~~~~~~~~~~~~~~-~ \frac{4051}{52488} \ell 
+ \frac{22}{405} \zeta(3) \ell - \frac{121}{17496} \ell^2 \right) C_A 
\right. \right. \nonumber \\
&& \left. \left. ~~~~~~~~~~~~~~~~~+~ \left( \frac{70}{729} \zeta(3)
- \frac{4}{27} \zeta(5) + \frac{2236871}{94478400} 
\right. \right. \right. \nonumber \\
&& \left. \left. \left. ~~~~~~~~~~~~~~~~~~~~~~~+~ \frac{187}{349920} \ell 
\right) C_F \right] a^2 \right] ~+~ O(a^3) \nonumber \\
\tilde{\Pi}^{W_3,\partial \partial W_3}_{(2)}(a) &=& 
\tilde{\Pi}^{W_3,\partial \partial W_3}_{(4)}(a) ~=~ O(a^3) 
\end{eqnarray}

\begin{eqnarray}
\tilde{\Pi}^{\partial W_3,\partial W_3}_{(1)}(a) &=& d(R) \left[ -~ \frac{7}{2916} 
- \frac{1}{972} \ell ~+~ C_F \left[ \frac{1}{81} \zeta(3) 
- \frac{3643}{291600} - \frac{11}{4860} \ell \right] a \right. \nonumber \\
&& \left. ~~~~~~~+~ C_F \left[ \left( \frac{9796}{164025} 
- \frac{38}{729} \zeta(3) + \frac{1961}{109350} \ell 
- \frac{4}{243} \zeta(3) \ell + \frac{11}{7290} \ell^2 \right) T_F \Nf 
\right. \right. \nonumber \\
&& \left. \left. ~~~~~~~~~~~~~~~~~+~ \left( \frac{1111}{7290} \zeta(3)
+ \frac{5}{243} \zeta(5) - \frac{237541}{1312200}
\right. \right. \right. \nonumber \\
&& \left. \left. \left. ~~~~~~~~~~~~~~~~~~~~~~~-~ \frac{21901}{437400} \ell 
+ \frac{11}{243} \zeta(3) \ell - \frac{121}{29160} \ell^2 \right) C_A 
\right. \right. \nonumber \\
&& \left. \left. ~~~~~~~~~~~~~~~~~+~ \left( \frac{169}{2430} \zeta(3)
- \frac{10}{81} \zeta(5) + \frac{54931}{3936600} 
\right. \right. \right. \nonumber \\
&& \left. \left. \left. ~~~~~~~~~~~~~~~~~~~~~~~-~ \frac{461}{48600} \ell 
- \frac{8}{3645} \ell^2 \right) C_F \right] a^2 \right] ~+~ O(a^3) \nonumber \\
\tilde{\Pi}^{\partial W_3,\partial W_3}_{(2)}(a) &=& O(a^3) \nonumber \\ 
\tilde{\Pi}^{\partial W_3,\partial W_3}_{(3)}(a) &=& d(R) \left[ \frac{1003}{72900}
+ \frac{23}{4860} \ell ~-~ C_F \left[ \frac{13351}{1458000} 
+ \frac{23}{405} \zeta(3) + \frac{931}{24300} \ell + \frac{4}{405} \ell^2 
\right] a \right. \nonumber \\
&& \left. ~~~~~~~+~ C_F \left[ \left( \frac{122}{405} \zeta(3) 
- \frac{104507}{4100625} + \frac{62801}{546750} \ell 
\right. \right. \right. \nonumber \\ 
&& \left. \left. \left. ~~~~~~~~~~~~~~~~~~+~ \frac{92}{1215} \zeta(3) \ell 
+ \frac{1571}{36450} \ell^2 + \frac{16}{3645} \ell^3 \right) T_F \Nf 
\right. \right. \nonumber \\
&& \left. \left. ~~~~~~~~~~~~~~~~~+~ \left( \frac{1110523}{10935000}
- \frac{6731}{12150} \zeta(3) + \frac{8}{135} \zeta(4) 
- \frac{23}{243} \zeta(5) \right. \right. \right. \nonumber \\
&& \left. \left. \left. ~~~~~~~~~~~~~~~~~~~~~~~-~ \frac{728341}{2187000} \ell
- \frac{109}{1215} \zeta(3) \ell - \frac{17761}{145800} \ell^2
- \frac{44}{3645} \ell^3 \right) C_A \right. \right. \nonumber \\
&& \left. \left. ~~~~~~~~~~~~~~~~~+~ \left(
\frac{51334453}{98415000} + \frac{46}{81} \zeta(5) - \frac{16}{135} \zeta(4) 
- \frac{3401}{7290} \zeta(3) + \frac{91499}{243000} \ell 
\right. \right. \right. \nonumber \\
&& \left. \left. \left. ~~~~~~~~~~~~~~~~~~~~~~~+~ \frac{764}{6075} \ell^2
+ \frac{64}{3645} \ell^3 \right) C_F \right] a^2 \right] ~+~ O(a^3) 
\end{eqnarray}

\begin{eqnarray}
\tilde{\Pi}^{\partial W_3,\partial \partial W_3}_{(1)}(a) 
&=& -~ \tilde{\Pi}^{\partial W_3,\partial \partial W_3}_{(3)}(a) ~=~
\frac{1}{2} \tilde{\Pi}^{\partial \partial W_3,\partial \partial W_3}_{(1)}(a) ~=~ 
-~ \frac{1}{2} \tilde{\Pi}^{\partial \partial W_3,\partial \partial W_3}_{(3)}(a) 
\nonumber \\
&=& d(R) \left[ -~ \frac{7}{1458} - \frac{1}{486} \ell ~-~ 
C_F \left[ \frac{59}{1944} - \frac{2}{81} \zeta(3) 
+ \frac{1}{162} \ell \right] a \right. \nonumber \\
&& \left. ~~~~~~~+~ C_F \left[ \left( \frac{3733}{26244} 
- \frac{76}{729} \zeta(3) + \frac{11}{243} \ell - \frac{8}{243} \zeta(3) \ell
+ \frac{1}{243} \ell^2 \right) T_F \Nf \right. \right. \nonumber \\
&& \left. \left. ~~~~~~~~~~~~~~~~~+~ \left( \frac{10}{243} \zeta(5)
+ \frac{227}{729} \zeta(3) - \frac{44615}{104976} \right. \right. \right. 
\nonumber \\
&& \left. \left. \left. ~~~~~~~~~~~~~~~~~~~~~~~-~ \frac{41}{324} \ell 
+ \frac{22}{243} \zeta(3) \ell - \frac{11}{972} \ell^2 \right) C_A 
\right. \right. \nonumber \\
&& \left. \left. ~~~~~~~~~~~~~~~~~+~ \left( \frac{145}{2916}
+ \frac{37}{243} \zeta(3) - \frac{20}{81} \zeta(5) 
+ \frac{1}{324} \ell \right) C_F \right] a^2 \right] ~+~ O(a^3) \nonumber \\
\tilde{\Pi}^{\partial W_3,\partial \partial W_3}_{(2)}(a) 
&=& \tilde{\Pi}^{\partial W_3,\partial \partial W_3}_{(4)}(a) ~=~ 
\tilde{\Pi}^{\partial \partial W_3,\partial \partial W_3}_{(2)}(a) 
~=~ \tilde{\Pi}^{\partial \partial W_3,\partial \partial W_3}_{(4)}(a) ~=~ O(a^3)
\end{eqnarray}

\begin{eqnarray}
\gamma^{W_3,W_3}_{(1)}(a) &=& d(R) \left[ \frac{1}{3780} ~+~ \frac{101}{59535} 
C_F a ~+~ \frac{C_F}{600112800} \left[ ( 19051200 \zeta(3) - 5588635 ) C_A 
\right. \right. \nonumber \\
&& \left. \left. ~~~~~~~+~ ( 30839406 - 38102400 \zeta(3) ) C_F 
+ 1443956 T_F \Nf \right] a^2 \right] ~+~ O(a^3) \nonumber \\ 
\gamma^{W_3,W_3}_{(2)}(a) &=& d(R) \left[ -~ \frac{1}{630} ~-~ 
\frac{139}{158760} C_F a ~+~ \frac{C_F}{800150400} \left[ ( 34630242
- 63504000 \zeta(3) ) C_A \right. \right. \nonumber \\
&& \left. \left. ~~~~~~~+~ ( 127008000 - 94674835 \zeta(3) ) C_F 
- 11046528 T_F \Nf \right] a^2 \right] ~+~ O(a^3) \nonumber \\ 
\gamma^{W_3,W_3}_{(3)}(a) &=& d(R) \left[ \frac{13}{3780} ~+~ 
\frac{13421}{793800} C_F a \right. \nonumber \\
&& \left. ~~~~~~~+~ \frac{C_F}{2000376000} \left[ ( 32330861
+ 173577600 \zeta(3) ) C_A \right. \right. \nonumber \\
&& \left. \left. ~~~~~~~+~ ( 200930804 - 347155200 \zeta(3) ) C_F 
- 7817764 T_F \Nf \right] a^2 \right] ~+~ O(a^3) \nonumber \\ 
\gamma^{W_3,W_3}_{(4)}(a) &=& d(R) \left[ -~ \frac{1}{252} ~-~ 
\frac{7649}{793800} C_F a \right. \nonumber \\
&& \left. ~~~~~~~+~ \frac{C_F}{4000752000} \left[ ( 51833522
- 393724800 \zeta(3) ) C_A \right. \right. \nonumber \\
&& \left. \left. ~~~~~~~+~ ( 787449600 \zeta(3) - 405771337 ) C_F 
- 24183088 T_F \Nf \right] a^2 \right] ~+~ O(a^3) \nonumber \\ 
\gamma^{W_3,\partial W_3}_{(1)}(a) &=& d(R) \left[ -~ \frac{1}{1620} ~-~ 
\frac{41}{29160} C_F a ~+~ \frac{C_F}{20995200} \left[ 68041 C_F 
\right. \right. \nonumber \\
&& \left. \left. ~~~~~~~-~ 40406 C_A + 13264 T_F \Nf \right] a^2 
\right] ~+~ O(a^3) \nonumber \\ 
\gamma^{W_3,\partial W_3}_{(2)}(a) &=& O(a^3) \nonumber \\ 
\gamma^{W_3,\partial W_3}_{(3)}(a) &=& d(R) \left[ \frac{7}{1620} ~+~ 
\frac{3121}{145800} C_F a ~+~ \frac{C_F}{104976000} \left[ ( 1527166
+ 12441600 \zeta(3) ) C_A \right. \right. \nonumber \\
&& \left. \left. ~~~~~~~+~ ( 16178419 - 24883200 \zeta(3) ) C_F 
- 461264 T_F \Nf \right] a^2 \right] ~+~ O(a^3) \nonumber \\ 
\gamma^{W_3,\partial W_3}_{(4)}(a) &=& d(R) \left[ -~ \frac{1}{180} ~-~ 
\frac{103}{8100} C_F a ~+~ \frac{C_F}{1458000} \left[ ( 65603 
- 259200 \zeta(3) ) C_A \right. \right. \nonumber \\
&& \left. \left. ~~~~~~~+~ ( 518400 \zeta(3) - 325498 ) C_F 
- 22612 T_F \Nf \right] a^2 \right] ~+~ O(a^3) \nonumber \\ 
\gamma^{W_3,\partial \partial W_3}_{(1)}(a) 
&=& -~ \gamma^{W_3,\partial \partial W_3}_{(3)}(a) \nonumber \\
&=& -~ d(R) \left[ \frac{1}{810} ~+~ 
\frac{13}{2916} C_F a ~+~ \frac{C_F}{2099520} \left[ 22222 C_A \right. \right.
\nonumber \\
&& \left. \left. ~~~~~~~~~~~+~ 139 C_F - 7376 T_F \Nf \right] a^2 
\right] ~+~ O(a^3) \nonumber \\ 
\gamma^{W_3,\partial \partial W_3}_{(2)}(a) 
&=& \gamma^{W_3,\partial \partial W_3}_{(4)}(a) ~=~ O(a^3) \nonumber \\ 
\gamma^{\partial W_3,\partial W_3}_{(1)}(a) &=& -~ d(R) \left[ 
\frac{1}{972} ~+~ \frac{11}{4860} C_F a \right. \nonumber \\
&&\left. ~~~~~~~~~~~+~ \frac{C_F}{874800} \left[ 3729 C_A - 4214 C_F 
- 1116 T_F \Nf \right] a^2 \right] ~+~ O(a^3) \nonumber \\ 
\gamma^{\partial W_3,\partial W_3}_{(2)}(a) &=& O(a^3) \nonumber \\
\gamma^{\partial W_3,\partial W_3}_{(3)}(a) &=& d(R) \left[ \frac{23}{4860} ~+~ 
\frac{541}{24300} C_F a ~+~ \frac{C_F}{4374000} \left[ ( 73859
+ 518400 \zeta(3) ) C_A \right. \right. \nonumber \\
&& \left. \left. ~~~~~~~+~ ( 667206 - 1036800 \zeta(3) ) C_F 
- 22036 T_F \Nf \right] a^2 \right] ~+~ O(a^3) \nonumber \\ 
\gamma^{\partial W_3,\partial W_3}_{(4)}(a) &=& d(R) \left[ -~ 
\frac{1}{180} ~-~ \frac{103}{8100} C_F a ~+~ \frac{C_F}{1458000} \left[ 
( 65603 - 259200 \zeta(3) ) C_A \right. \right. \nonumber \\
&& \left. \left. ~~~~~~~+~ ( 518400 \zeta(3) - 325498 ) C_F 
- 22612 T_F \Nf \right] a^2 \right] ~+~ O(a^3) \nonumber \\ 
\gamma^{\partial W_3,\partial \partial W_3}_{(1)}(a) 
&=& -~ \gamma^{\partial W_3,\partial \partial W_3}_{(3)}(a) ~=~ \frac{1}{2} 
\gamma^{\partial \partial W_3,\partial \partial W_3}_{(1)}(a) ~=~ -~ 
\frac{1}{2} \gamma^{\partial \partial W_3,\partial \partial W_3}_{(3)}(a) 
\nonumber \\
&=& -~ d(R) \left[ \frac{1}{486} ~+~ \frac{1}{162} C_F a ~+~ \frac{C_F}{5832} 
\left[ 89 C_A - 18 C_F - 28 T_F \Nf \right] a^2 \right] ~+~ O(a^3) 
\nonumber \\ 
\gamma^{\partial W_3,\partial \partial W_3}_{(2)}(a) &=& 
\gamma^{\partial W_3,\partial \partial W_3}_{(4)}(a) ~=~ 
\gamma^{\partial \partial W_3,\partial \partial W_3}_{(2)}(a) ~=~ 
\gamma^{\partial \partial W_3,\partial \partial W_3}_{(4)}(a) ~=~ O(a^3) ~. 
\end{eqnarray}

\subsection{Tensor-Transversity $2$.} 

\begin{equation}
\Pi^{T,T_2}_{(i)}(q) ~=~ (q^2)^2 \tilde{\Pi}_{(i)}^{T,T_2}(a)
\end{equation}

\begin{eqnarray}
\tilde{\Pi}^{T,T_2}_{(1)}(a) &=& d(R) \left[ \frac{2}{9} + \frac{1}{3} \ell ~+~ 
C_F \left[ \frac{491}{108} - 4 \zeta(3) + \frac{7}{9} \ell - \frac{1}{3} \ell^2
\right] a \right. \nonumber \\
&& \left. ~~~~~~~+~ C_F \left[ \left( \frac{512}{27} \zeta(3) 
- \frac{5336}{243} - \frac{383}{41} \ell + \frac{16}{3} \zeta(3) \ell
+ \frac{4}{9} \ell^2 + \frac{4}{27} \ell^3 \right) T_F \Nf \right. \right. 
\nonumber \\
&& \left. \left. ~~~~~~~~~~~~~~~~~+~ \left( \frac{19427}{324}
- \frac{1366}{27} \zeta(3) + \frac{7}{3} \zeta(4) - \frac{20}{3} \zeta(5) 
+ \frac{1771}{324} \ell \right. \right. \right. \nonumber \\
&& \left. \left. \left. ~~~~~~~~~~~~~~~~~~~~~~~-~ 10 \zeta(3) \ell
- \frac{10}{3} \ell^2 - \frac{11}{27} \ell^3 \right) C_A \right. \right. 
\nonumber \\
&& \left. \left. ~~~~~~~~~~~~~~~~~+~ \left( \frac{15973}{1944}
- \frac{304}{9} \zeta(3) - \frac{14}{3} \zeta(4) + 40 \zeta(5)
+ \frac{1075}{108} \ell - \frac{4}{3} \zeta(3) \ell
\right. \right. \right. \nonumber \\
&& \left. \left. \left. ~~~~~~~~~~~~~~~~~~~~~~~+~ \frac{1075}{108} \ell 
+ \frac{43}{18} \ell^2 + \frac{2}{9} \ell^3 \right) C_F \right] a^2 
\right] ~+~ O(a^3)
\end{eqnarray}

\begin{eqnarray}
\tilde{\Pi}^{T,T_2}_{(2)}(a) &=& -~ 2 \tilde{\Pi}^{T,T_2}_{(3)}(a) ~+~ O(a^3)
\nonumber \\ 
&=& d(R) \left[ \frac{20}{27} + \frac{2}{9} \ell ~+~ C_F \left[ \frac{803}{162}
- \frac{8}{3} \zeta(3) + \frac{14}{27} \ell - \frac{2}{9} \ell^2 \right] a 
\right. \nonumber \\
&& \left. ~~~~~~~+~ C_F \left[ \left( \frac{1024}{81} \zeta(3) 
- \frac{32368}{2187} - \frac{766}{243} \ell + \frac{32}{9} \zeta(3) \ell
+ \frac{8}{27} \ell^2 + \frac{8}{81} \ell^3 \right) T_F \Nf 
\right. \right. \nonumber \\
&& \left. \left. ~~~~~~~~~~~~~~~~~+~ \left( \frac{180043}{4374}
- \frac{836}{27} \zeta(3) + \frac{14}{9} \zeta(4) - \frac{40}{9} \zeta(5) 
+ \frac{1771}{486} \ell \right. \right. \right. \nonumber \\
&& \left. \left. \left. ~~~~~~~~~~~~~~~~~~~~~~~-~ \frac{20}{3} \zeta(3) \ell
- \frac{20}{9} \ell^2 - \frac{22}{81} \ell^3 \right) C_A \right. \right. 
\nonumber \\
&& \left. \left. ~~~~~~~~~~~~~~~~~+~ \left( \frac{9911}{972}
- \frac{2272}{81} \zeta(3) - \frac{28}{9} \zeta(4) + \frac{80}{3} \zeta(5) 
+ \frac{1075}{162} \ell \right. \right. \right. \nonumber \\
&& \left. \left. \left. ~~~~~~~~~~~~~~~~~~~~~~~-~ \frac{8}{9} \zeta(3) \ell
+ \frac{43}{27} \ell^2 + \frac{4}{27} \ell^3 \right) C_F \right] a^2 
\right] ~+~ O(a^3) 
\end{eqnarray}

\begin{equation}
\tilde{\Pi}^{T,T_2}_{(4)}(a) ~=~ ~+~ O(a^3) 
\end{equation} 

\begin{equation}
\tilde{\Pi}^{T,\partial T_2}_{(i)}(a) ~=~ 2 \tilde{\Pi}^{T,T_2}_{(i)}(a) ~+~ O(a^3) 
\end{equation} 

\begin{eqnarray}
\gamma^{T,T_2}_{(1)}(a) &=& d(R) \left[ \frac{1}{3} ~+~ \frac{11}{9} 
C_F a ~+~ \frac{C_F}{324} \left[ ( 1512 \zeta(3) - 1574 ) C_A 
\right. \right. \nonumber \\
&& \left. \left. ~~~~~~~+~ ( 4803 - 3024 \zeta(3) ) C_F 
+ 16 T_F \Nf \right] a^2 \right] ~+~ O(a^3) \nonumber \\ 
\gamma^{T,T_2}_{(2)}(a) &=& d(R) \left[ \frac{2}{9} ~+~ 2 C_F a ~+~ 
\frac{C_F}{486} \left[ ( 3218 + 1512 \zeta(3) ) C_A 
\right. \right. \nonumber \\
&& \left. \left. ~~~~~~~+~ ( 1203 - 3024 \zeta(3) ) C_F 
- 400 T_F \Nf \right] a^2 \right] ~+~ O(a^3) \nonumber \\ 
\gamma^{T,T_2}_{(3)}(a) &=& -~ d(R) \left[ \frac{1}{9} ~+~
C_F a ~+~ \frac{C_F}{972} \left[ ( 3218 + 1512 \zeta(3) ) C_A 
\right. \right. \nonumber \\
&& \left. \left. ~~~~~~~~~~+~ ( 1203 - 3024 \zeta(3) ) C_F 
- 400 T_F \Nf \right] a^2 \right] ~+~ O(a^3) \nonumber \\ 
\gamma^{T,T_2}_{(4)}(a) &=& O(a^3) \nonumber \\
\gamma^{T,\partial T_2}_{(1)}(a) &=& d(R) \left[ \frac{2}{3} ~+~ \frac{22}{9} 
C_F a ~+~ \frac{C_F}{162} \left[ ( 1512 \zeta(3) - 1574 ) C_A 
\right. \right. \nonumber \\
&& \left. \left. ~~~~~~~+~ ( 4803 - 3024 \zeta(3) ) C_F 
+ 16 T_F \Nf \right] a^2 \right] ~+~ O(a^3) \nonumber \\ 
\gamma^{T,\partial T_2}_{(2)}(a) &=& d(R) \left[ \frac{4}{9} ~+~ 4 C_F a ~+~ 
\frac{C_F}{243} \left[ ( 3218 + 1512 \zeta(3) ) C_A 
\right. \right. \nonumber \\
&& \left. \left. ~~~~~~~+~ ( 1203 - 3024 \zeta(3) ) C_F 
- 400 T_F \Nf \right] a^2 \right] ~+~ O(a^3) \nonumber \\ 
\gamma^{T,\partial T_2}_{(3)}(a) &=& -~ d(R) \left[ \frac{2}{9} ~+~
2 C_F a ~+~ \frac{C_F}{486} \left[ ( 3218 + 1512 \zeta(3) ) C_A 
\right. \right. \nonumber \\
&& \left. \left. ~~~~~~~~~~+~ ( 1203 - 3024 \zeta(3) ) C_F 
- 400 T_F \Nf \right] a^2 \right] ~+~ O(a^3) \nonumber \\ 
\gamma^{T,\partial T_2}_{(4)}(a) &=& O(a^3) ~.
\end{eqnarray} 

Finally, we note that in addition to the various checks we have mentioned so
far, our {\sc Form} code was written in such a way that only the Feynman rules 
for the operators and projectors needed to be input. The {\sc Mincer} 
integration code and its interface with the {\sc Qgraf} set of Feynman diagrams
forms the {\em same} central block module of the programme. In this way our 
approach was designed in order to minimize the potential places where errors 
could creep into the overall computer algebra computation. In this respect we 
have not in fact derived new Feynman rules for the parent operators $V$, $T$, 
$W_2$, $W_3$ or $T_2$ but imported those used in the progammes which underlay 
the results of \cite{21,22}. For the remaining total derivative operators it is
evident from the consistency, say, in the relations of their anomalous 
dimensions with those without the derivatives that their {\sc Form} Feynman 
rule module is not inconsistent. 

\sect{Tensor current $R$-ratio.}

Our final exercise is to derive the $R$-ratio for the tensor current to 
complete the evaluation for all the quark bilinear currents. As for $S$ and $V$
it can simply be derived from the current correlator by using  
\begin{equation}
R^T_{(i)}(a) ~=~ \frac{1}{2\pi s} \mbox{Im} \left( \Pi^T_{(i)}( - s - i
\varepsilon) \right)
\end{equation}
where $\varepsilon$ indicates the usual shift away from the real axis to avoid
ambiguity. Unlike $S$ there are two Lorentz tensor components and for the 
moment we assume there are two $R$-ratios. The case $V$ has in principle two
similar channels but due to gauge symmetry there is no contribution in the
longitudinal piece of the decomposition. With this definition and our results
(\ref{pitt1}) and (\ref{pitt2}) we find 
\begin{eqnarray} 
R^T_{(1)}(s) &=& -~ d(R) \left[ \frac{1}{3} ~+~ \left( \frac{7}{9} - \frac{2}{3}
\bar{\ell} \right) C_F a ~+~ \left( \left( \frac{16}{3} \zeta(3) 
- \frac{383}{81} - \frac{4}{27} \pi^2 + \frac{8}{9} \bar{\ell} 
+ \frac{4}{9} \bar{\ell}^2 \right) C_F \right. \right. \nonumber \\
&& \left. \left. ~~~~~~~~~~+~ \left( \frac{1771}{324} - 10 \zeta(3) 
+ \frac{11}{27} \pi^2 - \frac{20}{3} \bar{\ell} 
- \frac{11}{9} \bar{\ell}^2 \right) C_F C_A \right. \right. \nonumber \\
&& \left. \left. ~~~~~~~~~~+~ \left( \frac{1075}{108} - \frac{4}{3} \zeta(3) 
- \frac{2}{9} \pi^2 + \frac{43}{9} \bar{\ell} 
+ \frac{2}{3} \bar{\ell}^2 \right) C_F^2 \right) a^2 \right] ~+~ O(a^3) 
\nonumber \\ 
R^T_{(2)}(s) &=& d(R) \left[ \frac{2}{3} ~+~ \left( \frac{14}{9} 
- \frac{4}{3} \bar{\ell} \right) C_F a ~+~ \left( \left( \frac{32}{3} \zeta(3) 
- \frac{766}{81} - \frac{8}{27} \pi^2 + \frac{16}{9} \bar{\ell} 
+ \frac{8}{9} \bar{\ell}^2 \right) C_F \right. \right. \nonumber \\
&& \left. \left. ~~~~~~~+~ \left( \frac{1771}{162} - 20 \zeta(3) 
+ \frac{22}{27} \pi^2 - \frac{40}{3} \bar{\ell} 
- \frac{22}{9} \bar{\ell}^2 \right) C_F C_A \right. \right. \nonumber \\
&& \left. \left. ~~~~~~~+~ \left( \frac{1075}{54} - \frac{8}{3} \zeta(3) 
- \frac{4}{9} \pi^2 + \frac{86}{9} \bar{\ell} 
+ \frac{4}{3} \bar{\ell}^2 \right) C_F^2 \right) a^2 \right] ~+~ O(a^3) 
\end{eqnarray} 
where
\begin{equation}
\bar{\ell} ~=~ \ln \left( \frac{\mu^2}{s} \right) ~. 
\end{equation}
From these it is evident to see that there is a simple relationship to three
loops between both channels which is
\begin{equation}
R^T_{(2)}(s) ~=~ -~ 2 R^T_{(1)}(s) ~+~ O(a^3) ~. 
\label{rratrel}
\end{equation}
Whilst the full expressions for (\ref{pitt1}) and (\ref{pitt2}) are different
and do not satisfy an analogous relation, the behaviour of the $\ell$ terms do
which is the origin for the result (\ref{rratrel}). Consequently, if we now
include the Lorentz tensors of the projection basis we can write down the
Lorentz tensor dependence of the $R$-ratio as one would have derived it
directly from $\Pi^{T,T}_{\mu_1\mu_2\nu_1\nu_2}(q^2)$ if we had not had the
problem of focusing on scalar amplitudes in order to perform the {\sc Mincer}
calculations. Therefore, we have 
\begin{equation}
R^T_{\mu_1\mu_2\nu_1\nu_2}(s) ~=~ \left[ \tilde{P}_{\mu_1\nu_1}(q)
\tilde{P}_{\mu_2\nu_2}(q) ~-~ 
\tilde{P}_{\mu_1\nu_2}(q) \tilde{P}_{\mu_2\nu_1}(q) \right] R^T_{(1)}(s) 
\end{equation}
where we have introduced the common tensor structure 
\begin{equation}
\tilde{P}^{\mu\nu}(p) ~=~ \eta^{\mu\nu} ~-~ \frac{2p^\mu p^\nu}{p^2} ~.
\end{equation}
The appearance of this tensor structure is akin to that for case $V$ where the
longitudinal piece is absent. Put another way this form would have emerged
directly if we had chosen our Lorentz tensor basis in a more erudite fashion. 
For completeness, we have numerically evaluated the amplitude for the colour
group $SU(3)$ similar to Appendix B. We have 
\begin{eqnarray}
R^T_{(1)}(s) &=& -~ d(R) \left[ 0.333333 ~+~ \left[ 1.037037 - 0.888888 
\bar{\ell} \right] a \right. \nonumber \\
&& \left. ~~~~~~~~~~+~ \left[ 0.812771 + 0.146941 \Nf 
+ \left( -~ 18.172840 + 0.592593 \Nf \right) \bar{\ell} 
\right. \right. \nonumber \\
&& \left. \left. ~~~~~~~~~~~~~~~+~ \left( -~ 3.703704 + 0.296296 \Nf \right) 
\bar{\ell}^2 \right] a^2 \right] ~+~ O(a^3) ~.
\end{eqnarray}
For example, to see the convergence behaviour in relation to the expressions 
for $S$ and $V$ for three quark flavours when $s$~$=$~$\mu^2$ we have  
\begin{equation}
\left. \frac{}{} R^T_{(1)}(\mu^2) \right|_{\Nf=3} ~=~ -~3 \left[ 0.333333 ~+~ 
1.037037 a ~+~ 1.253594 a^2 \right] ~+~ O(a^3) ~.
\end{equation}

\sect{Discussion.}

We conclude with brief remarks since the main goal of the exercise to determine
the finite parts of various operator correlation function to $O(a^2)$ in the 
$\MSbar$ scheme has clearly been achieved. It extends the work of
\cite{26,27,28}. One novel feature was the need to properly account for the 
operator mixing into total derivative operators for the flavour non-singlet 
twist-$2$ operators used in deep inelastic scattering. The mixing matrix has 
been deduced for several low moments but to extend these to moments 
$n$~$\geq$~$4$ for arbitrary $n$ to even two loops would seem to be excluded at
this stage. For instance, the calculational machinery on a par with 
{\sc Mincer} is unfortunately not available. Whilst the main obstacle is the 
inability to disentangle the relations between counterterms one way through 
could be to embed the operators in higher leg Green's functions. Whilst this, 
in principle, will give more relations between the counterterms there is again 
the problem of lack of calculational machinery. Indeed with more legs with 
independent momenta any nullification of external momenta has the additional 
potential problem of introducing spurious infrared singularities. These would 
have to be properly treated using, say, infrared rearrangement to be confident 
in the correctness of the final counterterm relations. However, since the main
problem here was motivated by the need to provide only {\em low} moment flavour
non-singlet information for lattice computations, this is a problem which is 
left for future consideration. 
 
\vspace{1cm}
\noindent
{\bf Acknowledgements.} The author thanks Dr P.E.L. Rakow, Dr R. Horsley and
Prof. A. Vogt for valuable discussions and especially the former for a careful 
reading of the manuscript.

\appendix

\sect{Projectors.}

In this appendix we record the explicit forms of the tensors into which the
various correlation functions are decomposed. For each sector we also record 
the matrix ${\cal M}^{ij}$ used to project out each individual component of
the decomposition. The matrix ${\cal M}^{ij}_{kl}$ is derived by first 
constructing the matrix ${\cal N}^{ij}_{kl}$ where $k$ and $l$ label the
projectors, which is defined by
\begin{equation}
{\cal N}^{ij}_{kl} ~=~ 
{\cal P}^{ij}_{(k) \{ \mu_1 \ldots \mu_{n_i} | \nu_1 \ldots \nu_{n_j} \} }(q) 
{\cal P}^{ij ~ \{ \mu_1 \ldots \mu_{n_i} | \nu_1 \ldots \nu_{n_j} 
\} }_{(l)}(q) 
\end{equation} 
where there is no sum over the $i$ and $j$. The elements of this matrix are
polynomials in the dimension $d$ due to the contraction of the Lorentz indices.
Finally, ${\cal M}^{ij}$ is the inverse of ${\cal N}^{ij}$. Once 
${\cal M}^{ij}$ is specified then to project out, say, the $k$th piece of the 
tensor correlation function, one multiplies it by the projector 
\begin{equation}
\sum_{l=1}^{n_{ij}} {\cal M}^{ij}_{kl} 
{\cal P}^{ij}_{(l) \{ \mu_1 \ldots \mu_{n_i} | \nu_1 \ldots \nu_{n_j} 
\} }(q) 
\end{equation}
where there is no sum over the labels $\{ij\}$. The method we have used to 
construct the tensor basis, which of course is not unique, is to first write 
down the complete set of tensors built from the metric, $\eta_{\mu\nu}$, and 
the momentum, $q_\mu$, which have the same number of free indices as the 
operator correlation function of interest. Each of these independent tensors is
then multiplied by a different label and then the Lorentz symmetry properties 
of the two operators in the correlation function are enforced on the sum of all
independent tensors. This provides a set of linear equations for the labels 
which is fewer in number than the total number of original labels. Solving 
these equations reduces the number of independent labels, and hence independent 
combinations of the individual tensors, producing the tensor basis as 
enumerated in Table $1$. Therefore, it remains to list the relevant explicit 
expressions for the various sectors as:

\subsection{Vector-Vector.}

\begin{equation}
{\cal P}^{V,V}_{(1) \{\mu | \nu \} }(q) ~=~ \eta_{\mu\nu} ~-~  
\frac{q_\mu q_\nu}{q^2} ~~~,~~~
{\cal P}^{V,V}_{(2) \{\mu | \nu \} }(q) ~=~ \frac{q_\mu q_\nu}{q^2} 
\end{equation} 

\begin{equation}
{\cal M}^{V,V} ~=~ \frac{1}{(d-1)q^2} \left(
\begin{array}{cc}
1 & 0 \\
0 & d - 1 \\
\end{array}
\right) ~.
\end{equation} 

\subsection{Tensor-Tensor.}

\begin{eqnarray}
{\cal P}^{T,T}_{(1) \{\mu\nu | \sigma \rho \} }(q) &=& \eta_{\mu\sigma} 
\eta_{\nu\rho} ~-~ \eta_{\mu\rho} \eta_{\nu\sigma} \nonumber \\
{\cal P}^{T,T}_{(2) \{\mu\nu | \sigma \rho \} }(q) &=& 
\eta_{\mu\sigma} \frac{q_\nu q_\rho}{q^2} ~-~ 
\eta_{\mu\rho} \frac{q_\nu q_\sigma}{q^2} ~-~ 
\eta_{\nu\sigma} \frac{q_\mu q_\rho}{q^2} ~+~ 
\eta_{\nu\rho} \frac{q_\mu q_\sigma}{q^2} 
\end{eqnarray} 

\begin{equation}
{\cal M}^{T,T} ~=~ \frac{1}{4(d-1)(d-2)q^2} \left(
\begin{array}{cc}
2 & -~ 2 \\
-~ 2 & d \\
\end{array}
\right) ~.
\end{equation} 

\subsection{Vector-Wilson $2$.}

\begin{eqnarray}
{\cal P}^{V,W_2}_{(1) \{\mu | \sigma \rho \} }(q) &=& 
\left[ \eta_{\mu\sigma} q_\rho ~+~ \eta_{\mu\rho} q_\sigma ~-~
2 \frac{q_\mu q_\rho q_\sigma}{q^2} \right] \frac{1}{q^2} \nonumber \\
{\cal P}^{V,W_2}_{(2) \{\mu | \sigma \rho \} }(q) &=& 
\left[ \eta_{\sigma\rho} q_\mu ~-~ d \frac{q_\mu q_\rho q_\sigma}{q^2} \right]
\frac{1}{q^2} 
\end{eqnarray} 

\begin{equation}
{\cal M}^{V,W_2} ~=~ \frac{1}{2d(d-1)q^2} \left(
\begin{array}{cc}
d & 0 \\
0 & 2 \\
\end{array}
\right) ~.
\end{equation} 

\subsection{Vector-Wilson $3$.}

\begin{eqnarray}
{\cal P}^{V,W_3}_{(1) \{\mu | \sigma \rho \lambda \} }(q) &=& 
\eta_{\mu\sigma} \eta_{\rho\lambda} ~+~ \eta_{\mu\rho} \eta_{\sigma\lambda} ~+~
\eta_{\mu\lambda} \eta_{\rho\sigma} \nonumber \\
&& -~ (d+2) \left[ \eta_{\mu\sigma} q_\rho q_\lambda ~+~  
\eta_{\mu\rho} q_\sigma q_\lambda ~+~  
\eta_{\mu\lambda} q_\rho q_\sigma \right] \frac{1}{q^2} \nonumber \\
&& +~ 2 (d+2) \frac{q_\mu q_\sigma q_\rho q_\lambda}{(q^2)^2} \nonumber \\
{\cal P}^{V,W_3}_{(2) \{\mu | \sigma \rho \lambda \} }(q) &=& 
\left[ \eta_{\sigma\rho} q_\mu q_\lambda ~+~
\eta_{\sigma\lambda} q_\mu q_\rho ~+~ \eta_{\rho\lambda} q_\mu q_\sigma 
\right] \frac{1}{q^2} ~-~ (d+2) \frac{q_\mu q_\sigma q_\rho q_\lambda}{(q^2)^2}
\end{eqnarray} 

\begin{equation}
{\cal M}^{V,W_3} ~=~ \frac{1}{3(d-1)(d-1)(d+1)(q^2)^2} \left(
\begin{array}{cc}
1 & -~ 1 \\
-~ 1 & 3d + 4 \\
\end{array}
\right) ~.
\end{equation} 

\subsection{Wilson $2$-Wilson $2$.}

\begin{eqnarray}
{\cal P}^{W_2,W_2}_{(1) \{\mu \nu | \sigma \rho \} }(q) &=& 
\eta_{\mu\sigma} \eta_{\nu\rho} ~+~ \eta_{\mu\rho} \eta_{\nu\sigma} ~-~
\frac{2}{d} \eta_{\mu\nu} \eta_{\sigma\rho} \nonumber \\
{\cal P}^{W_2,W_2}_{(2) \{\mu \nu | \sigma \rho \} }(q) &=& 
-~ \frac{1}{d} \eta_{\mu\nu} \eta_{\sigma\rho} ~+~ \left[ \eta_{\mu\nu}
q_\sigma q_\rho ~+~ \eta_{\sigma\rho} q_\mu q_\nu \right] \frac{1}{q^2} ~-~
d \frac{q_\mu q_\nu q_\sigma q_\rho}{(q^2)^2} \nonumber \\
{\cal P}^{W_2,W_2}_{(3) \{\mu \nu | \sigma \rho \} }(q) &=& 
\left[ \eta_{\mu\sigma} q_\nu q_\rho ~+~ \eta_{\mu\rho} q_\nu q_\sigma ~+~ 
\eta_{\nu\sigma} q_\mu q_\rho ~+~ \eta_{\nu\rho} q_\mu q_\sigma \right.
\nonumber \\
&& \left. ~-~ 4 \frac{q_\mu q_\nu q_\sigma q_\rho}{q^2} \right] 
\frac{1}{q^2}
\end{eqnarray}

\begin{equation}
{\cal M}^{W_2,W_2} ~=~ \frac{1}{4(d-1)(d+1)(d-2)(q^2)^2} 
\left(
\begin{array}{ccc}
2 (d-1) & 4 & -~ 2 (d-1) \\
4 & 4 d & -~ 4 \\
-~ 2 (d-1) & -~ 4 & (d^2+d-4) \\
\end{array}
\right) ~.
\end{equation} 

\subsection{Wilson $3$-Wilson $3$.}

\begin{eqnarray}
{\cal P}^{W_3,W_3}_{(1) \{\mu \nu \sigma | \rho \lambda \psi \} }(q) &=& 
\eta_{\mu\nu} \eta_{\sigma\rho} \eta_{\lambda\psi} ~+~
\eta_{\mu\nu} \eta_{\sigma\lambda} \eta_{\rho\psi} ~+~
\eta_{\mu\nu} \eta_{\sigma\psi} \eta_{\rho\lambda} ~+~
\eta_{\mu\sigma} \eta_{\nu\rho} \eta_{\lambda\psi} ~+~
\eta_{\mu\sigma} \eta_{\nu\lambda} \eta_{\rho\psi} \nonumber \\
&& +~
\eta_{\mu\sigma} \eta_{\nu\psi} \eta_{\rho\lambda} ~+~
\eta_{\mu\rho} \eta_{\nu\sigma} \eta_{\lambda\psi} ~+~
\eta_{\mu\lambda} \eta_{\nu\sigma} \eta_{\rho\psi} ~+~
\eta_{\mu\psi} \eta_{\nu\sigma} \eta_{\rho\lambda} \nonumber \\
&& -~ \frac{(d+2)}{q^2} \left[ 
\eta_{\mu\nu} \eta_{\sigma\rho} q_\lambda q_\psi ~+~ 
\eta_{\mu\nu} \eta_{\sigma\lambda} q_\rho q_\psi ~+~ 
\eta_{\mu\nu} \eta_{\sigma\psi} q_\lambda q_\rho \right. \nonumber \\
&& \left. ~~~~~~~~~~~~~+~ 
\eta_{\mu\sigma} \eta_{\nu\rho} q_\lambda q_\psi ~+~ 
\eta_{\mu\sigma} \eta_{\nu\lambda} q_\rho q_\psi ~+~ 
\eta_{\mu\sigma} \eta_{\nu\psi} q_\lambda q_\rho \right. \nonumber \\
&& \left. ~~~~~~~~~~~~~+~ 
\eta_{\mu\rho} \eta_{\nu\sigma} q_\lambda q_\psi ~+~ 
\eta_{\mu\rho} \eta_{\lambda\psi} q_\nu q_\sigma ~+~ 
\eta_{\mu\lambda} \eta_{\nu\sigma} q_\rho q_\psi \right. \nonumber \\
&& \left. ~~~~~~~~~~~~~+~ 
\eta_{\mu\lambda} \eta_{\rho\psi} q_\nu q_\sigma ~+~ 
\eta_{\mu\psi} \eta_{\nu\sigma} q_\rho q_\lambda ~+~ 
\eta_{\mu\psi} \eta_{\rho\lambda} q_\nu q_\sigma \right. \nonumber \\
&& \left. ~~~~~~~~~~~~~+~ 
\eta_{\nu\rho} \eta_{\lambda\psi} q_\mu q_\sigma ~+~ 
\eta_{\nu\lambda} \eta_{\rho\psi} q_\mu q_\sigma ~+~ 
\eta_{\nu\psi} \eta_{\rho\lambda} q_\mu q_\sigma \right. \nonumber \\
&& \left. ~~~~~~~~~~~~~+~ 
\eta_{\sigma\rho} \eta_{\lambda\psi} q_\mu q_\nu ~+~ 
\eta_{\sigma\lambda} \eta_{\rho\psi} q_\mu q_\nu ~+~ 
\eta_{\sigma\psi} \eta_{\rho\lambda} q_\mu q_\nu \right] \nonumber \\
&& +~ \frac{2(d+2)}{(q^2)^2} \left[ 
\eta_{\mu\nu} q_\sigma q_\rho q_\lambda q_\psi ~+~ 
\eta_{\mu\sigma} q_\nu q_\rho q_\lambda q_\psi ~+~ 
\eta_{\nu\sigma} q_\mu q_\rho q_\lambda q_\psi \right. \nonumber \\
&& \left. ~~~~~~~~~~~~~~~+~ 
\eta_{\rho\lambda} q_\mu q_\nu q_\sigma q_\psi ~+~ 
\eta_{\rho\psi} q_\mu q_\nu q_\sigma q_\lambda ~+~ 
\eta_{\lambda\psi} q_\mu q_\nu q_\sigma q_\rho \right] \nonumber \\
&& +~ \frac{(d+2)^2}{(q^2)^2} \left[ 
\eta_{\mu\rho} q_\nu q_\sigma q_\lambda q_\psi ~+~ 
\eta_{\mu\lambda} q_\nu q_\sigma q_\rho q_\psi ~+~ 
\eta_{\mu\psi} q_\nu q_\sigma q_\rho q_\lambda \right. \nonumber \\
&& \left. ~~~~~~~~~~~~~~~+~ 
\eta_{\nu\rho} q_\mu q_\sigma q_\lambda q_\psi ~+~ 
\eta_{\nu\lambda} q_\mu q_\sigma q_\rho q_\psi ~+~ 
\eta_{\nu\psi} q_\mu q_\sigma q_\rho q_\lambda \right. \nonumber \\
&& \left. ~~~~~~~~~~~~~~~+~ 
\eta_{\sigma\rho} q_\mu q_\nu q_\lambda q_\psi ~+~ 
\eta_{\sigma\lambda} q_\mu q_\nu q_\rho q_\psi ~+~ 
\eta_{\sigma\psi} q_\mu q_\nu q_\rho q_\lambda \right] \nonumber \\
&& -~ \frac{8(d+2)^2}{(q^2)^3} q_\mu q_\nu q_\sigma q_\rho q_\lambda q_\psi 
\nonumber \\ 
{\cal P}^{W_3,W_3}_{(2) \{\mu \nu \sigma | \rho \lambda \psi \} }(q) &=& 
\eta_{\mu\rho} \eta_{\nu\lambda} \eta_{\sigma\psi} ~+~
\eta_{\mu\rho} \eta_{\nu\psi} \eta_{\sigma\lambda} ~+~
\eta_{\mu\lambda} \eta_{\nu\rho} \eta_{\sigma\psi} \nonumber \\
&& +~
\eta_{\mu\lambda} \eta_{\nu\psi} \eta_{\sigma\rho} ~+~
\eta_{\mu\psi} \eta_{\nu\rho} \eta_{\sigma\lambda} ~+~
\eta_{\mu\psi} \eta_{\nu\lambda} \eta_{\sigma\rho} \nonumber \\ 
&& -~ \frac{2}{q^2} \left[ 
\eta_{\mu\nu} \eta_{\sigma\rho} q_\lambda q_\psi ~+~ 
\eta_{\mu\nu} \eta_{\sigma\lambda} q_\rho q_\psi ~+~ 
\eta_{\mu\nu} \eta_{\sigma\psi} q_\lambda q_\rho \right. \nonumber \\
&& \left. ~~~~~~~~~+~ 
\eta_{\mu\sigma} \eta_{\nu\rho} q_\lambda q_\psi ~+~ 
\eta_{\mu\sigma} \eta_{\nu\lambda} q_\rho q_\psi ~+~ 
\eta_{\mu\sigma} \eta_{\nu\psi} q_\lambda q_\rho \right. \nonumber \\
&& \left. ~~~~~~~~~+~ 
\eta_{\mu\rho} \eta_{\nu\sigma} q_\lambda q_\psi ~+~ 
\eta_{\mu\rho} \eta_{\lambda\psi} q_\nu q_\sigma ~+~ 
\eta_{\mu\lambda} \eta_{\nu\sigma} q_\rho q_\psi \right. \nonumber \\
&& \left. ~~~~~~~~~+~ 
\eta_{\mu\lambda} \eta_{\rho\psi} q_\nu q_\sigma ~+~ 
\eta_{\mu\psi} \eta_{\nu\sigma} q_\rho q_\lambda ~+~ 
\eta_{\mu\psi} \eta_{\rho\lambda} q_\nu q_\sigma \right. \nonumber \\
&& \left. ~~~~~~~~~+~ 
\eta_{\nu\rho} \eta_{\lambda\psi} q_\mu q_\sigma ~+~ 
\eta_{\nu\lambda} \eta_{\rho\psi} q_\mu q_\sigma ~+~ 
\eta_{\nu\psi} \eta_{\rho\lambda} q_\mu q_\sigma \right. \nonumber \\
&& \left. ~~~~~~~~~+~ 
\eta_{\sigma\rho} \eta_{\lambda\psi} q_\mu q_\nu ~+~ 
\eta_{\sigma\lambda} \eta_{\rho\psi} q_\mu q_\nu ~+~ 
\eta_{\sigma\psi} \eta_{\rho\lambda} q_\mu q_\nu \right] \nonumber \\
&& +~ \frac{4}{(q^2)^2} \left[ 
\eta_{\mu\nu} q_\sigma q_\rho q_\lambda q_\psi ~+~ 
\eta_{\mu\sigma} q_\nu q_\rho q_\lambda q_\psi ~+~ 
\eta_{\nu\sigma} q_\mu q_\rho q_\lambda q_\psi \right. \nonumber \\
&& \left. ~~~~~~~~~~~~+~ 
\eta_{\rho\lambda} q_\mu q_\nu q_\sigma q_\psi ~+~ 
\eta_{\rho\psi} q_\mu q_\nu q_\sigma q_\lambda ~+~ 
\eta_{\lambda\psi} q_\mu q_\nu q_\sigma q_\rho \right] \nonumber \\
&& +~ \frac{2(d+2)}{(q^2)^2} \left[ 
\eta_{\mu\rho} q_\nu q_\sigma q_\lambda q_\psi ~+~ 
\eta_{\mu\lambda} q_\nu q_\sigma q_\rho q_\psi ~+~ 
\eta_{\mu\psi} q_\nu q_\sigma q_\rho q_\lambda \right. \nonumber \\
&& \left. ~~~~~~~~~~~~~~~+~ 
\eta_{\nu\rho} q_\mu q_\sigma q_\lambda q_\psi ~+~ 
\eta_{\nu\lambda} q_\mu q_\sigma q_\rho q_\psi ~+~ 
\eta_{\nu\psi} q_\mu q_\sigma q_\rho q_\lambda \right. \nonumber \\
&& \left. ~~~~~~~~~~~~~~~+~ 
\eta_{\sigma\rho} q_\mu q_\nu q_\lambda q_\psi ~+~ 
\eta_{\sigma\lambda} q_\mu q_\nu q_\rho q_\psi ~+~ 
\eta_{\sigma\psi} q_\mu q_\nu q_\rho q_\lambda \right] \nonumber \\
&& -~ \frac{16(d+2)}{(q^2)^3} q_\mu q_\nu q_\sigma q_\rho q_\lambda q_\psi 
\nonumber \\ 
{\cal P}^{W_3,W_3}_{(3) \{\mu \nu \sigma | \rho \lambda \psi \} }(q) &=& 
\frac{1}{q^2} \left[ 
\eta_{\mu\nu} \eta_{\rho\lambda} q_\sigma q_\psi ~+~ 
\eta_{\mu\nu} \eta_{\rho\psi} q_\sigma q_\lambda ~+~ 
\eta_{\mu\nu} \eta_{\lambda\psi} q_\sigma q_\rho \right. \nonumber \\
&& \left. ~~~~~+~ 
\eta_{\mu\sigma} \eta_{\rho\lambda} q_\nu q_\psi ~+~ 
\eta_{\mu\sigma} \eta_{\rho\psi} q_\nu q_\lambda ~+~ 
\eta_{\mu\sigma} \eta_{\lambda\psi} q_\nu q_\rho \right. \nonumber \\
&& \left. ~~~~~+~ 
\eta_{\nu\sigma} \eta_{\rho\lambda} q_\mu q_\psi ~+~ 
\eta_{\nu\sigma} \eta_{\rho\psi} q_\mu q_\lambda ~+~ 
\eta_{\nu\sigma} \eta_{\lambda\psi} q_\mu q_\rho \right] \nonumber \\
&& -~ \frac{(d+2)}{(q^2)^2} \left[ 
\eta_{\mu\nu} q_\sigma q_\rho q_\lambda q_\psi ~+~ 
\eta_{\mu\sigma} q_\nu q_\rho q_\lambda q_\psi ~+~ 
\eta_{\nu\sigma} q_\mu q_\rho q_\lambda q_\psi \right. \nonumber \\
&& \left. ~~~~~~~~~~~~~~+~ 
\eta_{\rho\lambda} q_\mu q_\nu q_\sigma q_\psi ~+~ 
\eta_{\rho\psi} q_\mu q_\nu q_\sigma q_\lambda ~+~ 
\eta_{\lambda\psi} q_\mu q_\nu q_\sigma q_\rho \right] \nonumber \\
&& +~ \frac{(d+2)^2}{(q^2)^3} q_\mu q_\nu q_\sigma q_\rho q_\lambda q_\psi 
\nonumber \\ 
{\cal P}^{W_3,W_3}_{(4) \{\mu \nu \sigma | \rho \lambda \psi \} }(q) &=& 
\frac{1}{q^2} \left[ 
\eta_{\mu\rho} \eta_{\nu\lambda} q_\sigma q_\psi ~+~ 
\eta_{\mu\rho} \eta_{\nu\psi} q_\sigma q_\lambda ~+~ 
\eta_{\mu\rho} \eta_{\sigma\lambda} q_\nu q_\psi \right. \nonumber \\
&& \left. ~~~~~+~ 
\eta_{\mu\rho} \eta_{\sigma\psi} q_\nu q_\lambda ~+~ 
\eta_{\mu\lambda} \eta_{\nu\rho} q_\sigma q_\psi ~+~ 
\eta_{\mu\lambda} \eta_{\nu\psi} q_\sigma q_\rho \right. \nonumber \\
&& \left. ~~~~~+~ 
\eta_{\mu\lambda} \eta_{\sigma\rho} q_\nu q_\psi ~+~ 
\eta_{\mu\lambda} \eta_{\sigma\psi} q_\nu q_\rho ~+~ 
\eta_{\mu\psi} \eta_{\nu\rho} q_\sigma q_\lambda \right. \nonumber \\
&& \left. ~~~~~+~ 
\eta_{\mu\psi} \eta_{\nu\lambda} q_\sigma q_\rho ~+~ 
\eta_{\mu\psi} \eta_{\sigma\rho} q_\nu q_\lambda ~+~ 
\eta_{\mu\psi} \eta_{\sigma\lambda} q_\nu q_\rho \right. \nonumber \\
&& \left. ~~~~~+~ 
\eta_{\nu\rho} \eta_{\sigma\lambda} q_\mu q_\psi ~+~ 
\eta_{\nu\rho} \eta_{\sigma\psi} q_\mu q_\lambda ~+~ 
\eta_{\nu\lambda} \eta_{\sigma\rho} q_\mu q_\psi \right. \nonumber \\
&& \left. ~~~~~+~ 
\eta_{\nu\lambda} \eta_{\sigma\psi} q_\mu q_\rho ~+~ 
\eta_{\nu\psi} \eta_{\sigma\rho} q_\mu q_\lambda ~+~ 
\eta_{\nu\psi} \eta_{\sigma\lambda} q_\mu q_\rho \right] \nonumber \\  
&& -~ \frac{2}{(q^2)^2} \left[ 
\eta_{\mu\nu} q_\sigma q_\rho q_\lambda q_\psi ~+~ 
\eta_{\mu\sigma} q_\nu q_\rho q_\lambda q_\psi ~+~ 
\eta_{\nu\sigma} q_\mu q_\rho q_\lambda q_\psi \right. \nonumber \\
&& \left. ~~~~~~~~~~~~+~ 
\eta_{\rho\lambda} q_\mu q_\nu q_\sigma q_\psi ~+~ 
\eta_{\rho\psi} q_\mu q_\nu q_\sigma q_\lambda ~+~ 
\eta_{\lambda\psi} q_\mu q_\nu q_\sigma q_\rho \right] \nonumber \\
&& -~ \frac{4}{(q^2)^2} \left[ 
\eta_{\mu\rho} q_\nu q_\sigma q_\lambda q_\psi ~+~ 
\eta_{\mu\lambda} q_\nu q_\sigma q_\rho q_\psi ~+~ 
\eta_{\mu\psi} q_\nu q_\sigma q_\rho q_\lambda \right. \nonumber \\
&& \left. ~~~~~~~~~~~~+~ 
\eta_{\nu\rho} q_\mu q_\sigma q_\lambda q_\psi ~+~ 
\eta_{\nu\lambda} q_\mu q_\sigma q_\rho q_\psi ~+~ 
\eta_{\nu\psi} q_\mu q_\sigma q_\rho q_\lambda \right. \nonumber \\
&& \left. ~~~~~~~~~~~~+~ 
\eta_{\sigma\rho} q_\mu q_\nu q_\lambda q_\psi ~+~ 
\eta_{\sigma\lambda} q_\mu q_\nu q_\rho q_\psi ~+~ 
\eta_{\sigma\psi} q_\mu q_\nu q_\rho q_\lambda \right] \nonumber \\
&& +~ \frac{2(d+14)}{(q^2)^3} q_\mu q_\nu q_\sigma q_\rho q_\lambda q_\psi 
\end{eqnarray}

\begin{eqnarray}
{\cal M}^{W_3,W_3} &=& \frac{1}{18(d^2-1)(d-2)(d+2)^2(d+3)(q^2)^3} \nonumber \\
&& \times \! \! 
\left(
\begin{array}{cccc}
2 (7d+18) & -\, 6 (d+2)^2 & -\, 2 (7d+18) & 6 (d+2)^2 \\
-\, 6(d+2)^2 & 3 (d+1) (d+2)^2 & 6 (d+2)^2 & -\, 3 (d+1) (d+2)^2 \! \! \\
-\, 2 (7d+18) & 6 (d+2)^2 & 2 (11d^2+50d+48) & -\, 2 (d+6) (d+2)^2 \! \! \\
6 (d+2)^2 & -\, 3 (d+1) (d+2)^2 & -\, 2 (d+6) (d+2)^2 & d (d+5) (d+2)^2 \\
\end{array}
\right) \nonumber \\ 
\end{eqnarray} 

\subsection{Tensor-Transversity $2$.} 

\begin{eqnarray}
{\cal P}^{T,T_2}_{(1) \{\mu \nu | \sigma \rho \lambda \} }(q) &=& 
\eta_{\mu\sigma} \eta_{\nu\rho} q_\lambda ~+~ 
\eta_{\mu\sigma} \eta_{\nu\lambda} q_\rho ~-~ 
\eta_{\mu\rho} \eta_{\nu\sigma} q_\lambda ~-~ 
\eta_{\mu\lambda} \eta_{\nu\sigma} q_\rho \nonumber \\
&& +~ \left[ \eta_{\mu\rho} q_\nu q_\sigma q_\lambda ~+~  
\eta_{\mu\lambda} q_\nu q_\sigma q_\rho ~-~  
\eta_{\nu\rho} q_\mu q_\sigma q_\lambda ~-~  
\eta_{\nu\lambda} q_\mu q_\sigma q_\rho \right. \nonumber \\
&& \left. ~~~~+~ 2 \eta_{\nu\sigma} q_\mu q_\rho q_\lambda ~-~  
2 \eta_{\mu\sigma} q_\nu q_\rho q_\lambda \right] \frac{1}{q^2} \nonumber \\ 
{\cal P}^{T,T_2}_{(2) \{\mu \nu | \sigma \rho \lambda \} }(q) &=& 
\eta_{\mu\sigma} \eta_{\rho\lambda} q_\nu ~-~ 
\eta_{\nu\sigma} \eta_{\rho\lambda} q_\mu \nonumber \\
&& +~ \left[  
\eta_{\nu\rho} q_\mu q_\sigma q_\lambda ~+~ 
\eta_{\nu\lambda} q_\mu q_\sigma q_\rho ~-~
\eta_{\mu\rho} q_\nu q_\sigma q_\lambda ~-~
\eta_{\mu\lambda} q_\nu q_\sigma q_\rho \right. \nonumber \\
&& \left. ~~~~+~
d \eta_{\nu\sigma} q_\mu q_\rho q_\lambda ~-~
d \eta_{\mu\sigma} q_\nu q_\rho q_\lambda \right] \frac{1}{q^2} \nonumber \\
{\cal P}^{T,T_2}_{(3) \{\mu \nu | \sigma \rho \lambda \} }(q) &=& 
\eta_{\mu\rho} \eta_{\sigma\lambda} q_\nu ~+~ 
\eta_{\mu\lambda} \eta_{\sigma\rho} q_\nu ~-~ 
\eta_{\nu\rho} \eta_{\sigma\lambda} q_\nu ~-~ 
\eta_{\nu\lambda} \eta_{\sigma\rho} q_\mu \nonumber \\
&& +~ \left[ (d+1) \left( 
\eta_{\nu\rho} q_\mu q_\sigma q_\lambda ~+~ 
\eta_{\nu\lambda} q_\mu q_\sigma q_\rho \, - \,
\eta_{\mu\rho} q_\nu q_\sigma q_\lambda ~-~
\eta_{\mu\lambda} q_\nu q_\sigma q_\rho \right) \right. \nonumber \\
&& \left. ~~~~+~ 
2 \eta_{\nu\sigma} q_\mu q_\rho q_\lambda ~-~
2 \eta_{\mu\sigma} q_\nu q_\rho q_\lambda \right] \frac{1}{q^2} \nonumber \\
{\cal P}^{T,T_2}_{(4) \{\mu \nu | \sigma \rho \lambda \} }(q) &=& 
\eta_{\mu\rho} \eta_{\nu\lambda} q_\sigma ~-~ 
\eta_{\mu\lambda} \eta_{\nu\rho} q_\sigma \nonumber \\
&& +~ \left[
\eta_{\nu\rho} q_\mu q_\sigma q_\lambda ~+~
\eta_{\nu\lambda} q_\mu q_\sigma q_\rho ~-~
\eta_{\mu\rho} q_\nu q_\sigma q_\lambda ~-~
\eta_{\mu\lambda} q_\nu q_\sigma q_\rho \right] \frac{1}{q^2} 
\end{eqnarray} 

\begin{equation}
{\cal M}^{T,T_2} ~=~ \frac{1}{4d^2(d^2-1)(d-2)(q^2)^2} 
\left(
\begin{array}{cccc}
d^2 (d+1) & 0 & 0 & 0 \\
0 & 2 (d^2+4) & -~ 4 d & 4 (d-2) \\
0 & -~ 4 d & d^2 & -~ 2 d (d-2) \\
0 & 4 (d-2) & -~ 2 d (d-2) & 2 (d-1) (d^2-4) \\
\end{array}
\right) ~.
\end{equation}

\sect{Expressions for $SU(3)$.}

For completeness and for practical use, we record the explicit numerical values
of the various amplitudes for the colour group $SU(3)$. We take the usual 
values for the Casimirs, $T_F$~$=$~$\frac{1}{2}$, $C_A$~$=$~$3$ and 
$C_F$~$=$~$\frac{4}{3}$ as well as $d(R)$~$=$~$3$ but leave the numbers of 
quarks unfixed. We only record those amplitudes which are non-zero. The 
remaining ones still satisfy the same relations to third order which were noted
in Section $4$. Thus, we have  

\subsection{Vector currents.} 

\begin{eqnarray}
\Pi^{S,S}(a) &=& 
3 \left[ 4.000000 + 2.000000 \ell ~+~ \left[ 48.867512 + 45.333333 \ell 
+ 8.000000 \ell^2 \right] a \right. \nonumber \\
&& \left. ~~~+~ \left[ 1925.894130 - 100.119646 \Nf 
+ \left( 1650.138715 - 61.022786 \Nf \right) \ell \right. \right. \nonumber \\
&& \left. \left. ~~~~~~~~+~ \left( 565.333333 - 19.555555 \Nf \right) 
\ell^2 \right. \right. \nonumber \\
&& \left. \left. ~~~~~~~~+~ \left( 50.666667 - 1.777778 \Nf \right) \ell^3 
\right] a^2 \right] ~+~ O(a^3) \nonumber \\
\Pi^{V,V}_{(1)}(a) &=& 
3 \left[ -~ 2.222222 - 1.333333 \ell ~+~ \left[ 1.199436 - 5.333333 \ell 
\right] a \right. \nonumber \\
&& \left. ~~~+~ \left[ 6.784729 \Nf - 38.534112 + \left( 2.459635 \Nf
- 42.361758 \right) \ell \right. \right. \nonumber \\
&& \left. \left. ~~~~~~~~+~ \left( 1.777778 \Nf - 29.333333 \right) \ell^2 
\right] a^2 \right] ~+~ O(a^3) \nonumber \\
\Pi^{T,T}_{(1)}(a) &=& 
3 \left[ -~ 0.444444 - 0.666667 \ell ~+~ \left[ 0.698484 - 2.074074 \ell 
+ 0.888889 \ell^2 \right] a \right. \nonumber \\
&& \left. ~~~+~ \left[ 27.577316 - 1.114284 \Nf 
+ \left( 22.743851 - 2.243433 \Nf \right) \ell \right. \right. \nonumber \\
&& \left. \left. ~~~~~~~~+~ \left( 18.172840 - 0.592593 \Nf \right) \ell^2 
\right. \right. \nonumber \\
&& \left. \left. ~~~~~~~~+~ \left( 2.469136 - 0.197531 \Nf \right) \ell^3 
\right] a^2 \right] ~+~ O(a^3) \nonumber \\
\Pi^{T,T}_{(2)}(a) &=& 
3 \left[ 2.222222 + 1.333333 \ell ~+~ \left[ 3.640070 + 4.148148 \ell 
- 1.777778 \ell^2 \right] a \right. \nonumber \\
&& \left. ~~~+~ \left[ -~ 32.988175 + 2.272463 \Nf 
+ \left( -~ 45.487702 + 4.486867 \Nf \right) \ell \right. \right. \nonumber \\
&& \left. \left. ~~~~~~~~+~ \left( -~ 36.345679 + 1.185185 \Nf \right) 
\ell^2 \right. \right. \nonumber \\
&& \left. \left. ~~~~~~~~+~ \left( -~ 4.938272 + 0.395062 \Nf \right) 
\ell^3 \right] a^2 \right] ~+~ O(a^3) 
\end{eqnarray}  

\subsection{Wilson moment $n$~$=$~$2$.}

\begin{eqnarray}  
\Pi^{W_2,W_2}_{(1)}(a) &=& 
3 \left[ 0.480000 + 0.200000 \ell ~+~ \left[ -~ 5.578236 - 2.802963 \ell 
- 0.711111 \ell^2 \right] a \right. \nonumber \\
&& \left. ~~~+~ \left[ -~ 88.800839 + 11.700261 \Nf 
+ \left( -~ 56.861337 + 7.125112 \Nf \right) \ell \right. \right. \nonumber \\
&& \left. \left. ~~~~~~~~+~ \left( -~ 15.116049 + 1.566420 \Nf \right) 
\ell^2 \right. \right. \nonumber \\
&& \left. \left. ~~~~~~~~+~ \left( -~ 0.921811 + 0.158025 \Nf \right) 
\ell^3 \right] a^2 \right] ~+~ O(a^3) \nonumber \\
\Pi^{W_2,W_2}_{(2)}(a) &=& 
3 \left[ 0.408889 + 0.133333 \ell ~+~ \left[ -~ 3.597014 - 1.947654 \ell 
- 0.474074 \ell^2 \right] a \right. \nonumber \\
&& \left. ~~~+~ \left[ -~ 57.728474 + 8.008477 \Nf 
+ \left( -~ 39.368555 + 4.896687 \Nf \right) \ell \right. \right. \nonumber \\
&& \left. \left. ~~~~~~~~+~ \left( -~ 10.231001 + 1.070617 \Nf \right) 
\ell^2 
\right. \right. \nonumber \\
&& \left. \left. ~~~~~~~~+~ \left( -~ 0.614540 + 0.105350 \Nf \right) 
\ell^3 \right] a^2 \right] ~+~ O(a^3) \nonumber \\
\Pi^{W_2,W_2}_{(3)}(a) &=& 
3 \left[ 0.075555 + 0.133333 \ell ~+~ \left[ 4.238377 + 3.780741 \ell 
+ 0.711111 \ell^2 \right] a \right. \nonumber \\
&& \left. ~~~+~ \left[ 75.027286 - 10.992767 \Nf 
+ \left( 56.696221 - 6.724712 \Nf \right) \ell \right. \right. \nonumber \\
&& \left. \left. ~~~~~~~~+~ \left( 21.758025 - 1.892346 \Nf \right) \ell^2 
\right. \right. \nonumber \\
&& \left. \left. ~~~~~~~~+~ \left( 0.921811 - 0.158025 \Nf \right) \ell^3 
\right] a^2 \right] ~+~ O(a^3) \nonumber \\
\Pi^{W_2,\partial W_2}_{(3)}(a) &=& 
3 \left[ 1.111111 + 0.666667 \ell ~+~ \left[ -~ 0.599718 + 2.666667 \ell 
\right] a \right. \nonumber \\
&& \left. ~~~+~ \left[ 19.267056 - 3.392365 \Nf 
+ \left( 21.180879 - 1.229818 \Nf \right) \ell \right. \right. \nonumber \\
&& \left. \left. ~~~~~~~~+~ \left( 14.666667 - 0.888889 \Nf \right) \ell^2 
\right] a^2 \right] ~+~ O(a^3)
\end{eqnarray}

\subsection{Wilson moment $n$~$=$~$3$.}

\begin{eqnarray}  
\Pi^{V,W_3}_{(1)}(a) &=& 
3 \left[ 0.045926 + 0.022222 \ell ~+~ \left[ 0.020544 + 0.098765 \ell 
\right] a \right. \nonumber \\
&& \left. ~~~+~ \left[ 0.917043 - 0.131672 \Nf 
+ \left( 0.838448 - 0.047350 \Nf \right) \ell \right. \right. \nonumber \\
&& \left. \left. ~~~~~~~~+~ \left( 0.497942 - 0.030178 \Nf \right) \ell^2 
\right] a^2 \right] ~+~ O(a^3) \nonumber \\
\Pi^{V,\partial W_3}_{(1)}(a) &=& 
3 \left[ 0.074074 + 0.037037 \ell ~+~ \left[ -~ 0.008626 + 0.148148 \ell 
\right] a \right. \nonumber \\
&& \left. ~~~+~ \left[ 1.211681 - 0.196695 \Nf 
+ \left( 1.176715 - 0.068323 \Nf \right) \ell \right. \right. \nonumber \\
&& \left. \left. ~~~~~~~~+~ \left( 0.814815 - 0.049383 \Nf \right) \ell^2 
\right] a^2 \right] ~+~ O(a^3) 
\end{eqnarray}

\begin{eqnarray}  
\Pi^{W_3,W_3}_{(1)}(a) &=& 
3 \left[ 0.001151 + 0.000265 \ell ~+~ \left[ -~ 0.037289 - 0.020408 \ell 
- 0.003527 \ell^2 \right] a \right. \nonumber \\
&& \left. ~~~+~ \left[ -~ 0.344878 + 0.084926 \Nf 
+ \left( -~ 0.203993 + 0.046931 \Nf \right) \ell \right. \right. \nonumber \\
&& \left. \left. ~~~~~~~~+~ \left( -~ 0.050228 + 0.009964 \Nf \right) 
\ell^2 \right. \right. \nonumber \\
&& \left. \left. ~~~~~~~~+~ \left( 0.000131 + 0.000784 \Nf \right) 
\ell^3 \right] a^2 \right] ~+~ O(a^3) \nonumber \\
\Pi^{W_3,W_3}_{(2)}(a) &=& 
3 \left[ -~ 0.004528 - 0.001587 \ell ~+~ \left[ 0.100640 + 0.049139 \ell 
+ 0.008818 \ell^2 \right] a \right. \nonumber \\
&& \left. ~~~+~ \left[ 1.022184 - 0.224027 \Nf 
+ \left( 0.551061 - 0.122691 \Nf \right) \ell \right. \right. \nonumber \\
&& \left. \left. ~~~~~~~~+~ \left( 0.109786 - 0.024512 \Nf \right) 
\ell^2 \right. \right. \nonumber \\
&& \left. \left. ~~~~~~~~+~ \left( -~ 0.000327 - 0.001960 \Nf \right) 
\ell^3 \right] a^2 \right] ~+~ O(a^3) \nonumber \\
\Pi^{W_3,W_3}_{(3)}(a) &=& 
3 \left[ 0.010207 + 0.003439 \ell ~+~ \left[ -~ 0.069547 - 0.035555 \ell 
- 0.009641 \ell^2 \right] a \right. \nonumber \\
&& \left. ~~~+~ \left[ -~ 1.354970 + 0.153442 \Nf 
+ \left( -~ 0.921737 + 0.095085 \Nf \right) \ell \right. \right. \nonumber \\
&& \left. \left. ~~~~~~~~+~ \left( -~ 0.233863 + 0.020396 \Nf \right) 
\ell^2 \right. \right. \nonumber \\
&& \left. \left. ~~~~~~~~+~ \left( -~ 0.017201 + 0.002143 \Nf \right) 
\ell^3 \right] a^2 \right] ~+~ O(a^3) \nonumber \\
\Pi^{W_3,W_3}_{(4)}(a) &=& 
3 \left[ -~ 0.008806 - 0.003968 \ell ~+~ \left[ 0.063950 + 0.031660 \ell 
+ 0.010935 \ell^2 \right] a \right. \nonumber \\
&& \left. ~~~+~ \left[ 1.580679 - 0.124987 \Nf 
+ \left( 1.078542 - 0.084549 \Nf \right) \ell \right. \right. \nonumber \\
&& \left. \left. ~~~~~~~~+~ \left( 0.309941 - 0.019979 \Nf \right) 
\ell^2 \right. \right. \nonumber \\
&& \left. \left. ~~~~~~~~+~ \left( 0.025932 - 0.002430 \Nf \right) 
\ell^3 \right] a^2 \right] ~+~ O(a^3) \nonumber \\
\Pi^{W_3,\partial W_3}_{(1)}(a) &=& 
3 \left[ -~ 0.001481 - 0.000617 \ell ~+~ \left[ 0.002510 - 0.001641 \ell 
\right] a \right. \nonumber \\
&& \left. ~~~+~ \left[ 0.048778 - 0.003811 \Nf 
+ \left( 0.009906 - 0.001818 \Nf \right) \ell \right. \right. \nonumber \\
&& \left. \left. ~~~~~~~~+~ \left( -~ 0.011698 + 0.000472 \Nf \right) 
\ell^2 \right] a^2 \right] ~+~ O(a^3) \nonumber \\
\Pi^{W_3,\partial W_3}_{(3)}(a) &=& 
3 \left[ 0.012840 + 0.004321 \ell ~+~ \left[ -~ 0.102610 - 0.052455 \ell 
- 0.013169 \ell^2 \right] a \right. \nonumber \\
&& \left. ~~~+~ \left[ -~ 1.653254 + 0.226264 \Nf 
+ \left( -~ 1.103477 + 0.137837 \Nf \right) \ell \right. \right. \nonumber \\
&& \left. \left. ~~~~~~~~+~ \left( -~ 0.272497 + 0.029267 \Nf \right) 
\ell^2 \right. \right. \nonumber \\
&& \left. \left. ~~~~~~~~+~ \left( -~ 0.017071 + 0.002926 \Nf \right) 
\ell^3 \right] a^2 \right] ~+~ O(a^3) \nonumber \\
\Pi^{W_3,\partial W_3}_{(4)}(a) &=& 
3 \left[ -~ 0.013333 - 0.005555 \ell ~+~ \left[ 0.154951 + 0.077860 \ell 
+ 0.019753 \ell^2 \right] a \right. \nonumber \\
&& \left. ~~~+~ \left[ 2.466690 - 0.325007 \Nf 
+ \left( 1.579482 - 0.197920 \Nf \right) \ell \right. \right. \nonumber \\
&& \left. \left. ~~~~~~~~+~ \left( 0.419890 - 0.043512 \Nf \right) 
\ell^2 \right. \right. \nonumber \\
&& \left. \left. ~~~~~~~~+~ \left( 0.025606 - 0.004390 \Nf \right) 
\ell^3 \right] a^2 \right] ~+~ O(a^3) \nonumber \\
\Pi^{W_3,\partial \partial W_3}_{(1)}(a) &=& 
3 \left[ -~ 0.002963 - 0.001235 \ell ~+~ \left[ -~ 0.002132 - 0.005487 \ell 
\right] a \right. \nonumber \\
&& \left. ~~~+~ \left[ -~ 0.055664 + 0.007575 \Nf 
+ \left( -~ 0.046580 + 0.002631 \Nf \right) \ell \right. \right. \nonumber \\
&& \left. \left. ~~~~~~~~+~ \left( -~ 0.027663 + 0.001677 \Nf \right) 
\ell^2 \right] a^2 \right] ~+~ O(a^3) \nonumber \\
\Pi^{\partial W_3,\partial W_3}_{(1)}(a) &=& 
3 \left[ -~ 0.002401 - 0.001029 \ell ~+~ \left[ 0.003129 - 0.003018 \ell 
\right] a \right. \nonumber \\
&& \left. ~~~+~ \left[ 0.039866 - 0.001957 \Nf 
+ \left( 0.000510 - 0.001236 \Nf \right) \ell \right. \right. \nonumber \\
&& \left. \left. ~~~~~~~~+~ \left( -~ 0.020450 + 0.001006 \Nf \right) 
\ell^2 \right] a^2 \right] ~+~ O(a^3) \nonumber \\
\Pi^{\partial W_3,\partial W_3}_{(3)}(a) &=& 
3 \left[ 0.013759 + 0.004733 \ell ~+~ \left[ -~ 0.103229 - 0.051084 \ell 
- 0.013169 \ell^2 \right] a \right. \nonumber \\
&& \left. ~~~+~ \left[ -~ 1.644342 + 0.224410 \Nf 
+ \left( -~ 1.094081 + 0.137255 \Nf \right) \ell \right. \right. \nonumber \\
&& \left. \left. ~~~~~~~~+~ \left( -~ 0.263695 + 0.028733 \Nf \right) 
\ell^2 \right. \right. \nonumber \\
&& \left. \left. ~~~~~~~~+~ \left( -~ 0.017071 + 0.002926 \Nf \right) 
\ell^3 \right] a^2 \right] ~+~ O(a^3) \nonumber \\
\Pi^{\partial W_3,\partial \partial W_3}_{(1)}(a) &=& 
3 \left[ -~ 0.004801 - 0.002058 \ell ~+~ \left[ -~ 0.000893 - 0.008230 \ell 
\right] a \right. \nonumber \\
&& \left. ~~~+~ \left[ -~ 0.073488 + 0.011283 \Nf 
+ \left( -~ 0.065373 + 0.003796 \Nf \right) \ell \right. \right. \nonumber \\
&& \left. \left. ~~~~~~~~+~ \left( -~ 0.045267 + 0.002743 \Nf \right) 
\ell^2 \right] a^2 \right] ~+~ O(a^3) 
\end{eqnarray}  

\subsection{Transversity moment $n$~$=$~$2$.}

\begin{eqnarray}  
\Pi^{T,T_2}_{(1)}(a) &=& 
3 \left[ 0.222222 + 0.333333 \ell ~+~ \left[ -~ 0.349242 + 1.037037 \ell 
- 0.444444 \ell^2 \right] a \right. \nonumber \\
&& \left. ~~~+~ \left[ -~ 13.788658 + 0.557142 \Nf 
+ \left( -~ 11.371925 + 1.121717 \Nf \right) \ell \right. \right. \nonumber \\
&& \left. \left. ~~~~~~~~+~ \left( -~ 9.086420 + 0.296296 \Nf \right) 
\ell^2 \right. \right. \nonumber \\
&& \left. \left. ~~~~~~~~+~ \left( -~ 1.234568 + 0.098765 \Nf \right) 
\ell^3 \right] a^2 \right] ~+~ O(a^3) \nonumber \\
\Pi^{T,T_2}_{(2)}(a) &=& 
3 \left[ 0.740741 + 0.222222 \ell ~+~ \left[ 2.335073 + 0.691358 \ell 
- 0.296296 \ell^2 \right] a \right. \nonumber \\
&& \left. ~~~+~ \left[ 5.429324 + 0.264127 \Nf 
+ \left( -~ 7.581284 + 0.747811 \Nf \right) \ell \right. \right. \nonumber \\
&& \left. \left. ~~~~~~~~+~ \left( -~ 6.057613 + 0.197531 \Nf \right) 
\ell^2 \right. \right. \nonumber \\
&& \left. \left. ~~~~~~~~+~ \left( -~ 0.823045 + 0.065844 \Nf \right) 
\ell^3 \right] a^2 \right] ~+~ O(a^3) ~.
\end{eqnarray}

\end{document}